\documentclass[aps,pra,superscriptaddress,twocolumn,longbibliography,tightenlines,notitlepage,10pt,floatfix,nofootinbib]{revtex4-1}
\RequirePackage{silence}
\RequirePackage{float}
\usepackage[paperwidth=210mm,paperheight=297mm,centering,hmargin=1cm,vmargin=2cm]{geometry}
\setcitestyle{compressed}
\usepackage[utf8]{inputenc}
\usepackage[USenglish]{babel}
\usepackage[T1]{fontenc}
\usepackage{graphicx,epsf}
\usepackage{subcaption}
\usepackage[toc,page,header]{appendix}
\usepackage{dsfont}

\usepackage{color}
\usepackage{breakurl}
\usepackage[bookmarksnumbered,hypertexnames=false,colorlinks,colorlinks=true,linkcolor=blue,urlcolor=black,citecolor=blue,anchorcolor=green]{hyperref}
\usepackage{bbm,braket,microtype,mathrsfs,amsmath,amssymb,color,amsthm,mathtools,fullpage,enumitem,ae,aecompl,bm,thm-restate}
\usepackage{tikz}
\usepackage{nameref}
\usepackage[font=small,labelfont=bf,
   justification=justified,
   format=plain]{caption}
\usepackage{color}
\usepackage[capitalize]{cleveref}
\usepackage{mathtools}

\allowdisplaybreaks
\newcommand{\be}{\begin{equation}}
\newcommand{\ee}{\end{equation}}
\newcommand{\ba}{\begin{equation}\begin{aligned}\hspace{0pt}}

\newcommand{\ea}{\end{aligned}\end{equation}}
\DeclareMathOperator{\tr}{tr}

\newcommand{\ignore}[1]{}
\newcommand{\st}[1]{\ket{#1}\!\!\bra{#1}}
\newcommand{\otoc}[0]{\text{OTOC}}
\newcommand{\p}[0]{\prime}

\newcommand{\expf}[1]{\mathrm{exp}\left ( {#1}\right )}

\newcommand{\poly}{\operatorname{poly}}

\newcommand{\epr}[0]{\operatorname{EPR}}

\def\CC{{\rm\kern.24em \vrule width.04em height1.46ex depth-.07ex
   \kern-.29em C}}
\def\P{{\rm I\kern-.25em P}}
\def\RR{{\rm
        \vrule width.04em height1.58ex depth-.0ex
        \kern-.04em R}}
\def\bbbone{\mathds{1}}
\def\bbbc{{\mathchoice {\setbox0=\hbox{$\displaystyle\rm C$}\hbox{\hbox
to0pt{\kern0.4\wd0\vrule height0.9\ht0\hss}\box0}}
{\setbox0=\hbox{$\textstyle\rm C$}\hbox{\hbox
to0pt{\kern0.4\wd0\vrule height0.9\ht0\hss}\box0}}
{\setbox0=\hbox{$\scriptstyle\rm C$}\hbox{\hbox
to0pt{\kern0.4\wd0\vrule height0.9\ht0\hss}\box0}}
{\setbox0=\hbox{$\scriptscriptstyle\rm C$}\hbox{\hbox
to0pt{\kern0.4\wd0\vrule height0.9\ht0\hss}\box0}}}}
\def\bbbz{{\mathchoice {\hbox{$\sf\textstyle Z\kern-0.4em Z$}}
{\hbox{$\sf\textstyle Z\kern-0.4em Z$}}
{\hbox{$\sf\scriptstyle Z\kern-0.3em Z$}}
{\hbox{$\sf\scriptscriptstyle Z\kern-0.2em Z$}}}}

\newlength{\fighskip} \fighskip=2pt
\newlength{\figvskip} \figvskip=1pt

\makeatletter
\def\namedlabel#1#2{\begingroup
   \def\@currentlabel{#2}%
   \label{#1}\endgroup
}
\makeatother
\newcommand\constructosum[3]{%
    \begin{tikzpicture}[baseline=(char.base), inner sep=0, outer sep=0]
        \draw (#1,0) circle (#2); 
        \node (char) at (0,0) {$#3\sum$}; 
    \end{tikzpicture}%
}

\newcommand{\modtwosum}{\mathop{\mathchoice
        {\constructosum{-0.3ex}{0.1}{\displaystyle}}
        {\constructosum{-0.3ex}{0.06}{\textstyle}}
        {\constructosum{-0.2ex}{0.05}{\scriptstyle}}
        {\constructosum{-0.15ex}{0.03}{\scriptscriptstyle}}
    }\displaylimits
}

\usepackage{hyperref}
\newtheorem{theorem}{Theorem}

\newtheorem{corollary}{Corollary}
\newtheorem{lemma}{Lemma}

\newtheorem{example}{Example}
\newtheorem{remark}{Remark}

\newtheorem{definition}{Definition}

\makeatletter
\def\namedlabel#1#2{\begingroup
   \def\@currentlabel{#2}%
   \label{#1}\endgroup
}
\makeatother
\newcommand{\sectionMain}[1]{
\let\oldaddcontentsline\addcontentsline
\renewcommand{\addcontentsline}[3]{}
\section{#1}
\let\addcontentsline\oldaddcontentsline
}
\begin{document}
\setcounter{secnumdepth}{3}
\title{Learning efficient decoders for quasichaotic quantum scramblers}

\author{Lorenzo Leone}\email{lorenzo.leone001@umb.edu}
\affiliation{Physics Department,  University of Massachusetts Boston, Massachusetts  02125, USA}
\affiliation{Theoretical Division (T-4), Los Alamos National Laboratory, Los Alamos, New Mexico 87545, USA}
\affiliation{Center for Nonlinear Studies, Los Alamos National Laboratory, Los Alamos, New Mexico 87545, USA}

\author{Salvatore F.E. Oliviero}\email{s.oliviero001@umb.edu}
\affiliation{Physics Department,  University of Massachusetts Boston,  02125, USA}
\affiliation{Theoretical Division (T-4), Los Alamos National Laboratory, Los Alamos, New Mexico 87545, USA}
\affiliation{Center for Nonlinear Studies, Los Alamos National Laboratory, Los Alamos, New Mexico 87545, USA}

\author{{Seth} {Lloyd}}\email{slloyd@mit.edu}
\affiliation{Department of Mechanical Engineering, Massachusetts Institute of Technology,  {Cambridge},  {Massachusetts}, {USA}}
\affiliation{Turing Inc., Brooklyn, NY, USA}

\author{Alioscia Hamma}\email{alioscia.hamma@unina.it}

\affiliation{Dipartimento di Fisica `Ettore Pancini', Universit\`a degli Studi di Napoli Federico II,
Via Cintia 80126,  Napoli, Italy}
\affiliation{INFN, Sezione di Napoli, Italy}


\begin{abstract}
Scrambling of quantum information is an important feature at the root of randomization and benchmarking protocols, the onset of quantum chaos, and black-hole physics. 
Unscrambling this information is possible given perfect knowledge of the scrambler~[\href{https://arxiv.org/abs/1710.03363}{arXiv:1710.03363}.]. 
We show that one can retrieve the scrambled information even without any previous knowledge of the scrambler, by a {\em learning} algorithm that allows the building of an efficient decoder. Remarkably, the decoder is classical in the sense that it can be efficiently represented on a classical computer as a Clifford operator. It is striking that a classical decoder can retrieve with fidelity one all the information scrambled by a random unitary that {\em cannot} be efficiently simulated on a classical computer, as long as there is no full-fledged quantum chaos. This result shows that one can learn the salient properties of quantum unitaries in a classical form, and sheds a new light on the meaning of quantum chaos. Furthermore, we obtain results concerning the algebraic structure of $t$-doped Clifford circuits, i.e., Clifford circuits containing t non-Clifford gates, their gate complexity, and learnability that are of independent interest. In particular, we show that a $t$-doped Clifford circuit $U_t$ can be decomposed into two Clifford circuits $U_{0},U^{\prime}_0$ that sandwich a local unitary operator $u_t$, i.e., $U_t=U_{0} u_{t}U_{0}^{\prime}$. The local unitary operator $u_t$ contains $t$ non-Clifford gates and acts nontrivially on at most $t$ qubits. As simple corollaries, the gate complexity of the $t$-doped Clifford circuit $U_t$ is $O(n^2+t^3)$, and it admits a efficient process tomography using $\poly(n,2^t)$ resources.

\end{abstract}
\maketitle
\setcounter{MaxMatrixCols}{20}
 \setlength{\intextsep}{0.8pt}
 \setlength{\abovedisplayskip}{4pt}
\setlength{\belowdisplayskip}{4pt}
\setlength{\abovedisplayshortskip}{0pt}
\setlength{\belowdisplayshortskip}{4pt}

\section{Introduction}

In quantum mechanics, learning an unknown quantum state or  process is a crucial problem. The applications of this task range from quantum information and benchmarking protocols~\cite{knill2008RandomizedBenchmarkingQuantum,wallman2014RandomizedBenchmarkingConfidence,roth2018RecoveringQuantumGates}, the understanding of quantum chaos~\cite{hosur2016ChaosQuantumChannels,cotler2018OutoftimeorderOperatorsButterfly,xu2020DoesScramblingEqual}, quantum chemistry~\cite{peruzzo2014VariationalEigenvalueSolver,mcclean2016TheoryVariationalHybrid,cao2019QuantumChemistryAge,huggins2021EfficientNoiseResilient, bauer2020QuantumAlgorithmsQuantum}, quantum cryptography~\cite{fuchs1999CryptographicDistinguishabilityMeasures,reichardt2013ClassicalCommandQuantum,mills2017InformationTheoreticallySecure,fitzsimons2017PrivateQuantumComputation,fitzsimons2017UnconditionallyVerifiableBlind,coladangelo2019VerifieronaLeashNewSchemes, mahadev2018ClassicalVerificationQuantuma,gheorghiu2019VerificationQuantumComputation,gheorghiu2019ComputationallysecureComposableRemote,supic2020SelftestingQuantumSystems,gisin2002QuantumCryptography}, and black-hole physics~\cite{hayden2007BlackHolesMirrors, yoshida2017EfficientDecodingHaydenPreskill}. 


If the quantum process to be investigated is modeled by a random unitary, its learning may prove an extremely daunting task. If one were able to learn a random unitary, one could do wonders: for example, to decode the information emitted in Hawking radiation without any previous knowledge of the black hole. In this paper, by learning, we mean learning enough features of the process so that scrambled information can be retrieved from it.

There is a special class of unitary operators, the Clifford group, that has been proven to be efficiently learnable with a polynomial effort~\cite{low2009LearningTestingAlgorithms,lai2022LearningQuantumCircuits}. It is not a coincidence that this class of unitary operators is the same that can be efficiently simulated by a classical computer~\cite{gottesman1998HeisenbergRepresentationQuantum}. From a quantum advantage point of view, the ability to learn only those unitaries that can be classically simulated is unsatisfactory. One wants to learn those quantum processes that {\em cannot} be efficiently simulated classically. Is it possible? How costly is it? 

In this paper, as well as in the companion Letter~\cite{oliviero2022BlackHoleComplexity}, we show that it is possible to learn a random unitary operator that cannot be efficiently simulated, as long as this process is not {\em fully} chaotic. Moreover, the learned features can be efficiently represented on a classical computer. The fact that something that cannot be efficiently simulated can then be learned in a form that is efficiently represented is so surprising that it almost sounds contradictory. It seems that then, after all, the random unitary {\em could} be efficiently simulated. Our result must be understood in terms of what we are learning. We are learning just the features that are enough to unscramble the information. The result is still very surprising, but at least it starts sounding more believable: the complex features are useless, and what is useful is efficiently representable. Moreover, the efficient representation is not \textit{always} possible: when quantum chaos kicks in, such a representation breaks down. At that point, in order to unscramble information, one also has to learn all the complex features of the unitary, and that requires an exponentially complex representation. In some sense, our result clarifies what quantum chaos is:  that feature of quantum evolutions that does not allow for any kind of classical representation~\cite{leone2021QuantumChaosQuantum}.

This work is organized as follows: in Sec.~\ref{overview} we present a more detailed overview of the problem and of our results, eschewing the heavy technical details; in Sec.~\ref{sec: review}, we briefly review previous results concerning the task of learning quantum processes; in Secs.~\ref{sec: scrambling} and~\ref{Sec: recoveryalgoYK}, we review the information scrambling setup introduced in Ref.~\cite{hayden2007BlackHolesMirrors} and the meaning of learning quantum information scrambled after a complex quantum dynamics; in Sec.~\ref{Sec: mainresult} we present the main results of the paper in a nontechnical fashion, while in Sec.~\ref{sec:stab}, after having introduced the technical tools needed for the proof of the main theorems, we present the quantum algorithm able to learn scrambled information.

\section{Overview  of the problem and results}\label{overview}

The notion of learning a quantum evolution is intimately connected with the notion of irreversibility. Quantum mechanics is unitary; one can, in principle, undo any quantum evolution by running it backward. Unfortunately, without any prior knowledge of the unitary $U$, the ability to undo a quantum evolution is almost never guaranteed. If one could learn $U$ by query accesses then one would be able to revert quantum evolutions. However, for the overwhelming majority of unitaries, the task of learning is exponentially hard~\cite{chuang1997PrescriptionExperimentalDetermination,childs2001RealizationQuantumProcess,altepeter2003AncillaAssistedQuantumProcess,mohseni2008QuantumprocessTomographyResource,merkel2013SelfconsistentQuantumProcess} and effectively reversibility is lost. This fact is related to the exponential growth of the Hilbert-space dimension with the number of degrees of freedom, which in turn would require exponentially small precision per exponentially many experiments.

As we mentioned above, there are some special quantum processes that do not feature complex behavior: Clifford circuits can be efficiently learned and simulated. On the opposite end of Clifford circuits, there are chaotic quantum circuits. These circuits can be obtained by random Clifford circuits on $n$ qubits with the addition of $c\,n$ ($c \ge 2$) non-Clifford resources, e.g., T-gates. They feature universal frame potentials~\cite{roberts2017ChaosComplexityDesign}, fluctuations of entanglement and of the higher-order out-of-time-order correlation functions (OTOCs)~\cite{leone2021QuantumChaosQuantum,leone2021IsospectralTwirlingQuantum,oliviero2021RandomMatrixTheory}. In view of the Gottesman-Knill theorem~\cite{gottesman1998HeisenbergRepresentationQuantum}, they also require an exponential number of resources to be simulated on a classical computer.

\begin{figure*}[t]
    \centering
    \includegraphics[width=\textwidth]{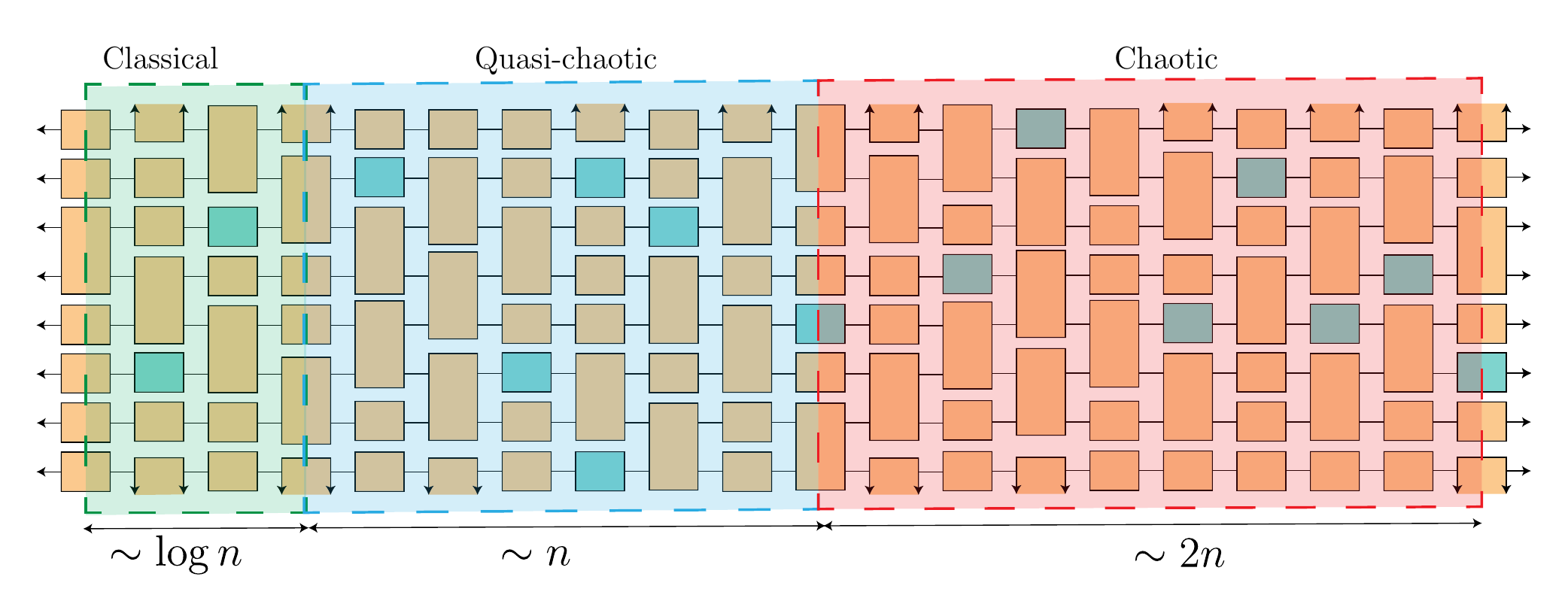}
    \caption{Sketch of a $t$-doped Clifford circuit with variable depth proportional to the doping $t$ counting the number of injected non-Clifford gates (blue). The first part of the circuit (green-dashed box) has depth within $\log n$ and corresponds to a  \textit{classical} regime, where the circuit can be efficiently simulated by a classical computer. As the depth reaches $t\sim n$ (blue-dashed box), the circuit reaches a quasichaotic regime, where classical simulation is exponentially hard but, at the same time, it cannot reproduce the universal properties of Haar-random circuits. However, quasichaotic circuits are learnable with an exponential effort in $t$. Quantum chaos is reached for $t\sim 2n$ (red-dashed box): then universal properties are reached and no learning is possible.}
    \label{fig1}
\end{figure*}

In this paper, we discuss the learnability for a wide class of unitary evolutions, that is, unitary operators obtained from a random Clifford circuit enriched with $t$ non-Clifford gates -  the so-called $t-$doped Clifford circuits $U_t$, see Fig.~\ref{fig1}. 
As shown in Ref.~\cite{leone2021QuantumChaosQuantum}, there is a gradual transition from Clifford circuits to quantum chaos. In the middle of the transition, that is, for a number of Clifford resources $c\,n$ with density $c<1$, one has not yet attained quantum chaos, although these circuits do require an exponential number of resources to be simulated. We call these circuits quasichaotic. 

As we shall see in Sec.~\ref{Sec: tdopedcircuits}, we constructively show that every $t$-doped Clifford circuit can be decomposed as
\be
U_{t}=U_{0}[\bbbone_{[n-t]}\otimes u_{[t]}]U_{0}^{\prime}\,,
\label{decompositionclifford}
\ee
i.e., as a product of two Clifford circuits $U_{0}, U_{0}^{\prime}$ and a local unitary $[\bbbone_{[n-t]}\otimes u_{[t]}]$ acting on at most $t$ qubits and containing $t$ non-Clifford gates, see Theorem~\ref{th2} (\textit{Compression Theorem}). The decomposition in Eq.~\eqref{decompositionclifford} is valid as long as $c<1$, i.e., for quasichaotic quantum circuits. It states that all the non-Cliffordness in $U_t$ can be \textit{compressed} in $t$ qubits only. Moreover, all the Clifford parts of $U_t$ can be learned by having query access to $U_t$. Here and throughout the work, we refer to \textit{query access} as the ability to perform the unitary transformation $U_t$ followed by a measurement on a quantum register consisting of $n$ qubits. We present an algorithm that learns the Clifford operations $U_{0}$ and $U_{0}^{\prime}$ by $\poly(n,2^t)$ query accesses to $U_t$, see Theorem~\ref{th3}. As a corollary, we show that with time complexity and query complexity both scaling as $\poly(n,2^t)$, it is possible to learn a full tomographic description of a general $t$-doped Clifford circuit, see Corollary~\ref{cor3}. Additionally, a straightforward consequence of Eq.~\eqref{decompositionclifford}, a $t$-doped stabilizer state $\ket{\psi_t}\equiv U_{t}\ket{0}^{\otimes n}$ can be \textit{compressed} as
\be
\ket{\psi_t}=U_{0}[\ket{0}^{\otimes (n-t)}\otimes \ket{\phi}_{t}]\label{decompositionstatedopedeq2}
\ee
i.e., to the computational basis state $\ket{0}^{\otimes (n-t)}$ and a nonstabilizer state $\ket{\phi}_t$ living on a $t$-qubit subsystem (see Corollary~\ref{Cor:dopedstabilizerstatedecomposition}). Again, a decomposition as in Eq.~\eqref{decompositionstatedopedeq2} is valid for $c<1$. What is more, the decomposition in Eq.~\eqref{decompositionclifford} shows that the gate complexity $\#(U_t)$---i.e., the minimum number of gates necessary to build $U_t$ from the identity~\cite{brown2018SecondLawQuantum}--- of Clifford$+T$ circuits obeys  $\#(U_t)=O(n^2+t^{3})$, that interpolates between $O(n^{2})$ for $t=o(n)$ to $O(n^3)$ for $t=\Omega(n)$, see Corollary~\ref{cor2}. Remarkably, Clifford circuits doped with no more than a logarithmic number of $T$ gates have the same gate complexity as Clifford circuits.

In this paper, we discuss the learnability of quantum evolutions in the context of information scrambling~\cite{hayden2007BlackHolesMirrors}: we present a quantum algorithm based on a constrained random Clifford Completion (CC)  that aims at learning a perfect and efficient decoder without any previous knowledge of the scrambler. The decoder is efficient in the sense that it can be efficiently represented on a classical computer, more precisely, it is a Clifford operator~\cite{gottesman1998HeisenbergRepresentationQuantum}. We show that such learning is efficient in terms of resources as long as the scrambler can be efficiently simulated, it is exponentially expensive in a quasichaotic regime and becomes impossible in the presence of full-fledged quantum chaos. 

The main result of the paper is the following: if ${U}_t$ is a  unitary obtained by a $t$-doped Clifford circuit, one can learn a {\em Clifford}  decoder ${V}$ by means of a probabilistic quantum algorithm based on a constrained random Clifford Completion algorithm, which employs 
$\poly(n,2^t)$ query accesses to ${U}_t$ and a time $\poly(n,2^t)$. 

The scrambler is a unitary $U_{t}^{AB}$ acting on the $|A|$ qubits of information  plus additional $|B|$ qubits with $n=|A|+|B|$. The output of the scrambler consists of $|C|+|D|=n$ qubits of which only $|D|$ can be accessed by the decoder. The decoder $V$ takes in the qubits in $D$ and returns an output that should contain the information initially present in $A$ with a fidelity :
\be\label{mainF}
\mathcal{F}_{V}(U_t) \ge \frac{1}{1+2^{2|A|+t-2|D|}}\,,
\ee
while the probability  of learning the decoder $V$ is 
\ba\label{mainP}
\mathscr{P}(V) \ge 1-2^{t-2(n-|D|)}\,.
\ea
Equations~\eqref{mainF} and \eqref{mainP} are the content of Theorem~\ref{th1}. It is important to highlight that the proposed CC algorithm searches for and implements a decoder $V$ belonging to the Clifford group. This means that the decoder, once (and if) found, can be efficiently represented in a classical computer: it is, in this sense, a {\em classical} decoder. In the companion Letter~\cite{oliviero2022BlackHoleComplexity}, we present the result of Theorem~\ref{th1} in a simplified fashion.

  When can such a decoder be found? If the scrambler $U_t$ is a  Clifford circuit (or Clifford circuits with doping up to $t\sim\log n$), the decoder exists and it can be learned by $\poly(n,2^t)$ resources. If the doping scales like  $t=cn$ with $c<1$, learning is still possible, but it requires exponential resources, as for simulability. We call this regime quasichaotic. As the density of non-Clifford resources increases, a transition to full quantum chaos is approached~\cite{leone2021QuantumChaosQuantum,oliviero2021TransitionsEntanglementComplexity} and, for $c\ge 2$, no learning is any longer possible, no matter the resources employed. 

As one can see from Eq.~\eqref{mainF}, as the number $t$ of $T$ gates in the circuit $U_{t}$ increases, the size of the subsystem $D$ the decoder should access in order to unscramble the information and reconstruct the state $\ket{\psi_A}$ must increase as well. Notably, the decoding via a Clifford operator breaks down only when the number $t$ of $T$ gates approaches $n$, $t\sim n$, i.e., in the quasichaotic regime (cfr. Fig.~\ref{fig1}). In this case, to obtain perfect fidelity, one has to acquire a number of qubits $|D|$ larger than half of the system, $|D|\ge n/2$, which in turn makes the probability of decoding scale as $\sim 1-2^{n-2|D|}$. As one increases $t$ beyond this threshold, the probability of decoding decays exponentially.  In other words, Clifford operations can unscramble  information as long as the dynamic is quasichaotic $(c<1)$ 
while unscrambling becomes impossible for fully chaotic dynamics $(c\ge 1)$.

\section{Review of previous results}\label{sec: review}
In this section, we provide a review of some known results on the problem of learning unitary dynamics and explain the advances of the present paper in the current literature.


The task of learning a unitary operator $U$ - defined on $n$ qubits - can be generally defined in two ways: $(i)$ either by the ability to learn and synthesize $U$ on a quantum device, or $(ii)$ by the ability to learn some problem-depending properties of $U$ and use them to solve a particular task. The latter includes tasks like disentangling a given quantum state~\cite{yang2017EntanglementComplexityQuantum,chamon2014EmergentIrreversibilityEntanglement}, unscrambling quantum information~\cite{leone2022RetrievingInformationBlack}, or learning expectation values. Intuitively, $(ii)$ is a weaker form of learning, and, as a matter of fact, $(i)$ implies $(ii)$ (up to computational challenges). Both approaches have been widely explored in the literature~\cite{holzapfel2015ScalableReconstructionUnitary,khatri2019QuantumassistedQuantumCompiling,tibbetts2012ExploringTradeoffFidelity,volkoff2021UniversalCompilingNo,chamon2014EmergentIrreversibilityEntanglement,marvian2016UniversalQuantumEmulator,yang2017EntanglementComplexityQuantum,true2022TransitionsEntanglementComplexity,leone2022RetrievingInformationBlack,piemontese2022EntanglementComplexityRokhsarKivelsonsign}. 


Naively, one can define the task of learning a unitary operator $U$ by knowing all the matrix elements $\braket{x|U|x^{\prime}}$ of $U$, where $\ket{x}, \ket{x^{\prime}}$ are basis vectors. The above task is immediately found to be inefficient because $O(2^{2n})$ matrix elements need to be determined and stored in a classical memory. The weaker requirement of applying an unknown unitary on a quantum state is more suitable: in Ref.~\cite{marvian2016UniversalQuantumEmulator}, the authors show that unitaries acting on $O(\poly\log(n))$ qubits can be emulated by having access to unknown samples of input-output states. While storing matrix elements of the unknown unitary would scale superpolynomially in the number of qubits, both the runtime and the query access of their algorithm scale polynomially. 
A more efficient way to define the learning task is knowing an efficient decomposition in elementary gates of a given unitary operator $U$. It is well known that a universal set of gates is given by CNOT, Hadamard gate that allows superposition in the computational basis, $S$ gate (a $\hat{z}$-rotation of $\pi/2$), and finally $T$-gate (a $\hat{z}$-rotation of $\pi/4$). The Solovay-Kitaev theorem ensures that any unitary operator can be approximated by an ordered product of elementary gates within any desired accuracy $\epsilon$~\cite{kitaev1997QuantumComputationsAlgorithms,nielsen2000QuantumComputationQuantum}. While learning the right order of gates to approximate a target unitary operator is, in general, a very hard combinatorial problem~\cite{nielsen2000QuantumComputationQuantum}, the decomposition in elementary gates allows an efficient classical representation of a quantum unitary operator. A third approach is to use the operator basis of Pauli operators. Thanks to the unitarity of $U$, only the action on $O(n)$ Pauli operators needs to be determined, but at the same time $(a)$ exponentially many measurements are typically required to learn each element of the map $P\mapsto U^{\dag}PU$ for $P$ being an element of operator basis of the Pauli group, and $(b)$ classical postprocessing requires, in general, exponentially many memory bits.

Clifford unitary operators constitute a particular example in which the latter approach is suitable. Indeed, Clifford operations map elements of the Pauli group to elements of the Pauli group. This means that it is just sufficient to learn the Pauli operators resulting from the adjoint action of $U$, $P\mapsto U^{\dag} PU\in\mathbb{P}$. The learning of Clifford circuits and states created by Clifford circuits has been studied in Refs.~\cite{low2009LearningTestingAlgorithms,montanaro2010QuantumBooleanFunctions,montanaro2017LearningStabilizerStates,lai2022LearningQuantumCircuits}. In Ref.~\cite{montanaro2017LearningStabilizerStates} it has been shown that with $O(n)$ queries to $2$ copies of a given stabilizer state $\ket{\psi}$ is possible to learn its tomographic decomposition.  While in Ref.~\cite{lai2022LearningQuantumCircuits}, generalizing the results of Refs.~\cite{low2009LearningTestingAlgorithms,montanaro2017LearningStabilizerStates}, it has been shown how $O(n)$ queries to a Clifford circuit $U$ are sufficient to learn and synthesize $U$ on a quantum computer. However, the algorithms developed in Refs.~\cite{low2009LearningTestingAlgorithms,montanaro2017LearningStabilizerStates,lai2022LearningQuantumCircuits} are specifically designed to learn a Clifford unitary, and therefore cannot be employed in the task of learning of a general unitary operator.

The task of learning quantum unitary dynamics is intimately connected to the problem of classical simulability of quantum computation. Clifford circuits admit an efficient classical representation. This means that computation made by states created by the action of Clifford unitary operators---the so-called stabilizer states---can be efficiently reproduced by a classical computer~\cite{gottesman1998HeisenbergRepresentationQuantum}. No quantum advantage can be achieved. At the same time, stabilizer states are typically highly entangled, a condition that has been believed to be the key property to unlock quantum computational advantage. 
Nevertheless, fine-grained properties in entanglement structure reveal the profound difference between the entanglement produced by Clifford circuits from that produced by universal unitary operators. Previous works~\cite{chamon2014EmergentIrreversibilityEntanglement,yang2017EntanglementComplexityQuantum,true2022TransitionsEntanglementComplexity} probed the operational difference between these two types of entanglement. It has been shown that, by employing a Montecarlo-Metropolis kind of algorithm, it is possible to completely disentangle a state evolved by Clifford gates. Conversely, the Metropolis algorithm fails at disentangling a state evolved by a universal circuit. At this level, the task of disentangling can be thought as that of finding a unitary operator $V$ that, applied on the evolved state $U\ket{\psi_0}$, makes the evolved state $VU\ket{\psi_0}$ nonentangled in a given bipartition. The success or failure of such a disentangling algorithm reveals the different entanglement structures produced by Clifford gates and universal gates. Besides connecting classical simulability and learnability of quantum dynamics, the above result naturally defines two complexity classes: (I) states that can be efficiently disentangled and (II) states that are not disentanglable.

It is noteworthy that, while it is possible to learn the classical representation of Clifford unitaries in terms of Pauli operators and to learn how to disentangle states entangled by Clifford circuits, there are other approaches to learning Clifford circuits that fail. One example is provided in the context of quantum machine learning by variational quantum algorithms  (VQAs). Compilation of a given unitary operator $U$ aims to find the minimum set of elementary gates that approximate the target $U$ (e.g., see Refs~\cite{chuang1997PrescriptionExperimentalDetermination,childs2001RealizationQuantumProcess,altepeter2003AncillaAssistedQuantumProcess,mohseni2008QuantumprocessTomographyResource,merkel2013SelfconsistentQuantumProcess,levy2021ClassicalShadowsQuantum,khatri2019QuantumassistedQuantumCompiling,volkoff2021UniversalCompilingNo}). In this context, the strategy of VQAs is to classically train a parametrized (fix depth) quantum circuit to minimize a problem-dependent expectation value $\tr[O V^{\dag}(\vec{\theta})U^{\dag}\psi_0U V(\vec{\theta})]$. Unfortunately, the above strategy fails in general: no-go theorems have been established that prevent the compilation of a unitary $U$ drawn from a unitary $k$-design with $k\ge 2$ (set of unitaries that reproduces up to the second moment of the full unitary group)~\cite{holmes2021BarrenPlateausPreclude}. It is noteworthy that the above VQA task fails even for the Clifford group, being a unitary $3$-design. At the same time, if in a VQA task, the learner accepts to spend exponentially many resources, in Ref.~\cite{kiani2020LearningUnitariesGradient} it has been shown that gradient descent can learn an arbitrary random unitary using an exponentially large parameter-landscape. Another example is provided by Probably-Approximately Correct (PAC)  learning~\cite{aaronson2007LearnabilityQuantumStates,aaronson2018ShadowTomographyQuantum,rocchetto2018StabiliserStatesAre,rocchetto2019ExperimentalLearningQuantum,gollakota2022HardnessPAClearningStabilizer}. The goal of PAC learning is to learn a function relative to a certain distribution of inputs. In the context of unitary circuits, the goal is to learn the output distribution of a set of observables $P_x$ (that can be reasonably be thought as being Pauli operators labeled by $x$) through $U$ given a set of input states $\rho_y$, i.e., $f(x,y):=\tr(P_xU\rho_y U^{\dag})$. Then, a PAC learner aims to design a function $\widetilde{f}(x,y)$ such that with probability at least $1-\delta$, obeys $\mathbb{E}_{x,y}[f(x,y)-\widetilde{f}(x,y)]^2<\epsilon$. Interestingly, Clifford unitary operators cannot be PAC learned without a collapse of polynomial hierarchies~\cite{liang2022CliffordCircuitsCan}.


After describing various results within the domain of learning Clifford circuits, let us now move beyond the confines of the Clifford group. Any Clifford circuit can be built out of $3$ elementary gates: CNOT, Hadamard, and Phase gate $S$. The addition of one non-Clifford gate makes the above set universal for quantum computation. In other words, Clifford circuits fail to be universal because of the lack of just one element, which is traditionally chosen to be the $T$ gate. The injection of non-Clifford gates into Clifford circuits gradually drives the circuits to feature universal properties~\cite{zhou2020SingleGateClifford}. This is reflected by the fact that the best-known classical simulation algorithm scales exponentially in the number of non-Clifford gates~\cite{bravyi2016ImprovedClassicalSimulation}. While for Clifford circuits the road map, of what can or cannot be learned and relative strategies, is sufficiently complete and, for universal circuits, the task of learning is believed to be unfeasible, regarding the gray area between these two "complexity classes", there are still many open questions. For example, how does the success of a learning task change for $t$-doped Clifford circuits, i.e., Clifford circuits polluted by $t$ non-Clifford gates? Does the learnability encounter a sharp transition or a continuous crossover driven by the amount $t$ of non-Cliffordness in the circuit?   

The doping of Clifford circuits is intimately connected with the concept of quantum pseudorandomness: a set of unitary operators is a unitary $k$-design if it reproduces up to the $k$-th moment of the Haar (uniform) distribution over the unitary group~\cite{divincenzo2002QuantumDataHiding,emerson2003PseudoRandomUnitaryOperators}. Clifford group has been proven to be a unitary $3$-design, and to fail to be a unitary $4$-design~\cite{zhu2016CliffordGroupFails}. In Ref.~\cite{haferkamp2020QuantumHomeopathyWorks}, one can see that a vanishing density $t/n$ of non-Clifford resources is sufficient to break the $4$-design barrier and reproduce any $k$-design (for $k<\log n$) with an error $\epsilon$. In other words, in the framework of unitary $k$-design, it is possible to \textit{homeopathically} dope Clifford circuits to obtain approximate $k$ designs within the desired accuracy $\epsilon$. Unfortunately, reproducing up to the $k$ moment of the distribution over the full unitary group within an error $\epsilon $ is not always sufficient to reproduce the complex universal behavior. In Refs.~\cite{leone2021QuantumChaosQuantum,oliviero2021TransitionsEntanglementComplexity}, it has been shown that to truly address the transition between the non-complex behavior of Clifford circuits and the complex Haar random behavior, an exponentially small error $\epsilon\sim 2^{-n}$ is required. Indeed in Ref.~\cite{roberts2017ChaosComplexityDesign}, it has been shown that a necessary and sufficient condition to form a $4$-design is to reproduce the universal value of the $8$-point OTOC
\be
\otoc_{8}(U)\!:=\!\frac{1}{d}\tr[P_1P_2(U)P_3P_4(U)P_1P_4(U)P_3P_2(U)]\,,
\label{otoc8def}
\ee
where $P_1,P_2,P_3,P_4$ are nonidentity Pauli operators, and $P_2(U)\equiv UP_2U^{\dag}$ and similarly for $P_4(U)$. The average $\otoc_{8}(U)$ for $t$-doped Clifford circuits $U\in\mathcal{C}_t$ is (proved in Ref.~\cite{leone2021QuantumChaosQuantum}):
\be
\braket{\otoc_{8}(U)}_{\mathcal{C}_t}=\Theta\left[\frac{1}{d^2}\left(\frac{3}{4}\right)^t+\frac{1}{d^4}\right]\,,
\label{transition}
\ee
that interpolates between the Clifford value $\braket{\otoc_{8}(U)}_{\mathcal{C}_0}=\Theta(d^{-2})$ and the Haar value $\braket{\otoc_{8}(U)}_{\mathcal{C}_\infty}=\Theta(d^{-4})$. As a result, the injection of $c\,n$ (with $c\ge 2$) non-Clifford gates in a Clifford circuit is both necessary and sufficient to drive the transition towards the universal behavior $\Theta(d^{-4})$. The value of the $8$ point OTOC discriminates between various regimes of interest of doped Clifford circuits. In particular, the injection of $t=\Theta(1)$ non-Clifford gates does not change at all the value of $\otoc_{8}$. The doping with $\Theta(\log n)$ non-Clifford resources---being part of the class of circuits that can be efficiently classically simulable---do not change the value of the OTOC up to a polynomial overhead, i.e., $\Theta(d^{-2}\poly^{-1}(n))$. Instead, the injection of a number $\Omega(\log n)<t< n$ of non-Clifford resources lies in the quasichaotic quantum circuit regime, i.e., a class of circuits that is transient between two universality classes (Clifford and Haar) that are non-universal but, at the same time, cannot be simulated by classical means. This transient regime is reflected by a value of the $8$-point OTOC of $\Theta(d^{-3})$ (cfr. Eq.~\eqref{transition}).

The above results thus suggest that the task of learning could, in principle, become unfeasible for universal (chaotic) circuits only, thus when the number of non-Clifford gates is $\approx 2n$.

The question about the learning of Clifford circuits polluted with $t$ non-Clifford gates has been explored in several ways. First, from the point of view of the disentangling algorithm, in Ref.~\cite{true2022TransitionsEntanglementComplexity}, it has been shown that the success of the disentangling algorithm is exponentially suppressed in the number $t$ of non-Clifford gates. In Ref.~\cite{lai2022LearningQuantumCircuits}, using techniques similar to those introduced by in Ref.~\cite{montanaro2017LearningStabilizerStates,low2009LearningTestingAlgorithms}, is proposed an efficient way to encode, learn and synthesize a particular class of $t$-doped Clifford circuit, i.e., circuit made as $U_{0}^{(1)} T_k U_{0}^{(2)}$, where $U_{0}^{(1)}$, $U_{0}^{(2)}$ are Clifford operations and $T_k$ are $k$ parallel single qubit $T$ gates. In the paper, it is also proven that the task of learning and synthesis is possible as long as the number of non-Clifford gates $t=O(\log n)$. Remarkably, this is the same threshold for a $t$-doped Clifford circuit to be efficiently simulated classically~\cite{bravyi2016ImprovedClassicalSimulation}.  Conversely, in Ref.~\cite{hinsche2022SingleGateMakes} the authors claim that, while the output distribution $P(x)=|\braket{x|U|0^{\otimes n}}|^2$ of a Clifford circuit can be efficiently learned, the injection of even a single $T$ gate in a Clifford circuit makes the task of learning the output distribution $P(x)$ hard [assuming the learning parties with noise (LPN) assumption~\cite{pietrzak2012CryptographyLearningParity}]. Their result provides a sharp separation between Clifford circuits and doped Clifford circuits, in contrast with the result previously discussed. Note that, the injection of a single $T$ gate in a Clifford circuit falls inevitably in the class of circuits that can be written as $U_{0}^{(1)} T U_{0}^{(2)}$, that can be efficiently encoded classically and learned, as shown in Ref.~\cite{lai2022LearningQuantumCircuits}. After all, it is well known the difference in performances of learning tasks with or without the possibility of having access to two copies of the target, being a unitary or a quantum state. At the same time, the question of whether the output distribution of a $1$-doped Clifford circuit can be learned when measurements in arbitrary single-qubit bases are available remains an open question.


Along these lines, in this work, the problem of learning doped Clifford circuits has been studied in the context of unscrambling quantum information. While the technicalities of the protocol will be discussed in the following sections, the concept of unscrambling is cognate to the one of disentangling: the task is to find a decoder unitary $V$ that, mocking the action of a unitary operator $U_t$, undoes the action of $U_t$ only on a subspace (say $A$) $VU_t\ket{\psi_A}\ket{\psi_B}$, retrieving quantum information $\ket{\psi_A}$ scrambled by $U_t$. In a previous work~\cite{leone2022RetrievingInformationBlack}, a Metropolis algorithm has been employed---similar to the one for disentangling---with the task of searching for the decoder $V$. By modeling the unitary $U_t$ as a $t$-doped Clifford circuit, it is numerically shown that the success rate of the algorithm, quantified by recovery fidelity $|\langle\psi_A|VU|\psi_A\otimes \psi_B\rangle|^2$, is exponentially decaying in $t$. In other words, the recovery fidelity is smaller than $\epsilon$ just after $t=\Omega(\log \epsilon^{-1})$ non-Clifford gates.

Is it possible to do better? The answer is yes, as the present paper shows. We show that the proposed CC algorithm is able to learn - with $\poly(n,2^t)$  resources - 
 a perfect decoder $V$ for a $t$-doped Clifford circuit, up to $t<n$ non-Clifford gates, as we set out to show starting from the next section. The main technical contribution, as mentioned earlier in the above section, is the development of the \textit{compression theorem} (Theorem~\ref{th2}) which reveals the existence of a compression method for $t$-doped Clifford circuits [as seen in Eq.~\eqref{decompositionclifford} above]. This compression effectively concentrates all the non-Clifford elements into a subsystem of $t$ qubits (thus independent from $n$), enabling the use of a brute-force tomographic algorithm to learn the non-Clifford components. Therefore, this task is feasible only up to $t = O(\log n)$. We then put this consideration in rigorous grounds in Corollary~\ref{cor3}, where we show that it is possible to learn an efficient classical description of a t-doped Clifford circuit using $\poly(n,2^t)$ resources, which includes both sample and computational complexity. In fact, this represents an advancement over the state-of-the-art algorithms for learning $t$-doped Clifford circuits, which had previously been constrained to specific circuit structures.
 
\section{Learning quantum information from an unknown scrambler}

\subsection{Information scrambling and decoupling theorem}\label{sec: scrambling}
In this section, we make a brief review of the decoupling theorem introduced by Hayden and Preskill in Ref.~\cite{hayden2007BlackHolesMirrors}, in the context of black-hole evaporation.  Consider the Hilbert space of $n=|A|+|B|=|C|+|D|$ qubits partitioned as 
\ba
\mathcal{H}=A\otimes {B}=C\otimes D
\ea
and a unitary map 
\ba
U_{AB}\,:\, A\otimes {B}\mapsto C\otimes D\,.
\ea
Denote as $\mathbb{P}(\Lambda)$ the Pauli group (modulo phases) on the subsystem $\Lambda$ composed of $|\Lambda|$ qubits with $\Lambda\in\{A,B,C,D\}$, $d_{\Lambda}\equiv 2^{|\Lambda|}$, and define the average four-point out-of-time-order correlation function $\Omega(U_{AB})$ as
\ba
\Omega(U_{AB})&:=\langle\otoc_4(U_{AB})\rangle_{P_A,P_D}
\\&\equiv\frac{1}{d}\braket{\tr(P_AP_{D}(U_{AB})P_AP_{D}(U_{AB}))}_{P_A,P_D}\,,
\label{otocdef}
\ea
where $P_{D}(U_{AB})\equiv U_{AB}^{\dag}P_{D}U_{AB}$ and $\braket{\cdot}_{P_A}\equiv \frac{1}{d_{A}^{2}}\sum_{P_A\in\mathbb{P}(A)}(\cdot)$ is the average over the Pauli group on $A$ and similarly for $P_D$.  The OTOC operationally quantifies how information initially encoded in $A$ is scrambled by $U$ through the output system $\mathcal{H}_{C}\otimes\mathcal{H}_{D}$, see Ref.~\cite{hosur2016ChaosQuantumChannels}. 
The function $\Omega(U_{AB})$ is a quantity related to the group commutator between the local Pauli group on $A$ and $D$: it attains the value one if the (average) support of  $P_D(U_{AB})$  commutes with Pauli operators in $A$ while it decreases as the support of $P_D$ grows in space, the so-called operator growth, which in turn defines scrambling behavior~\cite{nahum2018OperatorSpreadingRandom,khemani2018OperatorSpreadingEmergence,chamon2022QuantumStatisticalMechanics}: a unitary operator $U_{AB}$ is said to be a \textit{scrambler} if and only if~\cite{hosur2016ChaosQuantumChannels}
\be
\Omega(U_{AB})\simeq \frac{1}{d_{A}^{2}}+\frac{1}{d_{D}^{2}}-\frac{1}{d_{A}^{2}d_{D}^{2}}\,,
\label{otoc}
\ee
where $\simeq$ means \textit{up to an order $d^{-2}$}. 

Scrambling of quantum information is connected to that of the information retrieval~\cite{hayden2007BlackHolesMirrors}:  imagine Alice decides to encode some quantum information in $A$.  As this is quantum information, we need to possess a reference state on $R$ that is perfectly entangled with $A$. By denoting the EPR pair between two spaces of the same dimension $d_X$ by $\ket{XX'}=d_{X}^{-1/2}\sum_{i_X}\ket{i_{X}}\otimes\ket{i_{X'}}$, the quantum information possessed by Alice is encoded in the EPR pair $\ket{RA}$. At this point, Alice tosses her half of such EPR pair (A) in the scrambler. On the other hand, Bob wants to retrieve the information encoded by Alice and tossed into the scrambler by Alice by having access to part of the output state, namely $D$. If Bob initially possesses one half of an EPR pair $\ket{BB^{\prime}}$, the initial state of the system is $\ket{RA}\ket{BB^{\prime}}$, while after scrambling the total  state on $RB'CD$ is
\be
\ket{\Psi}_{RB^{\prime}CD}=U_{AB}\otimes I_{RB^{\prime}}\ket{RA}\ket{BB^{\prime}}\,.
\label{psi}
\ee

In the context of black-Hole evaporation, an old black hole $B$ is maximally entangled with the Hawking radiation $B^\prime$ possessed by Bob, while $D$ is the Hawking radiation emitted by the black hole after $U_{AB}$ has scrambled the quantum information tossed in it by Alice and $C$ represents the shrinking black-hole interior and is inaccessible for any observer, being beyond the event horizon. At this point, the question is: how much information, initially possessed by Alice, is, after the scrambling unitary, in Bob's possession? One quantifies the information shared by two parties, e.g., $R$ and $C$, by the quantum mutual information between $R$ and $C$, defined through von Neumann entropies
\be
I(R|C):=S(\rho_{R})+S(\rho_{C})-S(\rho_{RC})\, ,
\ee
where $S(\rho):=-\tr(\rho\log\rho)$ and $\rho_{\Lambda}:=\tr_{\bar{\Lambda}}(\ket{\Psi}\bra{\Psi})$ with $\bar{\Lambda}$ being the complement of $\Lambda$. Simple calculations~\cite{hosur2016ChaosQuantumChannels} show that for the state $\ket{\Psi}_{RB^{\prime}CD}$ (in Eq.~\eqref{psi}) one obtains $S(\rho_{R})=|A|$ and $S(\rho_{C})=|C|$. One can also show that the two R\'enyi entropy $S_{2}(\rho):=-\log \tr\rho^2$ obeys~\cite{hosur2016ChaosQuantumChannels} 
\be
S_{2}(\rho_{RC})=-\log \frac{d_{A}}{d_{C}}\Omega(U_{AB})\,.
\ee
From the hierarchy of R\'enyi entropies one then finds that, if $U_{AB}$ is a scrambler, the \textit{decoupling theorem} applies:
\be
I(R|C)=\mathcal{O}\left(2^{2|A|-2|D|}\right)
\ee
that is, only an $\epsilon$ amount of information is shared between Alice (R) and the output of the scrambler $C$ provided that $|D|=|A|+\log\epsilon^{-1}$. Thanks to the unitarity of the evolution, all the information is in Bob's possession, i.e., $DB^{\prime}$: the mutual information between $R$ and $B^{\prime}D$ is maximal
\ba
I(R|B^{\prime}D)=|A|-\mathcal{O}\left(2^{2|A|-2|D|}\right)\,.
\ea
Let us make some remarks concerning why the model can be applied in the context of black-hole evaporation. Let us take a step back, and review the Don Page calculations on the entropy production from a black hole, see Ref.~\cite{page1993AverageEntropySubsystem}. Indeed, modeling a black hole as a complex random unitary $U$, Page finds that the entanglement entropy between the black-hole interior $I$ and the emitted Hawking radiation $E$ is $S_I=-\log\frac{d_I+d_E}{d_Id_E+1}=-\log [d_{I}^{-1}+d_{E}^{-1}+O(d^{-1})]$. Thus, if $|I|=fn$ for $f<1/2$ one has $S_{I}=|I|+O(2^{(1-2f)n})$, i.e., maximal entropy up to a exponentially small error. Thus, following the Page reasoning, one has $(i)$ as long as $|I|\gg |E|$ the Hawking radiation $E$ does not contain any information about the black-hole interior $I$, but rather is the black-hole interior that knows all about $E$; $(ii)$ as soon as $|E|\gg |I|$ the Hawking radiation $E$ contains all the information about the black-hole interior $I$, being maximally entangled with it; while $(iii)$ between the two regimes there is a gray area where the entanglement is not maximal. In the context of the Hayden-Preskill thought experiment, the hypothesis that Bob $B^{\prime}$ shares an EPR pair with the black-hole initial interior $B$ relies exactly upon the Page reasoning: sharing an EPR pair means being maximally entangled with the initial black-hole interior, which is possible only if the black hole has emitted much more than half of the initial qubits, i.e., $|I|\ll |E|$. Thus, among all the radiation emitted $E$ in the history of this black hole, the qubits in Bob's possession are only a subset $B^{\prime}\subset E$. 

That said, the decoupling theorem, being a pure information-theoretic result, finds its own applications as a tool for, exempli gratia, quantum communication and quantum teleportation~\cite{yoshida2019DisentanglingScramblingDecoherence}. Thus, there is no need to specialize the discussion on black-hole physics.

\subsection{Recovery algorithm after a scrambling dynamics}\label{Sec: recoveryalgoYK}
The decoupling theorem says that the quantum information initially encoded in the input state in $A$ is completely transferred through the scrambling unitary dynamics to Bob,  i.e., the system $D$ and $B^{\prime}$. The very scrambling behavior of 
$U_{AB}$ has destroyed any correlation between the reference state in $R$ and the inaccessible part of the information in $C$. Since now the state in $R$ must be perfectly correlated with the state in the hands of Bob,
 there should exist a unitary operator $V$  on $B^{\prime}D$ able to recover all the information encoded in $A$. In other words, there should exist a unitary $V$ which enables Bob to distill an EPR pair between $R$ and a reference system of the same dimension of $R$, say $R^{\prime}$. One calls such operator a \textit{decoder}. In Ref.~\cite{yoshida2017EfficientDecodingHaydenPreskill}, it is shown how Bob can operate such distillation by picking as decoder the transpose of the scrambler $U_{AB}$: Bob needs a further EPR pair $\ket{A^\prime R^{\prime}}$ on auxiliary spaces $A'$ and $R'$ and appends it to the output of the scrambler, obtaining $\ket{\Psi}_{RCDB^{\prime}}\ket{A^{\prime}R^{\prime}}$. The dimension of $A'$ is chosen such that $A'\otimes B'$  is isomorphic to $A\otimes B$. Then Bob applies the operator $V^*_{B'A'} $ and finally projects onto an EPR pair on $D\otimes D'$ by 
 $\Pi_{DD^{\prime}}\equiv \ket{DD^{\prime}}\bra{DD^{\prime}}$.
 The final state after the algorithm performed by Bob is thus
\be
\ket{\Psi_{out}(V)}\equiv \frac{1}{\sqrt{P_{out}}}\Pi_{DD^{\prime}}V^{*}_{B^{\prime}A^{\prime}}\ket{\Psi}_{RCDB^{\prime}}\ket{A^{\prime}R^{\prime}}\,,
\label{recoverypsioutYoshida}
\ee
where $P_{out}$ is a normalization. The success of the algorithm, that is, $V^*_{B'A'} $ being a decoder, is guaranteed if the state  Eq.~\eqref{recoverypsioutYoshida} looks like $\ket{\Psi_{out} (V)}\simeq \ket{RR^{\prime}}\otimes \ket{rest}_{CC^{\prime}}\otimes \ket{DD^{\prime}}$, i.e., a factorized state with an EPR pair between Alice qubits $R$ and Bob qubits $R^{\prime}$. The factorization is possible only because no information is shared between $R$ and $CC^{\prime}$ thanks to the decoupling theorem. To check whether the algorithm is successful or not, one computes the fidelity between $\ket{\Psi_{out}(V)}$ and the \textit{target EPR pair} one wants to distill, i.e., $\ket{RR^{\prime}}$. The fidelity between the state in Eq.~\eqref{recoverypsioutYoshida} and $\ket{RR^{\prime}}$, $\mathcal{F}_{V}(U)\equiv \tr(\Pi_{RR^{\prime}}\st{\Psi_{out}(V)})$ , being a function of the scrambler $U_{AB}$ and the decoder $V$, can be recast as~\cite{leone2022RetrievingInformationBlack}
\be
\mathcal{F}_{V}(U)=\frac{1}{d_{A}^{2}}\frac{\braket{\tr(P_{D}(U)P_{D}(V))}_{P_D}}{\braket{\tr(P_{D}(U)P_{A}P_{D}(V)P_{A})}_{P_A,P_D}}\,,
\label{fidelityunknown}
\ee
where we dropped the subscript for both $U$ and $V$. Then one can see that if $V=U$ and the unitary $U$ is a {\em scrambler}, i.e., $\Omega(U)\simeq {d_{A}^{-2}}+{d_{D}^{-2}}-{d_{A}^{-2}d_{D}^{-2}}$ one obtains a  fidelity 
\ba
\mathcal{F}_{V}(U)= 1-O(4^{|A|-|D|})\,,
\ea
i.e., to have a fidelity $1-\epsilon$, one must have $|D|=|A|+\log\epsilon^{-1/2}$. In the context of black-hole physics, the radiation emitted by the black hole, after that Alice tosses their qubits in its interior, must contain $|D|=|A|+\log\epsilon^{-1/2}$ qubits to ensure a successful recovery by Bob~\cite{hayden2007BlackHolesMirrors}.

As we have seen, the decoder $V$ can be easily found if one knows perfectly $U_{AB}$. The main goal of this paper is to present a way of {\em learning} the decoder $V$ without any previous knowledge of $U_{AB}$.

In the following, we will drop the subscript $AB$ and denote the scrambler as $U_t$ as we will always be concerned with a $t-$doped Clifford circuit, that is, a Clifford circuit in which a number $t$ of single-qubit non-Clifford gates has been injected, see Fig.~\ref{fig1}.

 In the following sections, we present a learning quantum algorithm that aims at finding a decoder $V$ that maximizes the fidelity $\mathcal{F}_{V}(U_t)$. The main question of this paper is: can one learn the behavior of $U_{t}$ by limited access to it and limited resources? The answer is yes, provided that the scrambler is not too chaotic~\cite{leone2021QuantumChaosQuantum}. The learning quantum algorithm is a CC algorithm.
 
\subsection{Main Result}\label{Sec: mainresult}
In this section, we present the main result of the paper,  avoiding technical details of the CC algorithm, later presented in Sec.~\ref{Sec:alg}. We first present the main result as a main claim, and then make a rigorous statement in the form of Theorem~\ref{th1}.

\begin{quote}
{\bf Main claim:} If  $U_t$ is a $t-$doped Clifford circuit it is possible to build a perfect Clifford decoder $V$ using a quantum algorithm requiring $\poly(n,2^t)$ resources, provided that $t<n$, that is if $U_t$ is at most quasichaotic.
\end{quote}

{\em Remark M1.---} The Clifford decoder $V$ still satisfies the decoupling theorem (see Sec.~\ref{sec: scrambling}), as random Clifford unitaries are good scramblers~\cite{yoshida2021DecodingEntanglementStructure, leone2021QuantumChaosQuantum}. Surprisingly, a Clifford operator can decode a unitary $U_t$ that makes extensive use of non-Clifford resources. As stated above, a Clifford decoder exists as long as $U_t$ is quasichaotic. Beyond that threshold, $U_t$  finally becomes too complex to be decoded by a Clifford decoder $V$.

{\em Remark M2.---} For nonchaotic $t$-doped Clifford circuits, that is, for $t=O(\log n)$, the Clifford decoder exists and can be found with resources (time and sample complexity) both polynomial in $n$. The learning of the decoder is thus efficient. For quasichaotic circuits, i.e., for $t\lesssim n$, the efficient Clifford decoder can be found, but with a $\exp(n)$ amount of resources.

{\em Remark M3.---}  From the fidelity formula, Eq.~\eqref{fidelityunknown}, we can see that a perfect decoder (i.e., with fidelity $\mathcal{F}_V(U)\simeq1$) must reproduce the action of $U_t^{\dag}P_{D}U_t$ for any $P_{D}\in\mathbb{P}(D)$. For $t=0$, this requirement can obviously be fulfilled. However, for every $t>0$, it is not possible to {\em exactly} reproduce the action of a non-Clifford unitary $U_t$ on a Pauli operator $P_{D}$. How is it then possible that a decoder even exists? To understand this, let us explore the consequences of the fact that we only need to reproduce the behavior of $U_t$  on the Pauli operators in $D$. First, it might happen that for some Pauli operators, $U_t$ would send them again in Pauli operators, effectively behaving on them like a Clifford operator. Define the subgroup of the Pauli group on $D$
\ba
G_D(U):=\{P\in \mathbb{P}(D)\,|\, U^{\dag}P_DU\in\mathcal{P}\}\,,
\label{gddefinition}
\ea
where $\mathcal{P}$ is the Pauli group on $n$ qubits. If $U_t$ is a Clifford operator, then $G_D(U_t)\equiv \mathbb{P}(D)$. Similarly to Eq.~\eqref{otocdef}, we can define a \textit{truncated} OTOC by averaging over the group $G_D(U_t)$ instead of $\mathbb{P}(D)$
\be
\Omega_{G_D}(U_t):=\frac{1}{d}\braket{\tr(P_AP_{D}(U_t)P_AP_{D}(U_t))}_{\mathbb{P}(A), G_D(U_t)}\,.
\label{truncated}
\ee
If $U_t$ is a scrambler, one can easily see that~\cite{yoshida2019DisentanglingScramblingDecoherence} if $|G_{D}(U_t)|>1$
\ba
\Omega_{G_D}(U_t)\simeq \frac{1}{d_{A}^{2}}+\frac{1}{|G_{D}(U_t)|^{2}}-\frac{1}{d_{A}^{2}|G_{D}(U_t)|^{2}}\,.
\label{scramblerGD}
\ea
As far as the operators in $G_D(U_t)$ are concerned, a Clifford operator would still be a perfect decoder. In building the decoder $V$ then, we choose a Clifford operator with the constraints
\be
\forall P\in G_D(U_t), \quad V^{\dag}PV=U_t^{\dag}PU_t\,,
\label{constraint}
\ee
that is, $V$ equals the action of $U_t$ on the subgroup $G_D(U_t)$. The above requirement can be fulfilled because $U_t$ acts as a Clifford operator on $G_{D}(U_t)$ and because of the unitarity of both $U_t$ and $V$ or, equivalently, thanks to the group structure of $G_D(U_t)$. While a unitary operator $U_t$ is uniquely defined by its adjoint action on every Pauli operator (or, to be rigorous, on all the generators of $\mathbb{P}$), Eq.~\eqref{constraint} constraints the unitary $V$ only on the generators of the group $G_D(U_t)$, leaving the other degrees of freedom free. This will be the key insight for the success of the randomized algorithm presented in Sec.~\ref{Sec:alg}. The randomized algorithm builds a decoder $V$ by first imposing the constraints Eq.~\eqref{constraint} and then completing the Clifford operator in a {\em random} way.  We name this algorithm the {\em constrained random Clifford Completion (CC) algorithm}. We show that the fidelity attained by the decoder $V$ is
\be\label{eq:11}
\mathcal{F}_{V}(U_t)=\frac{1+R}{d_{A}^2\Omega_{G_D}(U_t)+R^\prime}
\ee
where 
\ba
R&:=(d|G_D(U_t)|)^{-1}\hspace{-0.8cm}\sum_{P_D\in \mathbb{P}(D)\setminus G_D(U_t)}\hspace{-0.7cm}\tr(P_{D}(U_t)P_{D}(V))\\
R^{\prime}&:=(d|G_D(U_t)|)^{-1}  \hspace{-1cm}\sum_{P_D\in \mathbb{P}(D)\setminus G_D(U_t),P_A}\hspace{-1cm}\tr(P_AP_{D}(U_t)P_AP_{D}(V))
\label{RR1}
\ea
(see proof in Appendix~\ref{App: proofs}). Is not surprising that if $\tr(P_D(U_t)P_D(V))=0$ for every $P_D\not\in G_D(U_t)$, then $R=R^{\prime}=0$. Whether $R,R'=0$ depends on both $U_t$ and $V$. Since $V$ is partially random, we can consider the probability of $R=R^\prime=0$. 
Remarkably, the unconstrained degrees of freedom in choosing the decoder $V$ allow finding, with an overwhelming probability, a decoder for which $R=R^{\prime}=0$. As we shall see, the size of the set of constrained degrees of freedom is of crucial importance. 

{\em Remark M4.---} 
The CC algorithm searches for and implements a decoder $V$ belonging to the Clifford group. There are two important consequences of this result: first, Clifford circuits admit an efficient classical representation and can be stored easily in a classical memory; second, synthesis of Clifford circuits is also efficient~\cite{aaronson2004ImprovedSimulationStabilizer}: starting from the classical representation of a Clifford unitary $V$, one needs $O(n^2)$ moves in terms of CNOT, Hadamard and Phase gate. Lastly, the implementation of Clifford circuits can be easily done fault-tolerantly, making the above algorithm not too expensive in terms of quantum resources.

The following theorem is the main result of the paper:
\begin{theorem}\label{th1}
Let $U_t$ be a $t$-doped Clifford scrambler. Let $\mathcal{V}_{U_t}^{D}:=\{V\in \mathcal{C}(n)\,|\, V^{\dag}PV=U_t^{\dag}PU_t\,\,\forall P\in G_D(U_t)\}\,$ the set of Clifford circuits obeying Eq.~\eqref{constraint}. The CC algorithm builds a Clifford decoder $V\in\mathcal{V}_{U_t}^{D}$ with time complexity and a number of query accesses scaling as  $\poly(n,2^t)$ such that, with probability 
\be
\underset{V\in{\mathcal{V}}_{U_t}}{\operatorname{Pr}}(R=0,R^\prime=0)\ge 1-2^{-(2|C|-t)}\,,
\ee yields a fidelity  obeying $R=R^{\prime}=0$. The decoder $V$ thus retrieves the information with a fidelity given by
\be\label{fid1}
\mathcal{F}_{V}(U_t)=\frac{1}{d_{A}^2\Omega_{G_D}(U_t)}\,.
\ee
If $U_t$ is a scrambler, then the fidelity reads 
\ba
\mathcal{F}_{V}(U_t)\simeq \frac{1}{1+\frac{d_{A}^{2}-1}{|G_{D}(U_t)|}}\ge \frac{1}{1+2^{2|A|+t-2|D|}}\,,
\ea
cfr. Eq.~\eqref{scramblerGD} and Lemma~\ref{lemmagd} in Appendix~\ref{App: proofs}.
\end{theorem}
\noindent
The above theorem says that a randomized decoder built according to the CC algorithm presented in Sec~\ref{Sec:alg}, recovers the information scrambled by $U_t$ with probability $\operatorname{Pr}(R=0,R^\prime=0)$ that converges to one exponentially fast with   $2|C|-t$, and success fidelity converging to one exponentially fast with  $2|D|-2|A|-t$. In Sec.~\ref{sec:numerics}, we provide numerical evidence of the success of the CC algorithm in finding a perfect decoder for quasichaotic scramblers.

{\em Remark T1.--} As later shown in Sec.~\ref{Sec:alg}, a query access to the unitary $U_t$ corresponds to the ability to apply the unitary $U_t$ on an $n$-qubit quantum register. We remark that querying the unitary $U_t$ twice enables the application of $U_{t}^{\otimes 2}$ on a $2n$-qubit quantum register.

{\em Remark T2.---} 
The key insight for the success of the algorithm   is that  the randomization over the unconstrained degrees of freedom in $V$  (which dictate the behavior of the decoder $V$  on the elements of $\mathbb{P}(C)\cup \mathbb{P}(D)\setminus G_D(U_t)$), yields, with high probability, a value 
$R=R^{\prime}=0$. First of all, this condition is not necessary to achieve perfect fidelity. What is needed is that $R,R^{\prime}\ll 1$. However, there is an intuitive explanation as to why the stronger condition $R=R^\prime=0$ is likely, given the assumptions. Both the quantities $R,R^\prime$ are proportional to the sum over Hilbert-Schmidt inner products. This sum depends on at most $2^t$ terms. Since $P_D(V)$ is still a Pauli operator (because $V$ is Clifford), it is the tensor product over $n$ qubits of single qubit Pauli matrices and can be represented by a $2n-$bit string. The Hilbert-Schmidt inner product then becomes the bit-string inner product. Of these, $2n$ bits, though, the constraints in Eq.~\eqref{constraint} fix  at least $2|D|-t$ bits leaving $2n-2|D|+t=2|C|+t$ bits free. The probability that this string is orthogonal to another $2n$ bit string is thus lower bounded by  $1- 2^{-(2|C|+t)}$.  However, the operator $P_D(U_t)$ is the linear combination of $2^t$ strings, because every $T$ gate evolved by a Clifford circuit produces two strings. In other words, a $t-$doped Clifford circuit produces string entropy~\cite{zhou2020SingleGateClifford,leone2022MagicHindersQuantum}. Finally, we can conclude that the probability that $2^t$ strings of type $P_D(V)$ are orthogonal to the corresponding $P_D(U)$ is then lower bounded by $1-2^{-(2|C|+t)}2^{2t}= 1-2^{-(2|C|-t)}$.

{\em Remark T3.---} From the above formulas, it can be easily checked that the number of $T$ gates increases the size of the subsystem $D$ that must be processed for a successful decoding. Indeed, the size of the subsystem $D$ that must be read by the decoder scales as $|D|=|A|+t/2+\log \epsilon^{-1/2}$ to ensure a decoding fidelity $\epsilon$-close to one. Notably, the decoding is still possible when the number of $T$ gates scales as $t\sim n$, while it becomes no longer possible as $t>n$, as the success probability becomes exponentially suppressed. 


{\em Remark T4.---} One of the reasons why the result of Theorem~\ref{th1} is surprising is that, if we read out too many bits $|D|$, for example, capturing too many bits of the Hawking radiation, then the algorithm fails. After all, one might think that the more one learns, the better it is. However, the fidelity crucially depends on the fact that we can imitate the unitary $U_t$, which is not Clifford, with a Clifford operator $V$. This is only possible if $V$ encodes away in $C$ all the differences between the two. If $|D|$ grows to become the full number of qubits $n$, the fidelity \eqref{eq:11} becomes the unitary fidelity $d^{-2} |\tr{(U_t^\dag V)}|^2$ which is obviously less than one even for a vanishing density of non-Clifford gates (see Lemma~\ref{lemmaunitaryfidelity} in Appendix~\ref{App: proofs}).


{\em Remark T5.---} 
The fact that, in order to achieve fidelity $\mathcal{F}_{V}(U_t)=1$,  the density $c$ of non-Clifford gates cannot exceed the unity is also important and it is connected to the transition in quantum complexity and crossover to quantum chaos driven by the doping by non-Clifford resources, see Refs.~\cite{leone2021QuantumChaosQuantum,oliviero2021TransitionsEntanglementComplexity}. To obtain universal purity fluctuations and universal behavior for the $8$-OTOC (Eq.~\eqref{otoc8def}),   the amount of non-Clifford gates must be greater than $2n$, cfr. Eq.~\eqref{transition}. The same result is obtained in Ref.~\cite{jiang2021LowerBoundTcount} with the tool of the unitary stabilizer nullity. Similar conclusions can be reached by looking at the stabilizer R\'enyi entropy $M(\ket{U_t})$ of the Choi state $\ket{U_t}$ associated with $U_t$. A necessary condition to obtain the universal (and maximal) value  $M(\ket{U_t})\simeq 2n$  is that the number of non-Clifford gates $t\ge 2n$~\cite{leone2022StabilizerRenyiEntropy,leone2022MagicHindersQuantum}). 

{\em Remark T6.---} The use of a Clifford circuit that learns a $t$-doped Clifford circuit allows efficient classical memory storage; indeed, Clifford operators can be efficiently encoded in classical memory using $O(n^2)$ parameter. Thus, although beyond $t=O(\log n)$ the algorithm becomes exponentially hard in $t$, the fact that Clifford operators are suitable decoders for quasichaotic quantum circuits implies that, at least from a memory-storage point of view, the algorithm remains efficient in terms of classical resources. Conversely, in the regime when the density $c$ of non-Clifford gates is $c\ge 2$, a chaotic circuit maps all the Pauli operators to a superposition of exponentially many Pauli strings that, preventing the possibility of finding a suitable Clifford decoder, leads to an exponential needs of classical memory and the consequent impossibility of the learning process, even provided an infinite measurement precision.

All the above considerations show why we call quasichaotic, doped Clifford circuits having finite densities less than one. As the density of non-Clifford resources overcomes $c=t/n>1$, a transition quantum complexity happens and eventually the dynamic reaches the Haar random behavior for $t/n\ge 2$ after which nothing can be reliably learned.

\section{The CC algorithm}

\subsection{Technical Preliminaries }\label{sec:stab}

In this section, we review well-known notions on the stabilizer formalism, as they are instrumental in proving the main result of the paper. We refer to Appendix~\ref{notations} for the list of notations used throughout the paper. Consider the Hilbert space of $n$ qubit $\mathcal{H}$ and let $d=2^n$  its dimension. Let us introduce the Pauli matrices $\bbbone_{[1]},\sigma^x,\sigma^y\mbox{ and }\sigma^z$:
\ba 
\bbbone_{[1]}&=
\begin{pmatrix}
    1 & 0 \\
    0 & 1
\end{pmatrix}\,,\quad
\sigma^x=
\begin{pmatrix}
    0 & 1 \\
    1 & 0
\end{pmatrix}\,
\\,\quad \sigma^y&=
\begin{pmatrix}
    0 & -i  \\
    i &  0
\end{pmatrix}\,,
\quad \sigma^z=
\begin{pmatrix}
    1 & 0 \\
    0 & -1 
\end{pmatrix}\,,
\ea
where $\bbbone_{[1]}$ is the identity on the space of  one qubit. Throughout the paper, we denote operators $O_{[m]}$ acting on a subsystem $[m]$ containing $m$ qubits with subscript $[m]$. The Pauli group $\mathcal{P}$ on $n$ qubits is defined as the $n$-fold tensor product of the single qubit group $\mathcal{P}([1])$ obtained by $\{\bbbone_{[1]},\sigma^x,\sigma^y,\sigma^z\}$ times a multiplicative factor of ${\pm 1,\pm i}$. Note that choosing two Pauli operators $P, Q\in \mathcal{P}$, they either commute $[P, Q]=0$ or anticommute $\{P, Q\}=0$. In what follows, we consider the quotient group of the Pauli group (that is, we ignore the global phases $\{\pm 1,\pm i\}$)
\be 
\mathbb{P}:=\mathcal{P}/\{\pm 1 ,\pm i \}\,.
\ee 

Note that $\mathbb{P}$, the group of Pauli strings, is an Abelian group with respect to the matrix multiplication modulo phases. In the following, we take the license to refer to both $\mathbb P$ and $\mathcal P$ as the Pauli group, but mind that the two different notations mean slightly different things. A set of generators for the Pauli group is given by the set $\mathfrak{l}=\{\sigma_{i}^{x},\sigma_{i}^{z}\}_{i=1}^{n}$, where $\sigma_{i}^{x,z}$ is the operator acting as $\sigma^x,\sigma^z$ on the $i$-th qubit and identically elsewhere. Otherwise, we denote as $\mathfrak{g}$ any other generators of the Pauli group. Thanks to the unitarity of $U\in\mathcal{U}(n)$, one can compute the adjoint action $U^{\dag}PU$ on every $P\in\mathbb{P}$ by knowing all the (adjoint) actions of $U$ on the set of generators $\mathfrak{g}$, i.e., $U^{\dag}\sigma U$ for any $\sigma\in \mathfrak{g}$. Thus, the knowledge of $U^{\dagger}PU$ for every $P\in\mathfrak{g}$ completely determines $U$ up to a global phase. This property is particularly useful because $|\mathfrak{g}|=2n$, i.e., the size of the set of generators scales linearly with $n$. Although, only $O(n)$ chunks of information are required to completely determine a $2^n\times 2^n$ matrix, for a general unitary operator, the knowledge of $U^{\dag}PU$ requires $4^n$ complex numbers. However, there exists a special class of unitary operators for which the knowledge of $U^{\dagger}PU$ requires just $O(n)$ bits of information: the Clifford group.

Denote as $\mathcal{C}(n)$ the Clifford group on $n$ qubit, i.e., a subgroup of the unitary group with the following property:
\be
\mathcal{C}(n):=\{ U_{0}\in\mathcal{U}(n)\,\,\, |\,\,\, U_{0}^{\dag}PU_{0}\in\mathcal{P}, \quad \forall P\in\mathcal{P}\}.
\ee
In other words, the Clifford group is the normalizer of the Pauli group $\mathcal{P}$. Thanks to the aforementioned property, quantum computation employing Clifford unitary operators can be simulated classically in a time scaling as $O(n^{3})$~\cite{gottesman1998HeisenbergRepresentationQuantum,aaronson2004ImprovedSimulationStabilizer}.
As we shall see, any Clifford operator $U_{0}$ can be encoded in a tableau~\cite{aaronson2004ImprovedSimulationStabilizer} $T_{U_{0}}$, which efficiently encodes the action of $U_{0}$ on a set of generators $\mathfrak{g}$, which is conventionally chosen to be the local set of generators $\mathfrak{l}$. Let us first introduce some technical notions. 
Let us first recall that $\mathfrak{l} =  \{\sigma_{i}^{x}, \sigma_{i}^{z}\}_{i=1}^{n}$ represents the set of generators for the Pauli operators. Hence, any Pauli operator $P$ in $\mathbb{P}$ can be expressed as
\begin{equation}\label{Paudecomposition}
\begin{aligned}
P=(&-i)^{\modtwosum_i^n x_{i}z_{i}} (\sigma^x_{1})^{x_1}(\sigma^z_1)^{z_2}\otimes(\sigma^x_{2})^{x_2}(\sigma^z_2)^{z_2}\\&\otimes\cdots\otimes(\sigma^x_{n})^{x_{n}}(\sigma^z_n)^{z_{n}}\,,
\end{aligned}
\end{equation}
where $\modtwosum$ denotes the sum modulo $2$, and where $(x_1,z_1,\ldots,x_n,z_n)$ belongs to $\mathbb{F}_2^{2n}$, with $\mathbb{F}_2$ being the finite field of integers with arithmetic modulo $2$. For sake of clarity, let us introduce the following notation $(P)_{\mathbf{x}\mathbf{z}}\equiv(x_1,z_1,\ldots,x_n,z_n)$. Thanks to Eq.~\eqref{Paudecomposition}, $(P)_{\mathbf{x}\mathbf{z}}$ completely characterizes the Pauli operator $P$, and so that there exists an isomorphism between $\mathbb{P}$ and the field $\mathbb{F}_2^{2n}$. 
\begin{example}
The single qubit Pauli group $\mathbb{P}([1])$ is isomorphic to $\mathbb{F}_{2}^{2}$:
\begin{equation*}
    \begin{aligned}
    (\bbbone_{[1]})_{xz}&=(00),\quad (\sigma^x)_{xz}= (10),\\
    \quad(\sigma^z)_{xz}&= (01),\quad (\sigma^y)_{xz}=(11)
    \end{aligned}
\end{equation*}
 \end{example}
The above example clearly shows how one can associate a pair of integers modulo $2$ with each Pauli matrix. In this isomorphism, the product of two Pauli operators is given by the XOR operation performed on the corresponding binary strings.
\begin{example}
Consider $P_1=\sigma_x$ and $P_2=\sigma_y$ their product is equal to:
\[\sigma^{x}\sigma^{y}\mapsto (10)\oplus (11)=(01)\mapsto \sigma^{z}\]
\end{example}
The final element to characterize the isomorphism between $\mathbb{P}$ and the field $\mathbb{F}_2^{2n}$ is given by the commutation relations of two Pauli operators. Given $P_1,P_2\in\mathbb{P}$ then\cite{zhu2016CliffordGroupFails}:
\be 
P_1P_2=(-1)^{\omega[(P_1)_{\mathbf{x}\mathbf{z}},(P_2)_{\mathbf{x}\mathbf{z}}]}P_2 P_1
\ee  
where $\omega[(P_1)_{\mathbf{x}\mathbf{z}},(P_2)_{\mathbf{x}\mathbf{z}}]\equiv(P_1)_{\mathbf{x}\mathbf{z}}^T\Omega(P_2)_{\mathbf{x}\mathbf{z}}$ is the symplectic form, with $\Omega$ a $2n\times 2n$ block-diagonal matrix with each block equal to $\left(\begin{smallmatrix}
0 &1\\ 1 &0
\end{smallmatrix}\right)$~\footnote{Note that the matrix $\Omega$ presented in our work deviates from the usual representation, which is commonly expressed as a block off-diagonal matrix $\left(\begin{smallmatrix}
0& I\\ I& 0\
\end{smallmatrix}\right)$, consequence of a different choice for the basis of $\mathbb{F}_2^{2n}$. The usual representation can be recovered by setting $(P)_{\mathbf{x}\mathbf{z}}\equiv(x_1,x_2,\ldots,x_n,z_1,z_2\ldots,z_n)$.}. In formulas
\be
\Omega:=\bigoplus_{i=1}^{n}\begin{pmatrix}
    0&1\\1&0
\end{pmatrix}
\ee
As a consequence $\omega[(P_1)_{\mathbf{x}\mathbf{z}},(P_2)_{\mathbf{x}\mathbf{z}}]$ is able to tells us if two Pauli operators $P_1$ and $P_2$ commutes or not. 
\begin{example}
Consider $n=1$ and $P=\sigma^x$, $P^{\prime}=\sigma^y$. Then $(P)_{xz}=(10)$ and $(P^\p)_{xz}=(11)$. Computing the symplectic form $\omega[\cdot,\cdot]$ , we have
\be
\omega[(P)_{xz},(P^\p)_{xz}]=[(1\cdot1)+(1\cdot0)]=1 
\ee
and therefore $\sigma_x$ anticommutes with $\sigma_y$ as expected. Now consider $n=2$ and $P\equiv \sigma^x\otimes\sigma^x$ and $P^\p\equiv \sigma^z\otimes \sigma^y$. One has $(P)_{\mathbf{xz}}=(1010)$ and $(P^\p)_{\mathbf{xz}}=(0111)$, thus
\be
\omega[(P)_{\mathbf{xz}},(P^\p)_{\mathbf{xz}}]=1\cdot1+0\cdot 0 +1\cdot 1+ 1\cdot 1=0
\ee
and therefore $\sigma^x\otimes\sigma^x$ commutes with $\sigma^z\otimes\sigma^y$ as expected.
\end{example}
The isomorphism between $\mathbb{P}$ and the field $\mathbb{F}_2^{2n}$ sets the first building block in the implementation of a $2n\times2n$ tableau that encodes all the Clifford information. Notably, with the efficient description of a Pauli operator $P$ in terms of a $2n$-dimensional vector $(P)_{\mathbf{x}\mathbf{z}}$, the possibility of implementing a classical representation of a Clifford operator becomes less surprising.
Consider the symplectic group $\mbox{Sp}(2n,\mathbb{F}_2^{2n})$, a group of $2n\times 2n$ matrices $M$ satisfying the following equation:
\be 
M\Omega M^T=\Omega
\ee 
It has been shown~\cite{dehaene2003CliffordGroupStabilizer} that for every Clifford operator $U_0\in\mathcal{C}(n)$, there exists a unique symplectic matrix $\tilde{T}_{U_0}\in\mbox{Sp}(2n,\mathbb{F}_2)$ such that for $U_0PU_0^{\dag}=\propto P^{\prime}$, one has
\be
\tilde{T}_{U_0}(P)_{\mathbf{x}\mathbf{z}}= (P^{\prime})_{\mathbf{x}\mathbf{z}}\,.
\ee
Conversely,  the opposite is also true, so every symplectic matrix is associated with a Clifford unitary $U_0\in\mathcal{C}(n)$. The aforementioned facts highlight that a Clifford unitary can be represented by a $2n\times 2n$ matrix, providing an efficient encoding scheme. The action of a Clifford operator on $\sigma\in \mathfrak{l}$ can be efficiently encoded---being $U_{0}^{\dag}\sigma U_{0}$ a Pauli operator---in a $2n+1-$bit string, where the first $2n$-bits encode $(U_{0}^{\dag}\sigma U_{0})_{\mathbf{x}\mathbf{z}}$, while the last bit encodes the phase of $U_{0}^{\dag}\sigma U_{0}$, that can be either $+1$ or $-1$. One can implement the tableau $T_{U_{0}}$ through a $(2n)\times (2n+1)$ Boolean matrix, where each row stores the action of $U_{0}$ on one of the $2n$ generators of $\mathbb{P}$ and, by convention, the set of generators is chosen to be $\mathfrak{l}$. A generic tableau $T_{U_{0}}$ can be written in the following way
\begin{widetext}
\ba \label{tableauTc}
&T_{U_{0}}\equiv (\widetilde{T}_{U_{0}}\, |\,\boldsymbol{\phi})\\ &=\begin{pmatrix}
    (U_{0}^\dag \sigma_{1}^{x} U_{0} )_{x_1 z_1}& (U_{0}^\dag \sigma_{1}^{x} U_{0} )_{x_2 z_2}& \ldots & (U_{0}^\dag \sigma_{1}^{x} U_{0} )_{x_n z_n}  &\vline& \phi_1\\
    (U_{0}^\dag \sigma_{1}^{z} U_{0} )_{x_1 z_1}& (U_{0}^\dag \sigma_{1}^{z} U_{0} )_{x_2 z_2}& \ldots & (U_{0}^\dag \sigma_{1}^{z} U_{0} )_{x_n z_n}  &\vline &\phi_2 \\
    \vdots & \vdots & \ddots&\vdots &\vline &\vdots \\
    (U_{0}^\dag \sigma_{n}^{z} U_{0} )_{x_{1} z_{1}}& (U_{0}^\dag \sigma_{n}^{z} U_{0} )_{x_2 z_2}& \ldots & (U_{0}^\dag \sigma_{n}^{z} U_{0} )_{x_n z_n}&\vline &\phi_{2n}\\
\end{pmatrix}\,,
\ea
\end{widetext}
where $\widetilde{T}_{U_{0}}\in \mbox{Sp}(2n, \mathbb{F}_{2}^{2n})$ is the \textit{partial tableau}~\cite{dehaene2003CliffordGroupStabilizer,gosset2021FastSimulationPlanar}, a $2n\times 2n$ symplectic matrix that encodes the action $\sigma\mapsto U_{0}^{\dag}\sigma U_{0}\in\mathbb{P}$; while $\boldsymbol{\phi}$ is a $2n\times 1$ matrix (vector) that encodes the phases of the adjoint action of $U_{0}$ on every $\sigma$. In the r.h.s. of Eq.~\eqref{tableauTc}, the notation $(U_{0}^\dag \sigma U_{0} )_{x_i z_i}$ stands for the two bits corresponding to the $i$-th component of the Pauli matrix on the $i$-th qubit, while $\phi_i$ stands for the phase of $U_{0}^{\dag}\sigma U_{0}$. For example, let $U_{0}^{\dag}\sigma U_{0}=\sigma^x\otimes \sigma^y\otimes \sigma^z\otimes\ldots\otimes\bbbone$, then $(U_{0}^{\dag}\sigma U_{0})_{x_1z_1}=(10)$, while $(U_{0}^{\dag}\sigma U_{0})_{x_3z_3}=(01)$, etc. According to the lighter notation for the $2n$ bit string introduced above, $(P)_{\mathbf{xz}}\equiv (x_1,z_1,x_2,z_2,\ldots, x_n,z_n)$, the partial tableau $\widetilde{T}_{U_{0}}$ in Eq.~\eqref{tableauTc} can be written as
\be
\tilde{T}_{U_{0}}=\begin{pmatrix}
    (U_{0}^{\dag}\sigma_{1}^{x}U_{0})_{\mathbf{xz}}\\
    (U_{0}^{\dag}\sigma_{1}^{z}U_{0})_{\mathbf{xz}}\\
    \vdots\\
        (U_{0}^{\dag}\sigma_{n}^{x}U_{0})_{\mathbf{xz}}\\
    (U_{0}^{\dag}\sigma_{n}^{z}U_{0})_{\mathbf{xz}}\\
\end{pmatrix}\,.
\ee
Note that the generators $\sigma_{i}^{x,z}\in \mathfrak{l}$ in the tableau $T_{U_{0}}$  are arranged so that $\{\sigma_{i}^{x},\sigma_{i}^{z}\}=0$ for any $i=1,\ldots,n$ and that $[\sigma_{i}^{x,z},\sigma_{j}^{x,z}]=0$ for any $i,j=1,\ldots,n$ and $j\neq i$. Let us set up the following notation for the rest of the paper: let $A$ be a square matrix, then $[A]_{\alpha}$ denotes the $\alpha$-th row vector of $A$, for example, $[\tilde{T}_{U_0}]_{1}=(U_{0}^{\dag}\sigma_{1}^{x}U_{0})_{\mathbf{xz}}$. Let us make an example.
\begin{example}
Consider the phase gate $S$, the partial tableau $\tilde{T}_{S}\equiv\left(\begin{smallmatrix}
    1&1\\
    0 & 1
\end{smallmatrix}\right)$ is symplectic
\be
T_{S}\Omega T_{S}^T=\begin{pmatrix} 1& 1\\
0&1
\end{pmatrix}\begin{pmatrix}
    0&1\\
    1&0
\end{pmatrix}\begin{pmatrix}
    1 & 0\\
    1 & 1 
\end{pmatrix}
 =\begin{pmatrix}
    0 & 1\\
    1 & 0   
\end{pmatrix}=\Omega
\ee
where we used that the arithmetic is modulo $2$. This simple example illustrates that the partial tableau $\tilde{T}_{S}$ corresponds to the unique symplectic matrix associated with the phase gate $S$.
\end{example}
The generating set of the Clifford group is given by the controlled $\text{NOT}$ gate $(\text{CNOT})$, the Hadamard gate $\text{H}$, and the Phase gate $\text{S}$. The action of these native gates is mapped to a matrix operation on the tableau $T_{U_{0}}$ by looking at their action on a Pauli string $P\in\mathbb{P}$. The Hadamard gate $\text{H}(i)$ acting on the qubit $i$ results in swapping the $x_{i}$th and the $z_i$th component on the entire column, namely:
\be
(P )_{x_i z_i}\xrightarrow[]{H(i)}(P )_{z_i x_i} 
\ee
for all $i=1,\ldots, n \quad \phi_i=\phi_i\oplus x_iz_i\,,$
while the phase gate $S(i)$ acting on the qubit $i$ results in a $\text{XOR}$ operation between the $x_i$s, and $z_i$s:
\be
(P )_{x_i z_i}\xrightarrow[]{S(i)}(P)_{x_i z_i\oplus x_i}\,,
\ee

for all $i=1,\ldots, n \quad \phi_i=\phi_i\oplus x_iz_i$.
Finally, the $\text{CNOT}(k,i)$ having control qubit $k$ and acting on the qubit $i$ reads:
\ba
(P )_{x_k z_k}(P )_{x_i z_i}\xrightarrow[]{\text{CNOT}(k,i)}(P )_{x_k z_i\oplus z_k}(P )_{x_i\oplus x_k z_i}\,.
\ea
for all $i=1,\ldots, n\quad \phi_i=\phi_i\oplus x_kz_i(x_i\oplus z_k\oplus 1)$.
Let us conclude this paragraph by introducing the concept of a symplectic transformation as a map denoted by $\mathcal{S}:\text{Sp}(2n,\mathbb{F}_2^{2n})\rightarrow\text{Sp}(2n,\mathbb{F}_2^{2n})$. This map transforms one tableau, $\tilde{T}{U_0}$, into another tableau, $\tilde{T}{U^\prime_0}$. Despite the general nature of this transformation, since it is a mapping between two symplectic matrices, there always exists an element $T_{U_0^{\prime\prime}}\in\text{Sp}(2n,\mathbb{F}_2^{2n})$ within the group such that $T_{U_0^{\prime\prime}}T_{U_0}=T_{U^\prime_0}$.In the paper, while we introduce other symplectic transformations, our focus will be on determining the specific symplectic matrix (Clifford unitary) that accomplishes the desired task. We will not pursue a general mapping approach but rather identify the particular symplectic matrix that achieves the desired transformation.

Let us define the diagonalizer transformation $\mathcal{D}_{\mathfrak{h}}(\cdot)$, a symplectic transformation, that will be the core of the CC algorithm in Sec.~\ref{Sec:alg}. We first need to define the following encoding on a subset of generators. 
\begin{definition}[$\tau_{\mathfrak{h}}$]\label{defn:tau} Consider a set $\mathfrak{g}$ of generators of the Pauli group, and a subset $\mathfrak{h}\equiv\{g_1,g_{2},\ldots g_h\}\subset \mathfrak{g}$ with $h$ elements. From a subset of generators $\mathfrak{h}$, we define the matrix $\tau_{\mathfrak{h}}$ as follows: 
\be\label{eq:bk}
\tau_{\mathfrak{h}}\equiv
\begin{pmatrix}
    (g_{1})_{\mathbf{xz}}\\
    (g_{2})_{\mathbf{xz}}\\
    \vdots\\
    (g_{i})_{\mathbf{xz}}\\
    \mathbf{0}\\
    \vdots\\
    (g_{h})_{\mathbf{xz}}\\
    \mathbf{0}\\
    \vdots\\
    \mathbf{0}\\
\end{pmatrix}\,,
\ee
where $(g_{i})_{\mathbf{xz}}$ corresponds to the $2n$-bit string $(x_{1}\, z_1\ldots x_n \, z_n)\in\mathbb{F}_2^{2n}$ encoding the generator $g_{i}$, and $\mathbf{0}$ is a $2n$-bit string of $0$s. The matrix $\tau_{\mathfrak{h}}$ is build from the subset of generators $\mathfrak{h}$ in the following way: if for a given $g_i\in \mathfrak{h}$ there is no $g_j\in \mathfrak{h}$ such that $\{g_i,g_j\}=0$, the generator $g_i$ is just followed by a null vector $\mathbf{0}$, otherwise $g_i$ and $g_j$ (s.t. $\{g_i,g_j\}=0$) are placed in two consecutive rows. There can be many ways to build the matrix $\tau_{\mathfrak{h}}$; however, in this paper, we adopt the following convention: \textit{paired} generators, which are couples of anticommuting generators, occupy the initial rows of the matrix $\tau_{\mathfrak{h}}$. The subsequent rows are filled with \textit{unpaired} generators, each followed by a null vector $\mathbf{0}$. The remaining part of the matrix is just filled by null vectors $\mathbf{0}$ (see \ref{rou:init} for a step-by-step algorithm detailing how to construct this matrix given a subset of generators denoted as $\mathfrak{h}$). 
\end{definition}
Let us provide a concrete example:
\begin{example}
Consider $n=2$ and a set of generators $\mathfrak{g}=\{\sigma^{x}\otimes\sigma^{x}, \sigma^{y}\otimes \sigma^{y}, \sigma^{z}\otimes \sigma^{x}, \sigma^{y}\otimes \sigma^{z}\}$. Let $\mathfrak{g}\supset\mathfrak{h}=\{\sigma^{x}\otimes\sigma^{x}, \sigma^{y}\otimes \sigma^{y}, \sigma^{z}\otimes \sigma^{x}\}$. To construct $\tau_{\mathfrak{h}}$, note that $[\sigma^{x}\otimes\sigma^{x},\sigma^{y}\otimes \sigma^{y}]=[\sigma^{z}\otimes\sigma^{x},\sigma^{y}\otimes \sigma^{y}]=0$, while $\{\sigma^{x}\otimes \sigma^{x}, \sigma^{z}\otimes \sigma^{x}\}=0$. Therefore we can assign $[\tau_{\mathfrak{h}}]_{1}=(\sigma^{x}\otimes\sigma^{x})\equiv (1010)$, $[\tau_{\mathfrak{h}}]_{2}=(\sigma^{z}\otimes\sigma^{x})\equiv (0110)$, $[\tau_{\mathfrak{h}}]_{3}=(\sigma^{y}\otimes\sigma^{y})\equiv (1111)$, and $[\tau_{\mathfrak{h}}]_{4}=\bold{0}\equiv (0000)$. Therefore, we can assign to $\mathfrak{h}$ the following  matrix $\tau_{\mathfrak{h}}$ 
\be
\tau_{\mathfrak{h}}=\begin{pmatrix}
    1&0&1&0\\
    0&1&1&0\\
    1&1&1&1\\
    0&0&0&0
\end{pmatrix}\label{tauhdems}
\ee
\end{example}

Note that, the way the matrix $\tau_\mathfrak{h}$ is filled is the same as the partial tableau $\widetilde{T}_{U_{0}}$ with the only difference that the partial tableau encodes all the $2n$ generators $\in \mathfrak{g}$ of $\mathbb{P}$, while $\tau_\mathfrak{h}$ encodes only a subset $\mathfrak{h}\subset \mathfrak{g}$. This fact motivates us to define the following set of Boolean matrices
\begin{definition}
Let $\mathcal{B}_{2n}$ the set of $2n\times 2n$ Boolean matrices. Define $\mathcal{T}_{2n}\subset\mathcal{B}_{2n}$ the set of matrices with the following properties:

\begin{itemize}
    \item $\forall \tau\in\mathcal{T}_{2n}$, $\tilde{\omega}([\tau]_{2i+1},[\tau]_{j})=0$ for $i=1,\ldots, n$ and $j=1,\ldots,2n$, $j\neq 2i+2$.
    \item $\forall \tau\in\mathcal{T}_{2n}$
    $
    \tilde{\omega}([\tau]_{2i+1},[\tau]_{2i+2})=\begin{cases}
        1,\quad \text{if}\quad [\tau]_{2i+2}\neq \bold{0}\\
        0,\quad \text{else}
    \end{cases}
    $ for $i=1,\ldots, n$.
\end{itemize}

where $\tilde{\omega}$ is an extension of the symplectic form $\omega$ applied to the rows of $\tau_{\mathfrak{h}}$.
\end{definition}
\begin{remark}
Given a subset $\mathfrak{h}\subset\mathfrak{g}$ of generators $\mathfrak{g}$, then $\tau_{\mathfrak{h}}\in\mathcal{T}_{2n}$. The partial tableau $\tilde{T}_{U_0}$ corresponding to a Clifford circuit $U_0$ belongs to $\mathcal{T}_{U_0}$. Moreover, the $2n\times 2n$ identity matrix $I_{2n}\in \mathcal{T}_{2n}$.
\end{remark}

Let us define the following $2n$-bit strings $e_\alpha\equiv(\delta_{\alpha1},\ldots,\delta_{\alpha2n})$, where $\delta_{\alpha\beta}$ is the Kronecker delta. 

\begin{definition}[Diagonalizer]\label{def:diag} Let $\mathfrak{h}$ be a subset of generators and $\tau_{\mathfrak{h}}\in\mathcal{T}_{2n}$ the corresponding matrix defined in Definition~\ref{defn:tau}. The Diagonalizer $\mathcal{D}_{\mathfrak{h}}(\cdot)$ is a map $\mathcal{D}_{\mathfrak{h}}\,\,:\,\,\mathcal{T}_{2n}\mapsto \mathcal{T}_{2n}$, whose action is defined as
\be
[\mathcal{D}_{\mathfrak{h}}(\tau_{\mathfrak{h}})]_{\alpha}=\begin{cases}
    e_{\alpha}, \quad \text{if}\,\, [\tau_{\mathfrak{h}}]_{\alpha}=(g_{i})_{\mathbf{xz}}\,\, \text{for some $g_i\in \mathfrak{h}$}\\
    \mathbf{0},\quad \text{if}\,\, [\tau_{\mathfrak{h}}]_{\alpha}=\mathbf{0}
\end{cases}
\label{diagonalizeraction}
\ee
\end{definition}
In other words, the diagonalizer $\mathcal{D}_{\mathfrak{h}}(\cdot)$ maps $\tau_{\mathfrak{h}}$ to a partial identity matrix belonging to $\mathcal{T}_{2n}\subset\mathcal{B}_{2n}$
\be \label{diagonalizer}
\mathcal{D}_{\mathfrak{h}}(\tau_\mathfrak{h}):=\begin{pmatrix}
 1 & 0 & \ldots &  0 & 0 & \ldots & 0 & 0 & \ldots & 0 & 0\\
 0 & 1 & \ldots &  0 & 0 & \ldots & 0 & 0 & \ldots & 0 & 0\\
\vdots & \vdots & \ddots & \vdots & \vdots & \vdots & \vdots & \vdots & \vdots & \vdots & \vdots\\
 0 & 0 & \ldots & 1 & 0 & \ldots &  0 & 0 & \ldots & 0 & 0 \\
 0 & 0 & \ldots & 0 & 0 & \ldots &  0 & 0 & \ldots & 0 & 0 \\
\vdots & \vdots & \vdots & \vdots & \vdots & \ddots & \vdots & \vdots & \vdots & \vdots & \vdots\\
0 & 0 & \ldots & 0 & 0 & \ldots & 1 & 0 & \ldots & 0 & 0 \\
0 & 0 & \ldots & 0 & 0 & \ldots & 0 & 0 & \ldots & 0 & 0 \\
\vdots & \vdots & \vdots & \vdots & \vdots & \vdots & \vdots & \vdots & \ddots & \vdots & \vdots\\
0 & 0 & \ldots & 0 & 0 & \ldots & 0 & 0 & \ldots & 0 & 0 
\end{pmatrix}\,.
\ee 
In Appendix~\ref{app:subrou}, we introduce the algorithm that, given any matrix $\tau_\mathfrak{h}\in\mathcal{T}_{2n}$ performs the Diagonalizer in time $O(n^2)$ in terms of symplectic transformations. As a consequence, the Diagonalizer $\mathcal{D}_{\mathfrak{h}}$ itself is a symplectic transformation and therefore equivalent to a Clifford operator, denoted as $\hat{\mathcal{D}}_{\mathfrak{h}}$, that maps the set $\mathfrak{h}$ in a subset of the set $\mathfrak{l}$ (the local generators of the Pauli group), i.e., $\hat{\mathcal{D}}_{\mathfrak{h}}^{\dag}\mathfrak{h}\hat{\mathcal{D}}_{\mathfrak{h}}\equiv\{\hat{\mathcal{D}}_{\mathfrak{h}}^{\dag} g_{1} \hat{\mathcal{D}}_{\mathfrak{h}},\hat{\mathcal{D}}_{\mathfrak{h}}^{\dag}g_2\hat{\mathcal{D}}_{\mathfrak{h}},\ldots ,\hat{\mathcal{D}}_{\mathfrak{h}}^{\dag}g_h\hat{\mathcal{D}}_{\mathfrak{h}}\}\subset \mathfrak{l}$. For example, consider the partial tableau $\widetilde{T}_{U_{0}}$ defined in Eq.~\eqref{tableauTc}, with set of generators $\mathfrak{g}:=\{\sigma_1,\sigma_2,\ldots,\sigma_{2n}\}$ such that $\{\sigma_{2i},\sigma_{2i+1}\}=0$ for all $i$, then the diagonalizer $\mathcal{D}_{\mathfrak g}$ on the partial tableau $\widetilde{T}_{U_{0}}$ acts as follows
\begin{widetext}
\be
\mathcal{D}_{\mathfrak{g}}(\widetilde{T}_{U_{0}})=\begin{pmatrix}
    (\hat{\mathcal{D}}^{\dag}U_{0}^\dag \sigma_1 U_{0}\hat{\mathcal{D}} )_{x_1 z_1}& (\hat{\mathcal{D}}^{\dag}U_{0}^\dag \sigma_1 U_{0}\hat{\mathcal{D}} )_{x_2 z_2}& \ldots & (\hat{\mathcal{D}}^{\dag}U_{0}^\dag \sigma_1 U_{0}\hat{\mathcal{D}} )_{x_n z_n}  \\
    (\hat{\mathcal{D}}^{\dag}U_{0}^\dag \sigma_2 U_{0}\hat{\mathcal{D}} )_{x_1 z_1}& (\hat{\mathcal{D}}^{\dag}U_{0}^\dag \sigma_2 U_{0} \hat{\mathcal{D}})_{x_2 z_2}& \ldots & (\hat{\mathcal{D}}^{\dag}U_{0}^\dag \sigma_2 U_{0} \hat{\mathcal{D}})_{x_n z_n}   \\
    \vdots & \vdots & \ddots&\vdots \\
    (\hat{\mathcal{D}}^{\dag}U_{0}^\dag \sigma_{2n} U_{0} \hat{\mathcal{D}})_{x_{1} z_{1}}& (\hat{\mathcal{D}}^{\dag}U_{0}^\dag \sigma_{2n} U_{0}\hat{\mathcal{D}} )_{x_2 z_2}& \ldots & (\hat{\mathcal{D}}^{\dag}U_{0}^\dag \sigma_{2n} U_{0}\hat{\mathcal{D}} )_{x_n z_n}\\
\end{pmatrix}
=
\begin{pmatrix}
    e_1\\
    e_2\\
    \vdots\\
    e_{2n}
\end{pmatrix}=I_{2n}\,,
\ee 
\end{widetext}
where $I_{2n}$ is the $2n\times 2n$ identity matrix. We remark here that in the above case, one has $\hat{\mathcal{D}}_{\mathfrak{g}}\equiv U_{0}^{\dag}$ and the diagonalizer is unique up to a phase. Before concluding the section, let us give a basic example of Diagonalizer.
\begin{example}
Let $n=2$. Consider the subset of generators $\mathfrak{h}=\{\sigma^{x}\otimes\sigma^{x}, \sigma^{y}\otimes \sigma^{y}, \sigma^{z}\otimes \sigma^{x}\}$. In Eq.~\eqref{tauhdems}, we computed the matrix $\tau_{\mathfrak{h}}\in\mathcal{T}_{4}$. The Diagonalizer $\mathcal{D}_{\mathfrak{h}}$ acting on $\tau_{\mathfrak{h}}$ results in
\be
\mathcal{D}_\mathfrak{h}\left[\begin{pmatrix}
    1&0&1&0\\
    0&1&1&0\\
    1&1&1&1\\
    0&0&0&0
\end{pmatrix}\right]=\begin{pmatrix}
    1&0&0&0\\
    0&1&0&0\\
    0&0&1&0\\
    0&0&0&0
\end{pmatrix}
\ee
and correspond to the Clifford operator $\hat{\mathcal{D}}_{\mathfrak{h}}=H(0)H(1)CNOT(1,2)H(1)CNOT(1,2)$.
\end{example}

\subsection{The structure of doped Clifford circuits: Compression Theorems, gate complexity, and learnability}\label{Sec: tdopedcircuits}
Equipped with the notions introduced in the previous section, here we discuss the structure of $t$-doped Clifford circuits, i.e., Clifford circuits doped with a finite number $t$ of (single qubit) non-Clifford gates, which for simplicity are considered $T$-gates~\footnote{The extension of this discussion to arbitrary single qubit non-Clifford gates is straightforward.}. Denote $U_t$ a $t$-doped Clifford circuit. Let $P\in\mathbb{P}$ be a Pauli operator. In general, $U_{t}^{\dag}PU_{t}\not\in\mathcal{P}$. However, there exists a subgroup $G(U_t)\subset\mathbb{P}$ of the Pauli group such that $U_{t}^{\dag}PU_{t}\in\mathcal{P}$ for every $P\in G(U_t)$. The group $G(U_t)$ is defined as
\be
G(U_t)=\{P\in\mathbb{P}~|~U_{t}^{\dag}PU_{t}\in\mathcal{P}\}\,,
\label{Gdefinition}
\ee
whose cardinality is lower bounded by $|G(U_t)|\ge 2^{2n-t}$~\cite{jiang2021LowerBoundTcount}, i.e., at most a fraction of $2^t$ Pauli operators gets not preserved by the action of $U_t$. Note that this notion is tied to the one introduced in Sec.~\ref{Sec: mainresult}, where we discussed the number of preserved Pauli operators with support on a subspace $D$. Let $G_{D}(U_t)$ be the group defined in Eq.~\eqref{gddefinition}, clearly we have $G_{D}(U_t)\subset G(U_t)$. Since $G(U_t)$ is a subgroup of the Abelian group $\mathbb{P}$, it is finally generated by a subset $g(U_t)\subset G(U_t)$ of generators, whose cardinality is lower bounded by $|g(U_t)|\ge 2n-t$. We denote with brackets $\braket{\cdot}$ the generating operation, e.g., $G(U_t)=\braket{g(U_t)}$. Thanks to the unitarity of $U_t$, there exists a set of Clifford operations $\mathcal{V}_{U_t}$ defined as~\footnote{The reason why there are Clifford unitaries capable of emulating a general unitary $U_t$ on $G(U_t)$ can be explained as follows: as shown in Sec.~\ref{sec:stab}, Clifford unitaries (as well as all the unitaries) are solely characterized by their action on Pauli operators. In particular, Clifford circuits map Pauli operators to Pauli operators with the only condition of preserving commutation relations between them. Since the conditions imposed by the unitary $U_t$ on $G(U_t)$ pertain to Pauli operators and, due to the unitarity of $U_t$, preserve their commutation relation.}
\be
\mathcal{V}_{U_t}:=\{V\in\mathcal{C}(n)\,|\,V^{\dag} P V=U_{t}^{\dag}P U_{t},\, \forall P\in G(U_t)\}.
\label{Cliffordizedversion}
\ee
After all, the action of $U_t$ on the group $G(U_t)$ is Clifford-like and can be replicated by some Clifford operations. Since $G(U_t)$ is a subgroup of the Pauli group with cardinality $|G(U_t)|$, there exists an integer $s$ and a Clifford operation $\hat{\mathcal{D}}_{g(U_t)}$ such that $\mathbb{P}([s])\subset \hat{\mathcal{D}}^{\dag}_{g(U_t)}G(U_t)\hat{\mathcal{D}}_{g(U_t)}$, where $\mathbb{P}([s])$ denote the local Pauli group on a system $[s]$ containing $s$ qubits. Note that the Clifford operation $\hat{\mathcal{D}}_{g(U_t)}$ introduced above is exactly the Clifford operation corresponding to the symplectic Diagonalizer operation $\mathcal{D}_{g(U_t)}$ defined in Definition~\ref{def:diag}. In particular, let $\tau_{g(U_t)}$ be the matrix corresponding to $g(U_t)$, then $[\mathcal{D}_{g(U_t)}(\tau_{g(U_t)})]_{\alpha}=e_{\alpha}$ if $[\tau_{g(U_t)}]_{\alpha}\neq \bold{0}$. Let $\hat{\mathcal{D}}_{g(U_t)}$ be the Clifford operator corresponding to the diagonalizer, then $\hat{\mathcal{D}}_{g(U_t)}$ acts on the subset $g(U_t)\subset\mathfrak{g}$ (for some set of generators $\mathfrak{g}$) and transforms it to a subset of the local generating set $\mathfrak{l}$ of the Pauli group. 
The integer $s$ obeys the following lower bound:
\be
s\ge n-t\,,
\label{lowerboundm}
\ee
i.e., in the worst case, only a local Pauli group on $t$ qubits is not preserved by the adjoint action of $U_t$. Equation~\eqref{lowerboundm} easily descends from the fact that $|g(U_t)|\ge 2n-t$: the set $g(U_t)$ is a subset of $\mathfrak{g}$, which consists of a set of generators containing $n$ pairs of anticommuting generators. As the cardinality of $g(U_t)$ is bounded from below by $2n-t$, it implies that $g(U_t)$ contains at least $n-t$ pairs of generators from $\mathfrak{g}$. These pairs, when diagonalized by $\mathcal{D}_{g(U_t)}$, correspond to the local generating set of the Pauli group $P([s])$ of a subsystem consisting of $s$ qubits, where $s$ is greater than or equal to $n-t$. The above considerations allow us to decompose any unitary $U_t$ in Clifford blocks plus a non-Clifford operation acting on (at most) $t$ qubits. To see this, let us show that for every $V \in\mathcal{V}_{U_t}$, the unitary operator given by the product $ \hat{\mathcal{D}}^{\dag}_{g(U_t)}U_{t}V ^{\dag}  \hat{\mathcal{D}}_{g(U_t)} $ acts identically on $s$ qubits. Let $P_s\in\mathbb{P}([s])\subset  \hat{\mathcal{D}}_{g(U_t)}^{\dag}G(U_t) \hat{\mathcal{D}}_{g(U_t)}$ a local Pauli operator on $s$ qubits, then
\be
 \hat{\mathcal{D}}_{g(U_t)}^{\dag}VU_{t}^{\dag} \hat{\mathcal{D}}_{g(U_t)}  P_s \hat{\mathcal{D}}_{g(U_t)}^{\dag}U_{t}V^{\dag}  \hat{\mathcal{D}}_{g(U_t)}=P_s\,;
\ee
indeed, $ \hat{\mathcal{D}}_{g(U_t)}   P_s  \hat{\mathcal{D}}_{g(U_t)}^{\dag}\in G(U_t)$ by definition, $V U_{t}^{\dag}PU_{t}V^{\dag} =P$ for every $P\in G(U_t)$ thanks to Eq.~\eqref{Cliffordizedversion} and $ \hat{\mathcal{D}}_{g(U_t)}$ sends $P$ back to $P_s$. This remarkable fact means that $ \hat{\mathcal{D}}_{g(U_t)}^{\dag}U_{t}V^{\dag}  \hat{\mathcal{D}}_{g(U_t)} = \bbbone_{[s]}\otimes u_{[n-s]}$ for some local unitary $u_{[n-s]}$ acting on $n-s$ qubits where $n-s\le t$ (see Eq.~\eqref{lowerboundm}). Moreover, note that the unitary $u_{[n-s]}$ contains all the $T$-gates, since $\hat{\mathcal{D}}_{g(U_t)}, V\in\mathcal{C}(n)$, and thus is a $t$-doped Clifford circuit on $(n-s)$ qubits. We denoted $\bbbone_{[s]}$ the identity matrix acting on the subsystem $[s]$ containing $s$ qubits. All the above considerations are summarized in Eq.~\eqref{decompositionclifford}, as well as in the following theorem, which is one of the main results of the paper.
\begin{theorem}[Compression theorem]\label{th2}
Let $U_t$ a $t$-doped Clifford circuit and $V \in\mathcal{V}_{U_t}$. There exists a integer $s\ge n-t$, a subset $[s]$ of $s$ qubits and a Clifford operation $ \hat{\mathcal{D}}_{g(U_t)}$---where $\mathcal{D}_{g(U_t)}$ is the Diagonalizer defined in Definition~\ref{def:diag}---that allows the following decomposition for $U_t$
\be
U_{t}=\hat{\mathcal{D}}_{g(U_t)}\left(\bbbone_{[s]}\otimes u_{[n-s]}\right)\hat{\mathcal{D}}_{g(U_t)}^{\dag}V\,.
\label{Cliffordblocks}
\ee
for some  unitary operator $u_{[n-s]}$ acting on $(n-s)$ qubit and containing at most $t$ non-Clifford $(T)$ gates. 
\end{theorem}
The above result, besides giving strong insights on the structure of $t$-doped Clifford circuit, as discussed in Sec.~\ref{overview}, allows us to bound gate complexity of $t$-doped Clifford circuits. The gate complexity $\#(U)$ of a unitary operator $U$ is defined as the minimum number of elementary gates, chosen from a certain universal subset (such as e.g., \{H, CNOT, T\}), necessary to build $U$ from the identity. Given that any Clifford unitary operator acting on $s$ qubits can be distilled using $O(n^2)$ gates~\cite{bravyi2021hadamard}, a simplistic upper bound for the gate complexity would be $O(t n^2)$. This is because the structure of a $t$-doped Clifford circuit can always be seen as Clifford circuits interleaved by $t$ non-Clifford $(T)$ gates, leading to the crude bound provided. If $t\le n$, the bound can be improved by employing the techniques of \textit{Clifford compression} derived in Theorem~\ref{th2}. We know that $\#(V)=O(n^2)$, $\#(\hat{\mathcal{D}}_{g(U_t)})=O(n^2)$, $\#(u_{[n-s]})=O(t^3)$, because $n-s\le t$. From this fact, the following corollary of Theorem~\ref{th2} readily descends.
\begin{corollary}\label{cor2}
Let $U_t$ be a $t$-doped Clifford circuit. The gate complexity $\#(U_t)=O(n^2 +t^3)$.
\end{corollary}

Moreover, as expected, the compression theorem above gives an analogous compression result for $t$-doped stabilizer states, i.e., states $\ket{\psi_t}$ obtained from a stabilizer initial state $\ket{\sigma}$ by the action of $U_t$.
\begin{corollary}\label{Cor:dopedstabilizerstatedecomposition}
Consider a $t$-doped Clifford circuit $U_t$ and the $t$-doped stabilizer state obtained $\ket{\psi_t}=U_t\ket{0}_{[n]}$. Then, there exists a choice of the diagonalizer of Eq.~\eqref{Cliffordblocks} of Theorem~\ref{th2}, denoted as $\tilde{\mathcal{D}}_{g(U_t)}$, such that the state $\ket{\psi_t}$ can be compressed as
\be
\ket{\psi_t}=\tilde{\mathcal{D}}_{g(U_t)}( \ket{0}_{[s]}\otimes \ket{\phi}_{[n-s]})\label{statecompression}
\ee
where $\ket{\phi}_{[n-s]}$ is a quantum state defined on the system $[n-s]$ of $n-s\le t$ qubits. 
\begin{proof} 
Call $G_{V\ket{0}}$ the stabilizer group of the stabilizer state $V\ket{0}$. The only thing to note is that, exploiting the freedom in defining the diagonalizer $\mathcal{D}_{g(U_t)}$, is possible to define a diagonalizer, denoted as $\tilde{\mathcal{D}}_{g(U_t)}$, such that the group $G(U_t)\cap G_{V\ket{0}}\subseteq G(U_t)$ is mapped in a local Pauli group $\mathbb{Z}[s]$, where $s\ge n-t$. In formulas,  $\tilde{\mathcal{D}}_{g(U_t)}^{\dag}G(U_t)\cap G_{V\ket{0}}\tilde{\mathcal{D}}_{g(U_t)}\subseteq \mathbb{Z}[s]$. This is because $G(U_t)\cap G_{V\ket{0}}$ is a commuting subgroup of $G(U_t)$ with cardinality lower bounded by $2^{n-t}$ and thus it is always possible to design a diagonalizer with this desired property (see Sec.~\ref{sec:stab}). Therefore, choosing $\tilde{\mathcal{D}}_{g(U_t)}$ we have the following chain of identities
\ba
U_t\ket{0}&=\tilde{\mathcal{D}}_{g(U_t)}\bbbone_{[s]}\otimes u_{[n-s]}\tilde{\mathcal{D}}_{g(U_t)}^{\dag}V\ket{0}\\&=\tilde{\mathcal{D}}_{g(U_t)}\bbbone_{[s]}\otimes u_{[n-s]}(\ket{0}_{[s]}\otimes \ket{\omega}_{[n-s]})\\&=\tilde{\mathcal{D}}_{g(U_t)}(\ket{0}_{[s]}\otimes \ket{\phi}_{[n-s]})
\ea
To conclude the proof, we remark that the choice of the diagonalizer $\tilde{\mathcal{D}}_{g(U_t)}$ that annihilates the action of $V$ on $\ket{0}_{[n]}$ strictly depends on the (stabilizer) input state.
\end{proof}
\end{corollary}

The following theorem and its subsequent corollary establish the methodology for learning a $t$-doped Clifford circuit $U_{t}$.
\begin{theorem}\label{th3}
Let $U_t$ be a $t$-doped Clifford circuit and let $G(U_t)$ the associated group, $g(U_t)$ its generating set, and $\mathcal{V}_{U_t}=\{V\in\mathcal{C}(n)\,|\,V^{\dag} P V=U_{t}^{\dag}P U_{t},\, \forall P\in G(U_t)\}$. Then, with $\poly(n,2^t)$ query accesses to $U_t$, the CC-algorithm finds and efficiently encodes $g(U_t)$, $V \in\mathcal{V}_{U_t}$ and $\hat{\mathcal{D}}_{g(U_t)}$ in a time $\poly(n,2^t)$. In particular, the CC algorithm finds the Clifford operations $U_0$ and $U_{0}^{\prime}$ in Eq.~\eqref{decompositionclifford}.
\begin{proof}
Call CC algorithm in Sec.~\ref{Sec:alg} with $m=0$. Thanks to Eq.~\eqref{Cliffordblocks}, then  $U_{0}= \hat{\mathcal{D}}_{g(U_t)}$ and $U_{0}^{\prime}=\hat{\mathcal{D}}_{g(U_t)}^{\dag}V$.
\end{proof}
\end{theorem}

The above theorems say that we can always decompose a $t$-doped Clifford circuit in a product of Clifford operations and a local unitary acting on (at most) $t$ qubits. Surprisingly, by employing a finite number of query accesses to $U_t$, one is able to isolate the non-Clifford gates and concentrate them into a local unitary acting on at most $t$ qubits, and learn Clifford blocks of the decomposition in Eq.~\eqref{decompositionclifford}. Once again, this fact discriminates circuits where the number of $T$-gates is less than or exceeds the number of qubits $n$: while for $t<n$ a \textit{Clifford compression} is possible, for $t>n$ (in general) the circuit cannot be compressed as in Eq.~\eqref{Cliffordblocks}. Let us conclude the section with the following corollary:
\begin{corollary}\label{cor3}
Let $U_t$ be a $t$-doped Clifford circuit, then using $\poly(n,2^t)$ total resources including time complexity and query complexity to the unitary $U_t$, is possible to learn a full tomographic description of $U_t$.
\begin{proof}
From Theorem~\ref{th2}, we know that $U_t=U_0(\bbbone_{[n-t]}\otimes u_{[t]})U_{0}^{\prime}$ for $U_0=\hat{\mathcal{D}}_{g(U_t)}$ and $U_{0}^{\prime}=\hat{\mathcal{D}}_{g(U_t)}V$. From Theorem~\ref{th3}, the CC algorithm learns and synthetizes $U_0$ and $U_0^{\prime}$ with $\poly(n,2^t)$ query accesses to $U_t$. As a consequence $U_{0}^{\dag}U_tU_{0}^{\dag\prime}$ acts nontrivially on at most $t$ qubits. This fact allows us to run a unitary process tomography that requires $\exp(t)$ resources~\cite{kiani2020LearningUnitariesGradient}.
\end{proof}
\end{corollary}

\subsection{The learning CC algorithm}\label{Sec:alg}
In this section, we present the Clifford Completion (CC) algorithm. Let $U_t$ be a $t$-doped Clifford circuit. Let $m$ be a integer, $0\le m\le n$, then one can define the following quantities for a $t$-doped Clifford circuit $U_t$:
 \ba
 &G_{[n-m]}(U_t)=\{P\in\mathbb{P}([n-m])~|~ U_{t}^{\dag}PU_{t}\in\mathcal{P}\}\,,\nonumber\\
&g_{[n-m]}(U_t)\subset G_{[n-m]}~|~ \braket{g_{[n-m]}}=G_{[n-m]}(U_t)\,,\nonumber\\
&\mathcal{V}_{U_t}^{[n-m]}\equiv \{V\in\mathcal{C}(n)~|~ V^{\dag}PV=U_{t}^{\dag}PU_{t}\,, \forall P\in G_{[n-m]}(U_t)\}.
 \label{learningquantities}
 \ea
Note that for $m=n$, one has $ G_{[n]}(U_t)\equiv G(U_t)$ defined in Eq.~\eqref{Gdefinition}, $g_{[n]}(U_t)\equiv g(U_t)$ and $\mathcal{V}_{U_t}^{[n]}\equiv \mathcal{V}_{U_t}$. From Lemma~\ref{lemmag}, we have that the following facts hold: 
\begin{itemize}
    \item $|G_{[n-m]}(U_t)|\ge 2^{2(n-m)-t}$;
    \item $| g_{[n-m]}(U_t)|\ge 2(n-m)-t$;
    \item $ G_{[n-m]}(U_t)\subset  G_{[n-m^{\prime}]}(U_t)$ for $m>m^{\prime}$;
    \item $ \mathcal{V}_{U_t}^{[n-m]}\subset  \mathcal{V}_{U_t}^{[n-m^{\prime}]}$for $m<m^{\prime}$.
\end{itemize} 
The CC algorithm is capable of learning $ g_{[n-m]}(U_t)$, $\hat{\mathcal{D}}_{g_{[n-m]}}$ and $V \in \mathcal{V}_{U_t}^{[n-m]}$ corresponding to a $t$-doped Clifford circuit $U_t$ by allowing query accesses to $U_t$ (which correspond to apply multiple times the unitary $U_t$ on a quantum register). In particular, the CC algorithm can learn $(i)$ for $m=0$ the decomposition of $U_t$ as $U_{t}=U_{0}[\bbbone_{n-t}\otimes u_t]U_{0}^{\prime}$ (see Theorem~\ref{th2}), and $(ii)$, as later discussed in Sec.~\ref{Sec:RCDGBCC}, for $m=|C|$ the Clifford decoder for the information recovery protocol.

\smallskip

{\em Tools.---}The algorithm will extensively utilize the tools presented in Sec.~\ref{sec:stab}. In particular, it will make use of the matrix $\tau_{\mathfrak{h}}\in\mathcal{T}_{2n}$, as defined in Definition~\ref{defn:tau}. This matrix can be systematically constructed from a subset $\mathfrak{h}$ of a generating set of the Pauli group, along with the Diagonalizer transformation, which can be built out from any matrix in $\mathcal{T}_{2n}$. Therefore, we recommend that interested readers first familiarize themselves with the formalism presented in Sec.~\ref{sec:stab}.


\smallskip

{\em Main Idea.---} Let us briefly explain the underlying idea behind the CC algorithm in a more technical fashion. First, define $\overline{g}_{[n-m]}(U_t)$ as the set of generators such that, together with $g_{[n-m]}(U_t)$, is able to generate all the Pauli group $\mathbb{P}([n-m])$ on $n-m$ qubits, i.e., $\mathbb{P}([n-m])=\braket{g_{[n-m]}(U_t)\cup \overline{g}_{[n-m]}(U_t)}$. Every operator $V\in \mathcal{V}_{U_t}^{[n-m]}$  mocks the action of $U_t$ on all $\sigma\in g_{[n-m]}(U_t)$, i.e., $V^{\dag}\sigma V =U_t^{\dag}\sigma U_t$; the action of $V$ on every other $\sigma\not\in g_{[n-m]}(U_t)$ is free and it is constrained only by the commutation relations with $\sigma \in g_{[n-m]}(U_t)$. This shows that the set $\mathcal{V}_{U_t}^{[n-m]}$ contains more than one element. Thus, the algorithm needs first to search the generating set $g_{[n-m]}(U_t)\cup \overline{g}_{[n-m]}(U_t)$, and then write the tableau corresponding to $V\in\mathcal{V}_{U_t}^{[n-m]}$ in such a generating basis, cfr. Sec.~\ref{sec:stab}. Naively, the search for this generating set is exponentially hard in $n-m$ because, in general, one should pick every Pauli operator $P$ in $\mathbb{P}([n-m])$ and check whether $P\in G_{[n-m]}(U_t)$ or not, i.e., one should check whether $P$ is preserved by the action of $U_t$ or not. In what follows, we describe how to sample Pauli operators in a way that allows us to find the generating set $g_{[n-m]}(U_t)\cup \overline{g}_{[n-m]}(U_t)$ in $\poly(n,2^t)$ steps.  Once that $g_{[n-m]}(U_t)\cup \overline{g}_{[n-m]}(U_t)$ is found, the algorithm learns the Clifford-like action of $U_t$ on every $\sigma\in g_{[n-m]}(U_t)$. The algorithm thus generates one instance of $V\in\mathcal{V}_{U_t}^{[n-m]}$ uniformly at random. Implementing the Diagonalizer $\mathcal{D}_{g_{[n-m]}}$, defined in Definition~\ref{def:diag}, on $\tau_{g_{[n-m]}}$ (where $g_{[n-m]}$ is a short notation for $g_{[n-m]}(U_t)$), defined in Definition~\ref{defn:tau}, the algorithm builds the Clifford operation $\hat{\mathcal{D}}_{g_{n-m}}$. In the case $m=0$, it thus finds $U_{0}\equiv \hat{\mathcal{D}}_{g(U_t)}$ and $U_{0}^{\prime}\equiv \hat{\mathcal{D}}^{\dag}_{g(U_t)}V$.

\smallskip

{\em Sampling Pauli operators.---} Let us describe the sampling method that allows us to find a generating set $g_{[n-m]}(U-t)$. Let $\mathfrak{g}_{k-1}\equiv\{g_1,\ldots, g_{N_{k-1}}\}\subset g_{[n-m]}(U_t)$ be the set of $N_{k-1}$ generators of $G_{[n-m]}(U_t)$ already found by the algorithm after $k-1$ steps. Note that, in general, $N_{k-1}\le 2(k-1)$. Consider the $2n\times 2n$ matrix $\tau_{\mathfrak{g}_{k-1}}$ corresponding to the subset of generators $\mathfrak{g}_{k-1}$. Let $\mathcal{D}_{\mathfrak{g}_{k-1}}$ be the diagonalizer acting on $\tau_{\mathfrak{g}_{k-1}}$ as $[\mathcal{D}_{\mathfrak{g}_{k-1}}(\mathfrak{g}_{k-1})]_{\alpha}=e_{\alpha}$ if $[\tau_{\mathfrak{g}_{k-1}}]_{\alpha}\in \mathfrak{g}_{k-1}$, otherwise $[\mathcal{D}_{\mathfrak{g}_{k-1}}(\tau_{\mathfrak{g}_{k-1}})]_{\alpha}=\mathbf{0}$ (see Definition~\ref{def:diag}). Let $\hat{\mathcal{D}}_{\mathfrak{g}_{k-1}}$ be the Clifford unitary operator associated with the diagonalizer $\mathcal{D}_{\mathfrak{g}_{k-1}}$. Define $\widetilde{G}_{{k-1}}:= G_{[n-m]}(U_t)/\braket{\mathfrak{g}_{k-1}}$ the quotient group of $G_{[n-m]}(U_t)$ respect to the normal subgroup $\braket{\mathfrak{g}_{k-1}}$. 
The operator $\hat{\mathcal{D}}_{\mathfrak{g}_{k-1}}$ maps the group $\widetilde{G}_{{k-1}}$ to a subgroup of the Pauli group $\mathbb{P}([n-m-k+1])$ on $n-m-k+1$ qubits, i.e., $\hat{\mathcal{D}}_{\mathfrak{g}_{k-1}}^{\dag}\widetilde{G}_{{k-1}}\hat{\mathcal{D}}_{\mathfrak{g}_{k-1}}\subset \mathbb{P}([n-m-k+1])$; this is because, by construction (see Eq.~\eqref{diagonalizer}), the diagonalizer maps the generators $\mathfrak{g}_{k-1}$ in a subset of the local generators $ \mathfrak{l}$ of the Pauli group on the first $k-1$ qubits. Therefore, to find the generators of the Pauli group $\mathbb{P}([n-m])$ containing $g_{[n-m]}(U_t)$, we sample a random Pauli operator on $n-m-k+1$ qubits, say $P^{x}$, and then check whether $\hat{\mathcal{D}}_{\mathfrak{g}_{k-1}}P^{x}\hat{\mathcal{D}}_{\mathfrak{g}_{k-1}}^{\dag}$ belongs to $G_{[n-m]}(U_t)$ or not. In this way, we are sure that $\hat{\mathcal{D}}_{\mathfrak{g}_{k-1}}P^{x}\hat{\mathcal{D}}_{\mathfrak{g}_{k-1}}^{\dag}\not \in \braket{\mathfrak{g}_{k-1}}$.

The diagonalizer thus allows us to extract independent Pauli operators at every step, making the effort to find the generators $g_{[n-m]}(U_t)$ exponentially hard in $t$, rather than in $n-m$. To see this, let us compute the probability of finding a Pauli operator belonging to $G_{[n-m]}(U_t)$ at the $k$-th step. The cardinality of the quotient group is 
\ba
|\widetilde{G}_{{k-1}}(U_t)|&=\frac{|G_{[n-m]}(U_t)|}{|\braket{\mathfrak{g}_{k-1}}|}\\&\ge \frac{2^{2n-2m-t}}{2^{N_{k-1}}}\\&\ge 2^{2(n-m-k+1)-t}\,.
\ea
\begin{table*}[t]
\sloppy
    \centering
    \begin{tabular}{cccc p{0.35\linewidth}}
\hline
\hline
    \emph{Subroutine} & {\em Input} &{\em Output}& \emph{Time Complexity}  & {\em Description} \\
\hline
    \ref{rou:Diagonalizer} & \centering $\tau_{\mathfrak{h}}$ & ${\mathcal{D}}_{\mathfrak{h}},\hat{\mathcal{D}}_{\mathfrak{h}}$ & $O(n^2)$ & It transforms the matrix $\tau_{\mathfrak{h}}$ to the partial identity, see Eq.~\eqref{diagonalizeraction}, saving the Clifford operator $\hat{\mathcal{D}}_{\mathfrak{h}}$.\\

    \ref{rou:constrainedcliff} &$ \tau_{\mathfrak{h}}, \bold{\phi}_{\mathfrak{h}}$ & $T_{U_0}$& $O(n^3)$ & It generate a tableau $T_{U_0}$ corresponding to the Clifford $U_0$ constrained by mapping some local generators to the generators $\mathfrak{h}$. The remaining local generators are mapped into random ones. \\

    \ref{rou:leapau} & $U,P\in\mathbb{P}$ & $Q\in\mathbb{P}$& $O(n^2M)$& It aims at learning $U^{\dag}PU$ using $O(nM)$ queries to $U$. It learns a Pauli string $Q$ regardless that $P$ is preserved by $U$. The algorithm fails with probability $O(n2^{-M})$. \\

    \ref{rou:checkkill} & $P, Q$& yes or no& $O(2^{5t})$ &  It checks whether the learned Pauli operator $Q$ via the \textit{Learning a Pauli string} subroutine is accurate and, consequently, whether $P$ is preserved.  \\

    \ref{rou:phasecheck}   & $P,Q\equiv \pm U^{\dag}PU$& $\pm 1$& $O(n^3)$& It learns the phase of the adjoint action of $U$ on a given Pauli operator $P$ using one single query to $U$.\\
\hline
\hline
    \end{tabular}
    \caption{Sketch of the subroutines used in the CC algorithm, see Appendix~\ref{app:subrou} for details.}
    \label{tableofsubroutine}
\end{table*}
The first inequality follows from Lemma~\ref{lemmagd}, and the second inequality follows from the fact that $|\braket{\mathfrak{g}_{k-1}}|\le 4^{k-1}$. The probability that the extracted Pauli operator belongs to $\widetilde{G}_{{k-1}}$ is therefore
    \ba
    \operatorname{Pr}[\hat{\mathcal{D}}_{\mathfrak{g}_{k-1}}P^{x}\hat{\mathcal{D}}_{\mathfrak{g}_{k-1}}^{\dag}\in \widetilde{G}_{{k-1}}]&=\frac{|\widetilde{G}_{{k-1}}|-1}{4^{n-k+1}}&\\\ge 2^{-t}-4^{-n+m+k-1}
    \ea
    i.e., the total dimension of the set we want to pick an element from (i.e., $\widetilde{G}_{{k-1}}\setminus \bbbone$) divided by the dimension of the set we are sampling Pauli operators from (i.e., $\mathbb{P}([n-m-k+1])$). Note that, the $-1$ is neglecting the identity. The (bound on the) above probability becomes zero for $2k=2n-2m-t+2$; that is because the algorithm already found a maximum number of generators (recall indeed that $g_{[n-m]}(U_t)\ge 2(n-m)-t$, see Lemma~\ref{lemmagd}). 
    
    In summary using the above sampling method, the probability of extracting a valid Pauli operator is lower bounded by $3/2^{-(t+2)}$ for every step and, sampling  $2^{t+2}/3\times n$ number of Pauli operators in $\mathbb{P}([n-m-k+1])$ (given that $k\le n-m-t/2$), one has an overwhelming probability, i.e., $\ge 1-\exp(-n)$, to extract a $P^{x}$ such that $\hat{\mathcal{D}}_{\mathfrak{g}_{k-1}}P^{x}\hat{\mathcal{D}}_{\mathfrak{g}_{k-1}}^{\dag}\in G_{[n-m]}(U_t)$.

{\em Subroutines.---} The upcoming algorithm will utilize several subroutines, namely \textit{Diagonalizer}, \textit{Constrained Random Clifford}, \textit{Learning a Pauli string}, \textit{Verification and Removal} and \textit{Phase Check}, which are described in the Appendix~\ref{app:subrou}. The decision to present these subroutines in the Appendix is to enhance the readability of the algorithm. Nonetheless, the input, output, and time complexity are summarized in Table~\ref{tableofsubroutine} for clarity. The algorithm is shown below.
\medskip
\hrule width \hsize \kern 0.5mm \hrule width \hsize

\begin{itemize}
\item {\bf Input:} $n,\, m<n,\, U_t$.
\item {\bf Output:} $g_{[n-m]}(U_t),\, V \in\mathcal{V}^{[n-m]}(U_t),\, \hat{\mathcal{D}}_{g_{[n-m]}}$.
\end{itemize}
Let $\widetilde{T}_{V}$ a $2n\times 2n$ Boolean matrix such that $[\widetilde{T}_{V}]_{\alpha}=\mathbf{0}$ for any $\alpha=1,\dots, 2n$, $\boldsymbol{\phi}=\mathbf{0}$, $\mathfrak{g}_0\equiv\emptyset$ and $\hat{\mathcal{D}}_{\mathfrak{g}_0}\equiv\bbbone$ and $k=1$.
\begin{enumerate}[label=(\Roman*)]

    \item\namedlabel{alg:cycle}{External Cycle} While $k\le n-m$, do:
\begin{enumerate}[label=(\roman*)]

\item For $M^{\prime\prime}$\footnote{Note that, if the algorithm does not find a valid $P_{D}^{x}$ such that $\hat{\mathcal{D}}_{k-1}P_{D}^{x}\hat{\mathcal{D}}_{k-1}^{\dag}\in G_D(U_t)$ after $ 2^{t+2}/3 \, n$, we can say that the algorithm found all the generators of $G_{D}(U_t)$ with probability $\ge 1-\exp(-n)$.} times do:
    \begin{enumerate}[label=(\alph*)]
    \item Extract a random Pauli operators on $(n-m)-k+1$ qubits $p^{x}\in \mathbb{P}([n-m-k+1])$, and define $P^{x}:=\bbbone_{[m+k-1]}\otimes p^{x}$;
    
    \item Use \textit{Learning a Pauli string} subroutine to learn the adjoint action of $U_t$ on $\hat{\mathcal{D}}_{\mathfrak{g}_{k-1}}^{\dag}P^{x}\hat{\mathcal{D}}_{\mathfrak{g}_{k-1}}$. Denote the learned Pauli string as $P^{x}(V)$, and the corresponding encoding string as $(P^{x}(V))_{\mathbf{xz}}$.
    \item Use \textit{Verification and Removal} subroutine, on $P_{x},\, P^{x}(V)$, to check whether $\hat{\mathcal{D}}_{\mathfrak{g}_{k-1}}P_{k}^{x}\hat{\mathcal{D}}_{\mathfrak{g}_{k-1}}^{\dag}\in G_{[n-m]}(U_t)$ or not. 
    \item If $\hat{\mathcal{D}}_{\mathfrak{g}_{k-1}}P_{k}^{x}\hat{\mathcal{D}}_{\mathfrak{g}_{k-1}}^{\dag}\in G_{[n-m]}(U_t)$ quit the while-loop and go to step (iii).
    \end{enumerate}
    \item go to step (II).
    \item  use the \textit{Phase Check} subroutine to read the phase $\phi_{2k-1}$ of $\hat{\mathcal{D}}_{\mathfrak{g}_{k-1}}^{\dag}P_{k}^{x}\hat{\mathcal{D}}_{\mathfrak{g}_{k-1}}$.
    \item \namedlabel{alg:update}{Rows update}Update:
    
    \ba
    [\widetilde{T}_{V}]_{2k-1}    &\mapsto& [\widetilde{T}_{V}]_{2k-1}\oplus (P_{k}^{x}(V ))_{\mathbf{xz}}\nonumber\\ 
\mathfrak{g}_{k-1}&\mapsto& \mathfrak{g}_{k-1}\cup \hat{\mathcal{D}}_{\mathfrak{g}_{k-1}}^{\dag}P_{k}^{x}\hat{\mathcal{D}}_{\mathfrak{g}_{k-1}}\\
    (\boldsymbol{\phi})_{2k-1}&\mapsto&(\boldsymbol{\phi})_{2k-1}\oplus\phi_{2k-1}\nonumber 
    \ea

    \item \namedlabel{alg:cycle2}{Inner Cycle} While $m<M^{\prime\prime}$, do: 
    \begin{enumerate}[label=(\alph*)]
    \item \namedlabel{alg:extract2}{Extract a random Pauli operator} Extract a random Pauli operator $p^{z}\in \mathbb{P}([n-m-k+1])$ such that $\{p^{x},p^{z}\}=0$ denote $P^z\equiv \bbbone_{[m+k-1]}\otimes p^z$, and do steps $(b)$ and $(c)$ of the algorithm with $P^{z}$ instead of $P^{x}$.\footnote{The probability of extraction of an anticommuting Pauli operator is $1/2$.} 
    \item \namedlabel{alg:condition}{Condition}if $\hat{\mathcal{D}}_{\mathfrak{g}_{k-1}} P^{z}\hat{\mathcal{D}}_{\mathfrak{g}_{k-1}}^{\dag}\in G_{[n-m]}(U_t)$, quit the for-cycle and go to step (vii)\footnote{ Let us compute the probability of success. There are two cases to discuss: $(i)$ either $\nexists\,\, Q\in G_D(U_t)$ such that $\{Q,\hat{\mathcal{D}}_{k-1} P_{D_k}^{x}\hat{\mathcal{D}}_{k-1}^\dag\}=0$, or $(ii)$ it needs to be found. It is easy to be convinced that, in the second case, if such a Pauli operator exists then the probability to be found is again lower bounded by $2^{-t}$, and choosing $M^{\prime\prime}=2^{t+2}/3 \, n$ one has $1-\expf{-4/3 n}$ probability to success after $M^{\prime\prime}$ steps.}.
       \end{enumerate}
       \item go to step (ix);
     \item use the \textit{Phase Check} subroutine to learn the phase $\phi_{2k}$ of $P_{k}^{z}(U_t)$.
    \item update:
    \ba
    [T_{V}]_{2k}    &\mapsto& [T_{V}]_{2k}\oplus (P_{k}^{z}(V ))_{\mathbf{xz}}\nonumber\\ 
\mathfrak{g}_{k-1}&\mapsto& \mathfrak{g}_{k-1}\cup \hat{D}_{\mathfrak{g}_{k-1}}^{\dag}P^{z}\hat{D}_{\mathfrak{g}_{k-1}}\\
    (\boldsymbol{\phi})_{2k}&\mapsto&(\boldsymbol{\phi})_{2k}+\phi_{2k}\nonumber
    \ea

    \item Use the \textit{Diagonalizer} subroutine on $\mathfrak{g}_{k-1}$, and denote $\hat{\mathcal{D}}_{\mathfrak{g}_{k-1}}$ the Clifford circuit corresponding to the diagonalizer ${\mathcal{D}}_{\mathfrak{g}_{k-1}}$ (cfr. Eq.~\eqref{diagonalizeraction})
    \item $k\rightarrow k+1$ 
\end{enumerate}
\item Use the \textit{Constrained Random Clifford} subroutine on the tableau $T_{V}\equiv (\widetilde{T}_{V}\,|\, \boldsymbol{\phi})$ to extract a random constrained Clifford $V$. Define
\ba
g_{[n-m]}(U_t)&\equiv& \mathfrak{g}_{k}\\
\hat{\mathcal{D}}_{g_{[n-m]}(U_t)}&\equiv&\hat{\mathcal{D}}_{\mathfrak{g}_{k}}\\
\mathcal{V}_{U_t}^{[n-m]}\ni V &\mapsto& \hat{\mathcal{D}}_{g_{[n-m]}}^{\dag}V
\ea
\end{enumerate}
\hrule width \hsize \kern 0.5mm \hrule width \hsize
\medskip
The above algorithm builds a random instance in $\mathcal{V}_{U_t}^{[n-m]}$, finds the generating set $g_{[n-m]}(U_t)$ and the diagonalizer $\mathcal{D}_{g_{n-m}}$. The algorithm runs in time $O(n^52^{6t})$. Specifically, the time complexity is $O(nM^{\prime\prime}(n^2M+2^{5t}))$, while the query complexity is $O(nM^{\prime\prime}(nM+2^{5t}))$, where $M$ is the number of shot measurements for \textit{Learning a Pauli string} subroutine which fails with probability $O(n2^{-M})$. Thus, choosing $M^{\prime\prime}=\frac{1}{3}2^{t+2}n$, and $M=n$, one has time complexity $O(n^52^{6t})$, query complexity scaling as $O(n^42^{6t})$ and exponentially small probability of failure $O(\poly(n)2^{-n})$, obtained by the union bound.

\subsection{Random Clifford decoder generation based on CC algorithm}\label{Sec:RCDGBCC}
In this section, we show the quantum algorithm based on CC  capable of learning a Clifford decoder $V$ for the information recovery protocol by means of $\poly(n,2^t)$ queries to $U_t$. 

We want to remark once again that the CC algorithm finds a Clifford decoder: we first construct the tableau $T_{V}$ corresponding to the decoder $V$, and subsequently, we employ the distillation algorithm described in Ref.~\cite{aaronson2004ImprovedSimulationStabilizer} to distill the circuit in terms of $\text{CNOT},\text{S}, \text{H}$.

Recall the definition of the group $G_D(U_t)$, given in Sec.~\ref{Sec: mainresult}, $G_D(U_t):=\{P_D\in \mathbb{P}(D)\,|\, U_t^\dag P_DU_t \in \mathcal{P} \}$, i.e., the set of all the Pauli operators (defined on $D$) sent by the adjoint action of $U_t$ in Pauli operators $\mathcal{P}$. We want the decoder $V$ to mock the action of $U_t$ on all $\sigma\in g_{D}(U_t)$, i.e., $V^{\dag}\sigma V=U_t^{\dag}\sigma U_t$ for any $\sigma \in g_{D}(U_t)$; for any other $\sigma\not\in g_{D}(U_t)$ we choose the action of $V$ at random. As in Eq~\eqref{learningquantities}, we define $\mathcal{V}_{U_t}^{D}$ (see Theorem~\ref{th1}) as
\be
\mathcal{V}_{U_t}^{D}=\{V\in\mathcal{C}(n)~|~ V^{\dag}PV=U_{t}^{\dag}PU_{t}\,, \forall P\in G_{D}(U_t)\}\,.
\ee
Since $G_{D}(U_t)\subset G(U_t)$, we clearly have $ \mathcal{V}_{U_t}\subset \mathcal{V}_{U_t}^{D}$. We can run the CC-algorithm, presented in Sec.~\ref{Sec:alg}, and learn a decoder $V\in \mathcal{V}_{U_t}^{D}$ for $m=|C|$. Once again, the reasons why we look for a decoder in $\mathcal{V}_{U_t}^{D}$, instead of one in $\mathcal{V}_{U_t}$, is twofold: 
\begin{itemize}
\item the entire CC algorithm can be run by an observer that has access to the subsystem $D$ only, thus making the whole CC algorithm suitable for the information unscrambling problem;
\item the set $\mathcal{V}_{U_t}^{D}$ is way larger than $\mathcal{V}_{U_t}$, having $2|C|+t$ unconstrained rows instead of $t$. The CC algorithm draws a decoder $V\in \mathcal{V}_{U_t}^{D}$ at random, making the probability of learning $\mathscr{P}(V)$ in Eq.~\eqref{mainP} exponentially close to one in the size of $C$.
\end{itemize}
 \begin{figure}[h]
\begin{subfigure}{0.50\textwidth}
  \includegraphics[width=\textwidth]{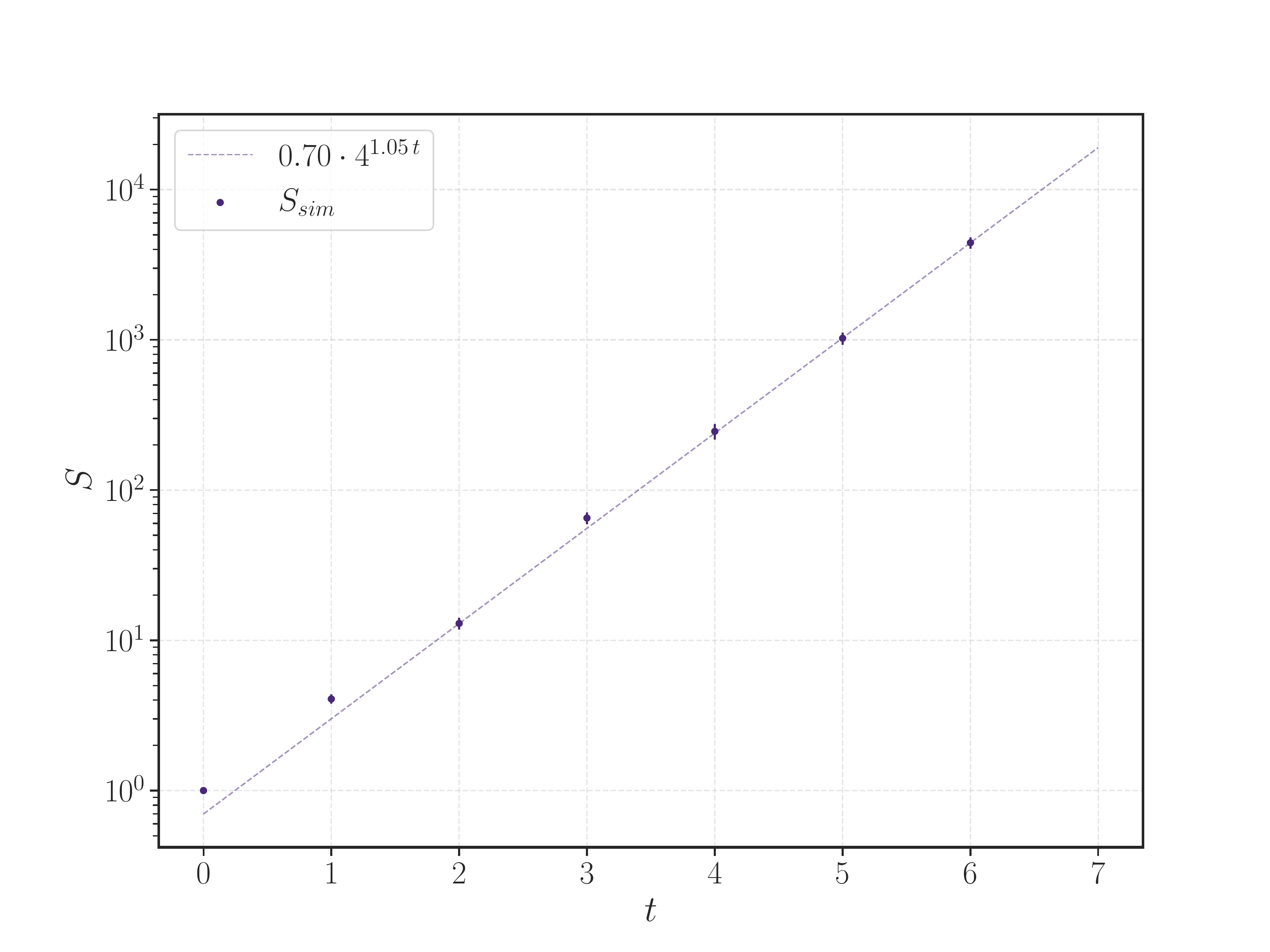}
  \caption{\kern-4em}
  \end{subfigure}
  \begin{subfigure}{0.50\textwidth}
    \includegraphics[width=\textwidth]{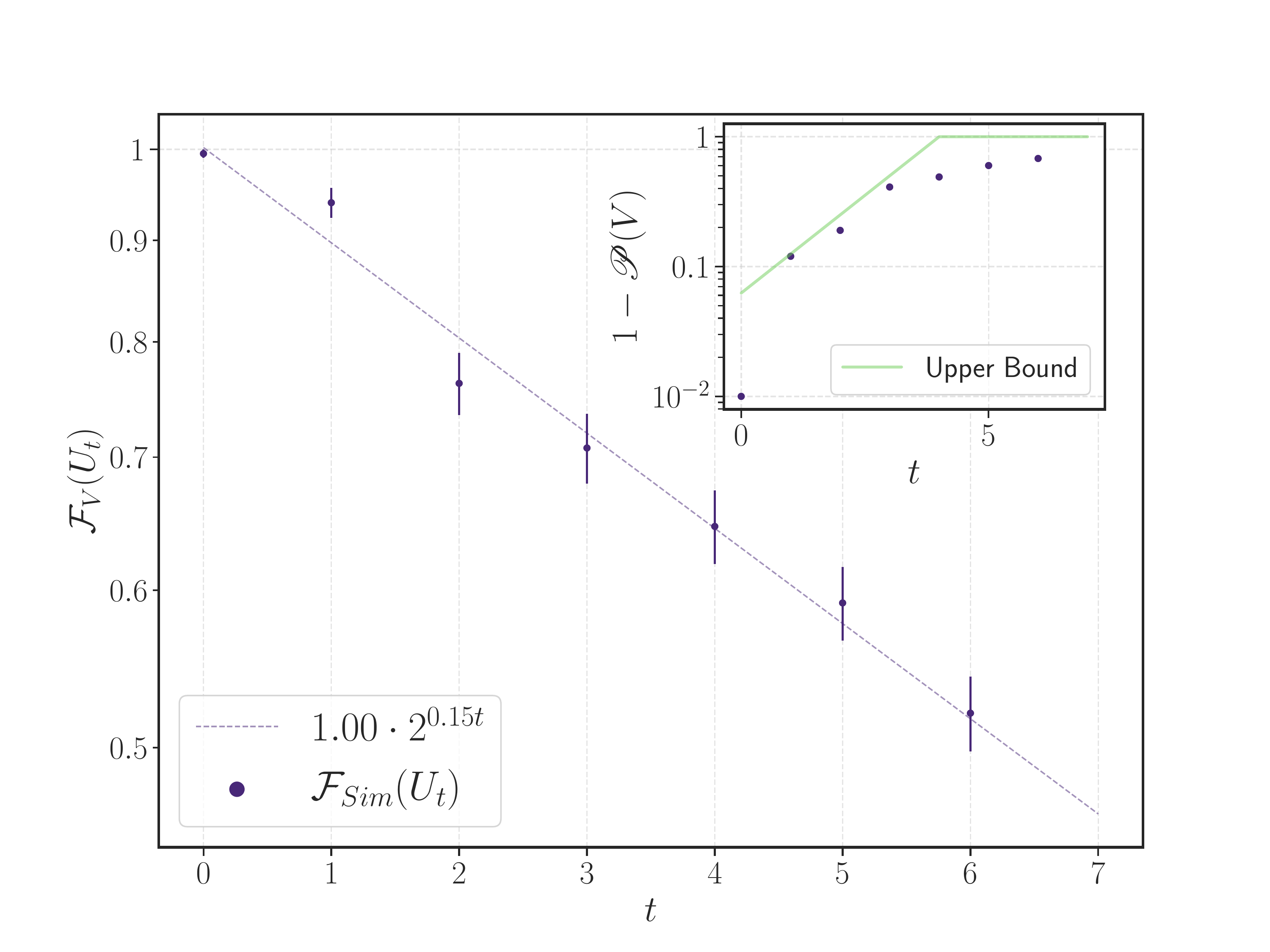}
\caption{\kern-4em}
  \end{subfigure}
    \caption{\raggedright{The results of the CC algorithm for decoding scramblers $U_t$ are depicted in this plot. The parameters used are $n=8$, $|D|=|C|=4$, $|A|=1$, and $|B|=7$. For each value of $t=0,1,\ldots, 6$, we collected $N_{sample}=100$ different samples of scramblers $U_t$. (a) illustrates the average number of steps $S$ over $N_{sample}$ taken by the CC algorithm to find a Clifford decoder $V$ for each scrambler $U_t$. The legend reveals that the average number of steps $S=O(4^t)$ is much more favorable compared with the worst case time complexity estimated as $O(2^{6t})$, see Sec.\ref{Sec:alg}. (b) displays the averaged fidelity $\mathcal{F}_{V}(U_t)$ over $N_{sample}$ realizations of the scrambler $U_t$. The inset highlights the failure probability, $1-\mathscr{P}(V)$, of finding perfect decoders that perfectly match the lower bound stated in Eq.~\eqref{mainP}, where $1-\mathscr{P}(V)\le 2^{t-8}$.}}
    \label{fig:numerics}
\end{figure}
\subsection{Numerics}\label{sec:numerics}
In this section, we perform numerical tests on the CC algorithm to probe the accuracy of the scalings we derived in Sec.~\ref{Sec:alg}. Due to the computational cost of the CC algorithm on a classical computer, we do not specifically test the scaling with the number of qubits. Instead, we focus on providing the scaling in terms of $t$, that is the number of non-Clifford gates used within a Clifford circuit. For the scrambler $U_t$, we utilize a T-depth 1 doped Clifford circuit model, which consists of a non-Clifford unitary acting on $t/2$ qubits, sandwiched between two deep Clifford circuits. We chose this specific architecture for the sake of numerical simulation convenience. However, it is important to note that, according to Theorem~\ref{th2}, such architecture is applicable in an (almost) general sense for $t$-doped Clifford circuits. The non-Clifford unitary circuit consists of t $T$-gates and is constructed as $\prod_{i=1}^{t/2}T_iH_iT_i$ for even $t$ while $\left(\prod_{i=1}^{(t-1)/2}T_iH_iT_i\right) T_{\frac{t+1}{2}}$ for odd $t$, where $T_i,H_i$ are Hadamard and $T$-gates acting on the $i$-th qubit. On the other hand, the deep Clifford circuits are randomly generated.
 The simulations were executed on a standard laptop, thus we set the values of $n=8$ and $|D|=|C|=4$, while $|A|=1$ and $|B|=7$. For each value of $t=0,1,\ldots, 6$, we collect $N_{sample}=100$ distinct samples of scramblers $U_t$. 

 In Fig.~\ref{fig:numerics} $(a)$, we plot the average time taken by the CC algorithm to find a Clifford decoder $V$ for the scrambler $U_t$. As shown in the inset, the estimated average time complexity is much more favorable with respect to the worst case algorithm's prediction (see Sec.~\ref{Sec:alg}), which is $O(2^{6t})$. In Fig.~\ref{fig:numerics} $(b)$, we plot the averaged fidelity $\mathcal{F}_{V}(U_t)$ over multiple realizations of the scrambler $U_t$. The inset showcases the failure probability of finding perfect decoders that align perfectly with the lower bound stated in Eq.~\eqref{mainP}, i.e., $1-\mathscr{P}(V)\le 2^{t-8}$.

\section{Conclusions} 

The possibility of learning relevant features of complex quantum dynamics from its observable behavior is of crucial importance for the understanding of quantum many-body systems away from equilibrium, loss of coherence and control in quantum devices, quantum chaos, criticality, and black-hole physics, and the general understanding of what quantum complexity is~\cite{brown2018SecondLawQuantum}. In particular, scrambling and information retrieval in quantum circuits pose a set of challenging questions in this context~\cite{hosur2016ChaosQuantumChannels,ding2016ConditionalMutualInformation,cotler2017BlackHolesRandom,roberts2017ChaosComplexityDesign,cotler2017ChaosComplexityRandom,chen2018OperatorScramblingQuantuma,zhang2019InformationScramblingChaotic,xu2020DoesScramblingEqual,yan2020InformationScramblingLoschmidt,mi2021InformationScramblingComputationally,zhuang2019ScramblingComplexityPhase,styliaris2021InformationScramblingBipartitions,touil2021InformationScramblingDecoherence,garcia2021QuantumScramblingClassical}, the most relevant of which is to what extent one can learn how to unscramble information having no previous knowledge of the scrambling dynamics and limited access (e.g., Hawking radiation) to the system. Unscrambling is achieved by means of a unitary operator called the {\em decoder}. One asks: is there always a decoder? Can it be found, and under which conditions? Is the decoder in itself efficient? The general answer to the existence of a decoder was given in Ref.~\cite{hayden2007BlackHolesMirrors}. The decoder exists provided the information is properly scrambled and one has sufficient access to the system. A suitable decoder requires complete knowledge of the scrambling dynamics. It is important to highlight that, even with this knowledge, this decoder is a complex quantum unitary operator and cannot in itself be efficiently simulated classically. This is hardly surprising: complex quantum dynamics must not be unscrambled classically, after all. 

In this work, we explore the very ambitious problem of  {\em learning an efficient decoder}. This means that, at the same time, we want to build the decoder with limited access to the system, without any previous knowledge of the internal dynamics, {\em and} have a decoder that is efficiently represented on a classical computer. One may think that this might be only possible if the scrambling dynamics is, in itself, classically simulable, for instance, the dynamics described by a random Clifford circuit: in this case, we indeed show an algorithm (CC) that can learn a perfect and efficient decoder with only polynomial resources. As the Clifford circuit is polluted by $t$ non-Clifford gates, the cost of simulation grows exponentially in $t$. As $t$ reaches a scaling with the number of qubits $n$, there is no efficient simulation. At this point, one would think that the system is chaotic, and one needs an exponential number of resources on a classical computer to perform the decoding even if the decoder is given~\cite{yoshida2017EfficientDecodingHaydenPreskill}. The surprising result presented here is that the CC algorithm, in spite of requiring exponential resources to build the decoder, does build a decoder that is itself classical: it is a Clifford unitary that can be efficiently encoded in a classical computer. We see then that we are in a gray area where simulation and finding the decoder is hard, while the decoding itself is efficiently represented even on a classical computer and still achieves perfect recovery. Only when the doping crosses over to $t\simeq 2n$, the decoder loses its capability of retrieving information. At that point, there is real quantum chaos~\cite{leone2021QuantumChaosQuantum,oliviero2021TransitionsEntanglementComplexity}: learning is hard, and what one has learned is hard to keep in a classical memory. 

Why is this possible? How can a classical decoder be so good at retrieving quantum information that has been scrambled by a unitary dynamic that cannot be represented classically? Is this a contradiction in terms? First of all, this is not a contradiction. We do not learn the full scrambling dynamics. The gate fidelity between the decoder and the scrambler is strictly less than one. Indeed, one cannot turn a complex quantum unitary into a classical one. However, what we learn are the relevant features of the dynamics, defined in terms of being able to decode the scrambled information. And we find that, to some extent, these can be represented in a classical operator even for very complex (but not fully chaotic) quantum dynamics. 

Again, how this is possible requires an explanation. The decoder might be acting like a quantum correction code, encoding away the non-Cliffordness in the part of the system that is inaccessible and this process is found to be possible until the onset of full-fledged quantum chaos.  If this is true, then the amount of non-Cliffordness (i.e., {\em magic}~\cite{campbell2010BoundStatesMagic,campbell2012MagicStateDistillationAll,veitch2014ResourceTheoryStabilizer,koukoulekidis2022ConstraintsMagicState,saxena2022QuantifyingMultiqubitMagic,leone2022StabilizerRenyiEntropy,oliviero2022MeasuringMagicQuantum,oliviero2022MagicstateResourceTheory,hahn2022QuantifyingQubitMagic,haug2023ScalableMeasuresMagic}) shoved in the inaccessible part by the decoder must increase after the decoding.

Finally, one can ask: can one improve access to resources needed to find the decoder in the low doping case, by looking for decoders that are not Clifford? These operators must have, after all, better global gate fidelity. We believe the answer is no. Looking outside the Clifford group will pollute the search and make the search more complex, as it has been shown in the case of disentangling algorithms~\cite{chamon2014EmergentIrreversibilityEntanglement,yang2017EntanglementComplexityQuantum,true2022TransitionsEntanglementComplexity,leone2022RetrievingInformationBlack,piemontese2022EntanglementComplexityRokhsarKivelsonsign}. This kind of effect is also at play in the appearance of barren plateaus in VQAs even in the case of Clifford circuits as one tries to learn Clifford operations by using non-Clifford resources~\cite{holmes2021BarrenPlateausPreclude}. 

Quantum complexity is thus not necessarily featured in the number of elementary gates needed to decompose a unitary~\cite{brown2018SecondLawQuantum}, but in the hardness of search problems, i.e., in the size of neighborhoods of target quantum states and processes, a point of view that is more reminiscent of Boltzmann's entropy~\cite{goldstein2019GibbsBoltzmannEntropy}. All the above questions beg for an answer, and we believe they will be the source of very exciting future works.

\smallskip

\textbf{Acknowledgements}

The authors thank Claudio Chamon for enlightening conversations that have been at the inception of this work and Bin Yan for a very useful remark about how to compute the success probability. We are also grateful to Dylan Lewis for helpful and long discussions in examining some technical details of the algorithm. Special thanks go to Lennart Bittel for fundamental inceptions for the proof of Lemma~\ref{lemma:finite_resolution}. L.L and S.F.E.O acknowledge support from NSF award number 2014000. A.H. acknowledges financial support from PNRR MUR project PE0000023-NQSTI and PNRR MUR project CN 00000013-ICSC. The work of L.L. and S.F.E.O. was supported in part by the U.S. Department of Energy (DOE) through a quantum computing program sponsored by the Los Alamos National Laboratory Information Science \& Technology Institute, and by the  Center for Nonlinear Studies at Los Alamos National Laboratory (LANL). L.L. and S.F.E.O. contributed equally to this work.

\clearpage
\appendix
\onecolumngrid

\setcounter{secnumdepth}{1}

\begin{center}

\textbf{\large Appendix}
\end{center}
\setcounter{equation}{0}
\setcounter{figure}{0}
\setcounter{table}{0}

\makeatletter
\renewcommand{\theequation}{A\arabic{equation}}
\renewcommand{\thefigure}{A\arabic{figure}}
\newtheorem{thmS}{Theorem A\ignorespaces}

\newtheorem{claimS}{Claim A\ignorespaces}

\makeatletter

\section{Notations}\label{notations}
In the following, we list the notations used throughout the paper:
\begin{itemize}
\item $n$: total number of qubits.
\item $\mathcal{H}$: Hilbert space of $n$ qubits.
\item $\mathcal{P}$: Pauli group on $n$ qubits.
\item $\mathbb{P}:=\mathcal{P}/\{\pm 1, \pm i\}$, is the group of Pauli strings, also referred to, with license, as just Pauli group. Note that $\mathbb P\ne \mathcal P$.
\item $\mathcal{C}(n)$: Clifford group on $n$ qubits.
\item $U_{t}$: Clifford unitary operator doped with a number $t$ of single qubit non-Clifford gates.
\item $\mathcal{C}_t$ set of Clifford circuits doped with $t$ single qubit non-Clifford gates.
\item $O(U)\equiv U^{\dag}OU$, where $O$ is a operator on $n$ qubits and $U$ a unitary operator on $n$ qubits.
\item $|A|$: number of qubits in the subsystem $A$.
\item $A,B,C,D$: subsystem of qubits such that $|A|+|B|=|C|+|D|$.
\item $[m]$: subsystem of $m$ qubits.
\item $u_{[m]}$: local unitary acting on a subsystem $[m]$ of $m$ qubits.
\item $\mathbb{P}([m])$: a local Pauli group on a region $[m]$ of $m$ qubits.
\item $\braket{f(P_A)}_{P_A}$: average over the local Pauli group $P(A)$.
\item Let $G$ be a group. Then if $\mathfrak{G}\subset G$ is a generating set, we write $G=\braket{\mathfrak{G}}$.
\item $\mathfrak{g}\subset \mathbb{P}$ denotes a generating set of the Pauli group $\mathbb{P}$.
\item $\mathfrak{l}\equiv \{\sigma_{i}^{x},\sigma_{i}^{z}\}_{i=1}^{n}$ is the local generating set of $\mathbb{P}$.
\item Let $\mathfrak{h}\subset \mathfrak{g}$. Then $\tau_{\mathfrak{h}}$ is the $2n\times 2n$ Boolean matrix defined in Definition~\ref{defn:tau}.
\item Let $\mathfrak{h}\subset \mathfrak{g}$. Then $\mathcal{D}_{\mathfrak{h}}$ is the diagonalizer acting on $\mathfrak{h}$ defined in Definition~\ref{def:diag}.
\item  $\hat{\mathcal{D}}_{\mathfrak{h}}$: Clifford operator corresponding to the diagonalizer $\mathcal{D}_{\mathfrak{h}}$.
\end{itemize}

\section{Subroutines}\label{app:subrou}
In this section, we present the subroutines used throughout the paper. 
\begin{enumerate}[label=\arabic*.]
    \item \ref{rou:init}
    \item \ref{rou:sweeping}
    \item  \ref{rou:randomclifford}
    \item  \ref{rou:Diagonalizer}
    \item \ref{rou:constrainedcliff}
    \item \ref{rou:leapau}
    \item \ref{rou:checkkill}
    \item \ref{rou:phasecheck}
\end{enumerate}
 Note that the algorithms for the subroutines \ref{rou:sweeping} and \ref{rou:randomclifford} have been introduced in Ref.~\cite{berg2021SimpleMethodSampling}.
\subsection{Matrix initializer}\namedlabel{rou:init}{\textit{Matrix initializer}}
This section is devoted to the subroutine required to initialize the matrix $\tau_{\mathfrak h}$ given a subset of generators $\mathfrak h$
 Below, we give an algorithm that builds the matrix $\tau_{\mathfrak{h}}$ from the subset $\mathfrak{h}$.
 
 \medskip
 
\hrule width \hsize \kern 0.5mm \hrule width \hsize
 \begin{flushleft}
 \begin{itemize}
     \item \textbf{Input:} $\mathfrak{h}$
     \item \textbf{Output:} $\tau_{\mathfrak{h}}$
 \end{itemize}
\begin{enumerate}
    \item Initialize $\tau_{\mathfrak{h}}^{\prime},\tau_{\mathfrak{h}} $ as two $2n\times 2n$ matrices filled by zeros;
    \item let $h=\operatorname{card}(\mathfrak{h})$;
    \item {\bf for} $i\in (0,h-1)$ {\bf do}:
    \begin{enumerate}
    \item $[\tau_{\mathfrak{h}}^{\prime}]_{2i+1}\mapsto[\tau_{\mathfrak{h}}^{\prime}]_{2i+1}\oplus(\mathfrak{h}[1])_{\mathbf{xz}}$;
    \item $\mathfrak{h}\mapsto\mathfrak{h}\setminus \{\mathfrak{h}[1]\}$;
    \item {\bf for} $j\in (0,\operatorname{card}(\mathfrak{h})-1)$ {\bf do}:
    \begin{enumerate}
        \item {\bf if} $\tilde{\omega}([\tau_{\mathfrak{h}}]_{2i+1},(\mathfrak{h}[j+1])_{\mathbf{xz}})=1$:
    \begin{enumerate}
    \item $[\tau_{\mathfrak{h}}^{\prime}]_{2i+2}\mapsto [\tau_{\mathfrak{h}}^{\prime}]_{2i+2}\oplus (\mathfrak{h}[j+1])_{\mathbf{xz}}$;
    \item $\mathfrak{h}\mapsto\mathfrak{h}\setminus \{\mathfrak{h}[j+1]\}$;
    \end{enumerate}
    \end{enumerate}
    \end{enumerate}
    \item $k=0$;
    \item {\bf for} $i \in (0,h-1)$ {\bf do}:
    \begin{enumerate}
        \item {\bf if} $[\tau_{\mathfrak{h}}^{\p}]_{2i+2}\neq \bold{0}$:
        \begin{enumerate}
            \item $[\tau_{\mathfrak{h}}]_{2k+1}\mapsto [\tau_{\mathfrak{h}}]_{2k+1}\oplus [\tau_{\mathfrak{h}}^{\p}]_{2i+1}$;
            \item $[\tau_{\mathfrak{h}}]_{2k+2}\mapsto [\tau_{\mathfrak{h}}]_{2k+2}\oplus [\tau_{\mathfrak{h}}^{\p}]_{2i+2}$
            \item $k\mapsto k+1$;
        \end{enumerate}
        
    \end{enumerate}
    \item {\bf for} $i\in (0,h-1)$ {\bf do}:
    \begin{enumerate}
        \item {\bf if} $[\tau_{\mathfrak{h}}^{\p}]_{2i+1}\neq \bold{0}$ \textbf{and} $[\tau_{\mathfrak{h}}^{\p}]_{2i+2}= \bold{0}$:
        \begin{enumerate}
            \item $[\tau_{\mathfrak{h}}]_{2k+1}\mapsto [\tau_{\mathfrak{h}}]_{2k+1}\oplus [\tau_{\mathfrak{h}}^{\p}]_{2i+1}$;
            \item $k\mapsto k+1$;
        \end{enumerate}
        
    \end{enumerate}
\end{enumerate}
\end{flushleft}
\hrule width \hsize \kern 0.5mm \hrule width \hsize
\medskip
Let us briefly explain the content of the algorithm. The first \textbf{for} cycle fills the matrix with generators. Each generator is followed either by an anticommuting generator or by $\bold{0}$ if there is no anticommuting generator in $\mathfrak{h}$. The other two \textbf{for} cycles are used to order the matrix according to the convention outlined in Sec. \ref{sec:stab}: paired generators are placed first, then followed by unpaired generators, and then by zeros. The time complexity of the matrix initializer is $O(nh)$
\subsection{Sweeping}\namedlabel{rou:sweeping}{\textit{Sweeping}}
\begin{figure}[H]
    \centering
    \includegraphics[scale=.5]{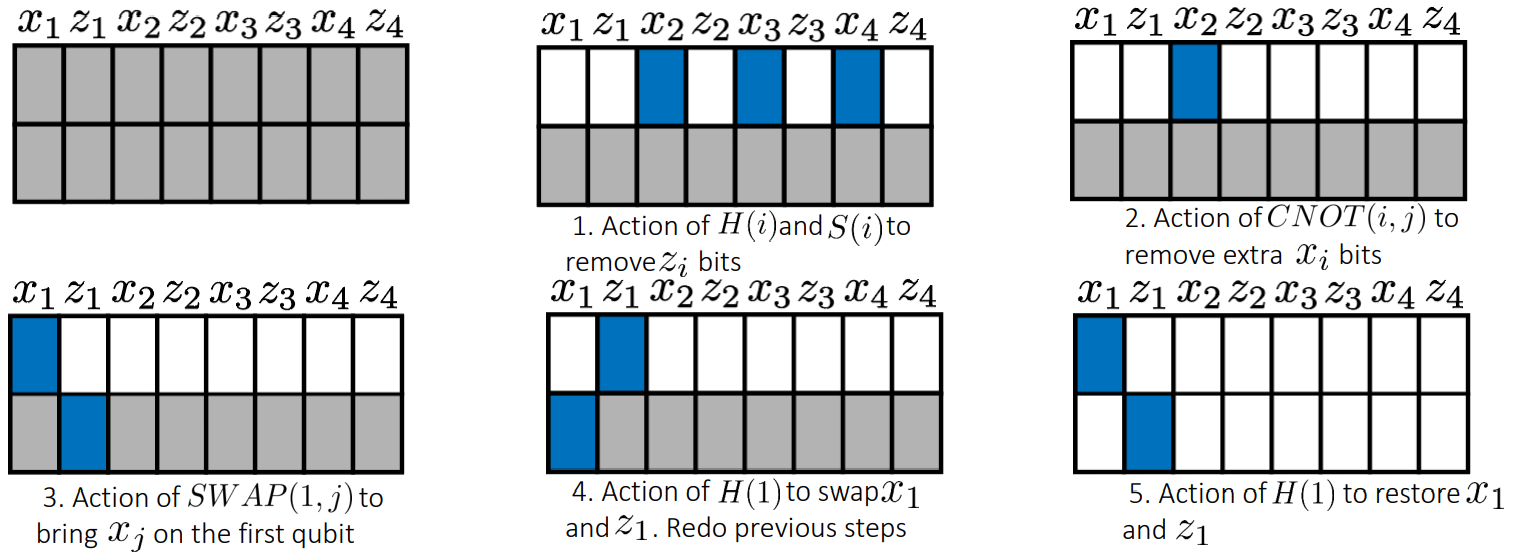}
    \caption{Sketch of the Sweeping subroutine summarized in $5$ steps.}
    \label{fig:my_label}
\end{figure}

The following algorithm is called as a subroutine for both the \ref{rou:Diagonalizer} and \ref{rou:randomclifford} subroutines. 
The sweeping is a symplectic transformation, denoted as $Sw$, whose task is to manipulate two rows of $2k$ bits  (odd and even) at a time and map them to the $2k$-bit string encoding $\sigma_1^x/\sigma_1^z$, i.e.,
\be
\begin{pmatrix}
    1& 0 & 0& 1&\ldots &0&1\\
    0& 1 & 1& 0&\ldots &1&1\\
\end{pmatrix}\xrightarrow[]{Sw} \begin{pmatrix}
    1& 0 & 0& 0&\ldots &0&0\\
    0& 1 & 0& 0&\ldots &0&0\\
\end{pmatrix}
\label{cassdasd}
\ee
through symplectic transformations (equivalent to elementary Clifford gates). A sketch of the algorithm is the following. Denote $P_{1}, P_{2}$ the Pauli operators corresponding to the odd and the even row, respectively. First, it maps $P_{1}$ in a Pauli string of $\sigma^{x}$s. Such mapping can be realized by the repetitive application of Hadamard and Phase gates. Then, it cancels the redundant $\sigma^x$s, leaving just the $\sigma^x$ on the most significant qubit by using CNOTs, and results in $P_{1}\mapsto \tilde{P}_{1}=\sigma^{x}_{1}$. After that, the task of the algorithm is to map  $P_{2}$ to $\sigma^z$. By applying the Hadamard gate on the first qubit, it transforms $\tilde{P}_{1}$ to $\sigma^{z}_{1}$. This operation allows to map $P_{2}$ to $\sigma^{x}_{1}$ following the same procedure described above for $P_{1}$. The subsequent application of a Hadamard gate on the first qubit restores the right order; in this way, one has the encoding of $\sigma_{1}^{x}$ on the odd row and the one of $\sigma_{1}^{z}$ on the even row (see Fig.~\ref{fig:my_label} for a sketch of the Sweeping subroutine).  
The algorithm proceeds as follows:
\medskip
\hrule width \hsize \kern 0.5mm \hrule width \hsize
\begin{flushleft}
\begin{itemize}
\item {\bf Input:} $k, P_1,\, P_2 \in \mathbb{P}\,|\, \{P_1,P_2\}=0$
\item {\bf Output:} $\sigma^x_1,\sigma^z_1$
\end{itemize}

\begin{enumerate}
    \item  Look at the odd row: if $z_i=0$, do nothing; if instead $z_i=1$ and $x_i=0$ apply a Hadamard gate $\text{H}$ to the circuit on $i$-th qubit, otherwise if $z_i=1$ and $x_i=1$ apply a phase gate $\text{S}$ to the circuit on $i$-th qubit.
    \item  Search for the $x_i=1$ and build a sorted list of indices $\mathcal{J}=\{i|x_i=1\}$ then apply $\text{CNOT}$ gates on the different indices of $\mathcal{J}$ as follows:
    \begin{enumerate}
    \item Pair the indices of $\mathcal{J}$ from the less significant qubit to the most significant qubits, then apply in parallel $\text{CNOT}(\mathcal{J}_{i},\mathcal{J}_{i+1})$ on the different pairs.
    \item \label{srou:remzi}Update the set $\mathcal{J}$ removing the last element of the pairs from it.
    \item Repeat step~\ref{srou:remzi} until only the most significant qubit index is contained in the set $\mathcal{J}$.
\end{enumerate}
    \item Apply a $\text{SWAP}$ between the $i$-th location and the first location. 
    \item Apply a Hadamard gate on the first qubit.
    \item Repeat the steps from $(i)$ to $(iii)$ on the even row.
    \item Apply a Hadamard gate on the first qubit.
\end{enumerate}   
\end{flushleft}
\hrule width \hsize \kern 0.5mm \hrule width \hsize
\medskip
The time complexity of the sweeping is $O(n^2)$~\cite{berg2021SimpleMethodSampling}.
\subsection{Random Clifford sampling}\namedlabel{rou:randomclifford}{\textit{Random Clifford Sampling}}

The task of the algorithm is to build a random Clifford operator (following ~\cite{berg2021SimpleMethodSampling,aaronson2004ImprovedSimulationStabilizer}). As shown in Sec.~\ref{sec:stab}, a tableau $T_{U_{0}}$ can efficiently encode a Clifford operator, and thus sampling $T_{U_{0}}$ is equivalent to sampling a Clifford operator. The algorithm to sample a tableau $T_{U_{0}}$ works on two rows per time, namely the $i$-th and the $i+1$-th row, and proceeds as follows: 
\medskip
\hrule width \hsize \kern 0.5mm \hrule width \hsize
\begin{flushleft}
\begin{itemize}
    \item {\bf Input:} $n$
    \item {\bf Output:} $T_{U_0}$
\end{itemize}

Let $\widetilde{T}_{U_{0}}$ be a two $2n\times 2n$ Boolean matrix such that $[\widetilde{T}_{V}]_{\alpha}=\mathbf{0}$ for any $\alpha=1,\dots, 2n$, $\boldsymbol{\phi}_{U_{0}}=\mathbf{0}$. While $j\le n$
\begin{enumerate}[label=(\roman*)]
\item Extract a  random Pauli operators acting on $n-j+1$ qubits $P_{n-j+1}^{x}\in \mathbb{P}([n-j+1])$
    \item Extract a  random Pauli operators acting $n-j+1$ qubits $P_{n-j+1}^{z}\in \mathbb{P}([n-j+1])$, such that $\{P_{n-j+1}^{x},P_{n-j+1}^{z}\}=0$
    \item Update: 
    \ba
    \left[\widetilde{T}_{U_{0}}\right]_{2j-1}    &\mapsto& [\widetilde{T}_{U_{0}}]_{2j-1}\oplus (\bbbone_{j-1}\otimes P_{n-j+1}^{x})_{\mathbf{xz}}\\
    \left[\widetilde{T}_{U_{0}}\right]_{2j}    &\mapsto& [\widetilde{T}_{U_{0}}]_{2j}\oplus (\bbbone_{j-1}\otimes P_{n-j+1}^{z})_{\mathbf{xz}}
    \ea
    \item Perform a Sweeping $Sw_j$ on the rows $[\widetilde{T}_{U_{0}}]_{2j-1}$ and $[\widetilde{T}_{U_{0}}]_{2j-1}$
    \item Sample $[\boldsymbol{\phi}_{U_{0}}]_{2j-1}\in\{0,1\}$ 
    \item Sample $[\boldsymbol{\phi}_{U_{0}}]_{2j}\in\{0,1\}$ 
\end{enumerate}
\end{flushleft}
\hrule width \hsize \kern 0.5mm \hrule width \hsize
\medskip
The Tableau $T_{U_{0}}$ is the $2n\times 2n+1$ matrix obtained as:
\be
T_{U_{0}}=((Sw_{n}Sw_{n-1}\cdots Sw_{1})^{-1}(I_{2n})|\boldsymbol{\phi}_{U_0})
\ee
where $I_{2n}$ is the identity $2n\times 2n$ matrix. $Sw(A)$ denotes the action of the symplectic transformations $Sw_{j}$ for $j=1,\ldots, n$ on a symplectic matrix $A$. The algorithm has a time complexity of $O(n^2)$. To synthesize a Clifford circuit from the random tableau built out from the algorithm, one can employ the algorithm described in Ref.~\cite{aaronson2004ImprovedSimulationStabilizer} with time complexity of $O(n^3)$.

Let us comment on the above algorithm. In steps $(i)$ and $(ii)$, we perform the sampling of an anticommuting pair of Pauli operators. In step $(i)$, the task is to sample a non-identity Pauli operator whose sampling probability is equal to $(1-4^{-k})$. In step $(ii)$, the task is to sample a Pauli operator anticommuting with the one obtained from step $(i)$. The probability of such sampling is equal to $1/2$ since a $k$-qubit Pauli operator anticommutes with $4^{k}/2$ Pauli operators. Thus, the probability to sample a pair of anticommuting Pauli operators $P_1,P_2\in\mathbb{P}([k])$, having nontrivial support on $k$ qubits, reads
\be 
\text{Pr}\left(P_1,P_2\,|\,\{P_1,P_2\}=0\right)=\frac{(1-4^{-k})}{2}\ge \frac{3}{8}
\ee 
  The time complexity of a single iteration of this subroutine is $O(n)$. The reason why the algorithm works on $2n-2j$ bits each $j$-th step of the \textit{while} loop is due to the constraints given by the commutation relations between the Pauli operators of a tableau (cfr. Sec.~\ref{sec:stab}). In more detail, at the $j+1$-th step of the while loop, the algorithm has already sampled $j$ pairs of generators. Via the sweeping tranformation the algorithm tranforms the $j$ pairs of generators to $\{\sigma_{i}^{x},\sigma_{i}^{z}\}_{i=1}^{j}$. A new pair of anticommuting generators must commute with the already found $j$ pairs of generators. The only way to sample a pair of generators, that commutes with $\{\sigma_{i}^{x},\sigma_{i}^{z}\}_{i=1}^{j}$, is that it acts identically on the first $j$ qubits, which in turn implies that the first $2j$ bits are zeros. 
\subsection{Diagonalizer}\namedlabel{rou:Diagonalizer}{\textit{Diagonalizer}}
\noindent
This section describes the subroutine that maps an incomplete tableau to a partial identity, i.e., the diagonalizer $\mathcal{D}_{\mathfrak{h}}(\cdot)$, where $\mathfrak{h\subset \mathfrak{g}}$ a generating set of $\mathbb{P}$, introduced in Sec.~\ref{sec:stab}. Let $\tau_\mathfrak{h}$ be the matrix corresponding to the subset $\mathfrak{h}$, as the one introduced in Eq.~\eqref{eq:bk}. 
\be
\tau_{\mathfrak{h}}\equiv
\begin{pmatrix}
    (g_{1})_{\mathbf{xz}}\\
    (g_{2})_{\mathbf{xz}}\\
    \vdots\\
    (g_{i})_{\mathbf{xz}}\\
    \mathbf{0}\\
    \vdots\\
    (g_{h})_{\mathbf{xz}}\\
    \mathbf{0}\\
    \vdots\\
    \mathbf{0}\\
\end{pmatrix}
\ee
The algorithm can then be written in the following way:
\medskip
\hrule width \hsize \kern 0.5mm \hrule width \hsize
\begin{flushleft}
\begin{itemize}
    \item {\bf Input:} $n,\tau_{\mathfrak{h}}$
    \item {\bf Output:} ${\mathcal{D}}_\mathfrak{h},\hat{\mathcal{D}}_\mathfrak{h}$
\end{itemize}
\begin{enumerate}[label=(\roman*)]
\item For each $j=1,\ldots, n$, do
\begin{enumerate}[label=(\alph*)]
     \item Perform the \ref{rou:sweeping} of the pair $([\tau_{\mathfrak{h}}]_{2j-1},[\tau_\mathfrak{h}]_{2j})$.
\end{enumerate}
\item For each $j=1,\ldots, n$, do
\begin{enumerate}[label=(\alph*)]
     \item Check all the elements corresponding to the columns $2j-1$, $2j$, and the rows $2j< r< 2n$.
     \item Locate the extra $1$ bits, and store the row index $r$ and the column index $c$. If $c$ is even, apply a Hadamard gate on the $j$-th qubit, while if $r$ is even, apply a Hadamard gate on the $r/2$-th qubit.
     \item If $r$ is even, apply $\text{CNOT}(r/2,j)$, else apply $\text{CNOT}((r-1)/2,j)$.
     \item In the end, if $c$ is even, apply a Hadamard gate on the $j$-th qubit, while if $r$ is even, apply a Hadamard gate on the $r/2$-th qubit.
\end{enumerate}
\end{enumerate}
\end{flushleft}
\hrule width \hsize \kern 0.5mm \hrule width \hsize
\medskip
The algorithm after step (i) maps the matrix $\tau_\mathfrak{h}$ in a new matrix $\widetilde{\tau}_\mathfrak{h}$, that reads:
\be\label{sweepcon}
\widetilde{\tau}_{\mathfrak{h}}:=
\begin{pmatrix}
 1 & 0 & \ldots &  0 & 0 & \ldots & 0 & 0 & \ldots & 0 & 0\\
 0 & 1 & \ldots &  0 & 0 & \ldots & 0 & 0 & \ldots & 0 & 0\\
\vdots & \vdots & \ddots & \vdots & \vdots & \vdots & \vdots & \vdots & \vdots & \vdots & \vdots\\
 0 & 0 & \ldots & 1 & 0 & \ldots &  0 & 0 & \ldots & 0 & 0 \\
 0 & 0 & \ldots & 0 & 0 & \ldots &  0 & 0 & \ldots & 0 & 0 \\
\vdots & \vdots & \vdots & \vdots & \vdots & \ddots & \vdots & \vdots & \vdots & \vdots & \vdots\\
0 & 0 & \ldots & 0 & z_1 & \ldots & 1 & 0 & \ldots & 0 & 0 \\
0 & 0 & \ldots & 0 & 0 & \ldots & 0 & 0 & \ldots & 0 & 0 \\
\vdots & \vdots & \vdots & \vdots & \vdots & \vdots & \vdots & \vdots & \ddots & \vdots & \vdots\\
0 & 0 & \ldots & 0 & 0 & \ldots & 0 & 0 & \ldots & 0 & 0 
\end{pmatrix}
\ee
As also shown in Eq.~\ref{sweepcon}, the action of the sweeping on two anticommuting rows cancel all the $1$ bits in the most significant columns.
Instead, when the sweeping is performed on unpaired rows, the cancellation of $1$ bits is partial since it is not constrained by the commutation and anticommutation relations. Step $(ii)$ addresses this issue: it locates all the spurious $\sigma^{z}$s in the string and converts them to $\sigma^x$s through the action of Hadamard gates $(\text{H}(\sigma^z )\text{H}=\sigma^x)$. Then it removes the redundant $\sigma^x$s through the action of CNOT gates $(\text{CNOT}(\sigma^x\otimes \sigma^x) \text{CNOT}=\sigma^x\otimes \bbbone_{[1]})$. In the end, a layer of Hadamard gates (equal to the one applied before) is applied to restore the most significant $\sigma^z$. After the steps $(ii)$ the matrix $\widetilde{\tau}_\mathfrak{h}$ is then mapped in:
\be
[\mathcal{D}_\mathfrak{h}(\tau_\mathfrak{h})]_{\alpha}=\begin{cases}
    e_{\alpha}, \quad \text{if}\,\, [\tau_\mathfrak{h}]_{\alpha}=(g_{i})_{\mathbf{xz}}\,\, \text{for some $g_i\in \mathfrak{h}$}\\
    \mathbf{0},\quad \text{if}\,\, [\tau_\mathfrak{h}]_{\alpha}=\mathbf{0}
\end{cases}
\ee
Note that, at the end of the algorithm, both the Diagonalizer, as a $2n \times 2n$ symplectic matrix, and the corresponding Clifford circuit are revealed. The time complexity of the diagonalizer subroutine is $O(n^2)$ due to the time complexities of the two main steps: step $(i)$ and $(ii)$. Both steps can be shown to possess time complexity $O(n^2)$.

\subsection{Constrained random Clifford completion}\namedlabel{rou:constrainedcliff}{\textit{Constrained random Clifford}}
This section describes the novel algorithm to generate a random Clifford when some rows of the tableau are already fixed.  Such random Clifford is \textit{constrained} because its adjoint action on some Pauli operators is fixed by the given rows of the tableau. As explained in Sec.~\ref{sec:stab}, the tableau is organized as a list of $2n$-bit strings (ignoring the last phase-bit) describing the map $\sigma\mapsto U_{0}^\dag \sigma U_{0}$ for $\sigma\in \{\sigma_1^{x},\sigma_{1}^{z},\ldots,\sigma_{n}^{x},\sigma_{n}^{z}\}$. The fixed rows of the given incomplete tableau can be of two types: $(i)$ paired, i.e., the tableau contains the information about the map $\sigma_{j}^{x}\mapsto U_{0}^\dag\sigma_{j}^{x}U_{0}$ and $\sigma_{j}^{z}\mapsto U_{0}^\dag\sigma_{j}^{z}U_{0}$ for some $j$; or, unpaired, i.e., the tableau just contains the information about either $\sigma_{j}^{x}\mapsto U_{0}^\dag\sigma_{j}^{x}U_{0}$ or $\sigma_{j}^{z}\mapsto U_{0}^\dag\sigma_{j}^{z}U_{0}$ for some $j$. Therefore, without loss of generality, one can consider an incomplete tableau $\tau_\mathfrak{h}$ as the one introduced in Definition~\ref{def:diag}, and a phase vector $\boldsymbol{\phi}_\mathfrak{h}=(\phi_1,\phi_2,\ldots,\phi_i,0,\ldots,\phi_h,0\ldots 0)$. The matrix $\tau_\mathfrak{h}$, will be written as
\be
\tau_\mathfrak{h}=
\begin{pmatrix}
    (g_1)_{\mathbf{xz}}\\
    (g_2)_{\mathbf{xz}}\\
    \vdots \\
    (g_i)_{\mathbf{xz}}\\
    \mathbf{0}\\
    \vdots \\
    (g_{h})_{\mathbf{xz}}\\
    \mathbf{0}\\
    \vdots\\
    \mathbf{0}
\end{pmatrix}=\begin{pmatrix}
P=\text{Paired rows}\\
N=\text{Unpaired rows}\\
\mathbf{0}_{2(n-n_P-n_N)}\\
\end{pmatrix}
\label{tableau:inc}
\ee
where $P$ is a matrix of dimension $2n_P\times 2n$ that encodes the set of anticommuting pairs of Pauli operators, $N$ is a matrix of dimension $2 n_N\times 2n$ that encodes the set of unpaired Pauli operators; $\mathbf{0}_{2(n-n_P-n_N)}$ is a matrix of dimension $2(n-n_P-n_N)\times 2n$ that encodes the set of unconstrained rows. Note that the total number of constrained rows is $2n_P+n_N$.
The task is to fill the missing rows uniformly at random and implement one of the corresponding Clifford operators. The algorithm proceeds as follows:
\medskip
\hrule width \hsize \kern 0.5mm \hrule width \hsize
\begin{flushleft}
\begin{itemize}
    \item {\bf Input:} $n,\tau_\mathfrak{h},\boldsymbol{\phi}_\mathfrak{h}$
    \item {\bf Output:} $T_{U_0}$
\end{itemize}
   \begin{enumerate}[label=(\roman*)]
    \item For $i=1,\ldots,n_p$ do:
    \begin{enumerate}[label=(\alph*)]
    \item Perform the \ref{rou:sweeping} on the pair $([\tau_\mathfrak{h}]_{2i-1},[\tau_\mathfrak{h}]_{2i})$.
    \end{enumerate}
    \item Apply a random symplectic transformations $Sym$ on the last $n-n_P$ qubits. Denote with $\widetilde{\tau_\mathfrak{h}}$ the tableau after the action of $Sym$
    \item For $i=n_p+1,\ldots,n_N+n_P$ do:
    \begin{enumerate}[label=(\alph*)]
    \item Perform the \ref{rou:sweeping} on the pair $([\widetilde{\tau}_\mathfrak{h}]_{2i-1},[\widetilde{\tau}_\mathfrak{h}]_{2i})$. 
    \end{enumerate}
    \item For $j=n_P+1,\ldots,n_N+n_P$
    \begin{enumerate}[label=(\alph*)]
     \item Check all the elements corresponding to the columns $2j-1$, $2j$, and the rows $2j< r< 2(n_P+n_N)$.
     \item Locate the extra $1$ bits, and store the row index $r$ and the column index $c$. If $c$ is even, apply a Hadamard gate on the $j$-th qubit, while if $r$ is even, apply a Hadamard gate on the $r/2$-th qubit.
     \item If $r$ is even, apply $\text{CNOT}(r/2,j)$, else apply $\text{CNOT}((r-1)/2,j)$.
     \item In the end, if $c$ is even, apply a Hadamard gate on the $j$-th qubit, while if $r$ is even, apply a Hadamard gate on the $r/2$-th qubit.
     \item Sample $[\boldsymbol{\phi}]_{2j}\in\{0,1\}$
\end{enumerate}
\item Fill the incomplete rows by completing the identity.
\item Fill the last $2(n-n_P-n_N)$ columns of $\mathbf{0}_{2(n-n_P-n_N)}$ with a random tableau (See~\ref{rou:randomclifford}).
\item for $j=n_N+n_P,\ldots,n$
\begin{enumerate}
    \item Sample $[\boldsymbol{\phi}_\mathfrak{h}]_{2j-1}\in\{0,1\}$
    \item Sample $[\boldsymbol{\phi}_\mathfrak{h}]_{2j}\in\{0,1\}$
\end{enumerate}
\end{enumerate} 
\end{flushleft}
\hrule width \hsize \kern 0.5mm \hrule width \hsize
\medskip
The final tableau is then given by
\be 
T_{U_0}=(\tau_\mathfrak{h}|\boldsymbol{\phi}_\mathfrak{h})=((\widetilde{Sw}_2 Sw_2 Sym Sw_1)^{-1}T_{U_{0}^{(1)}}(I_{2n})|\boldsymbol{\phi}_\mathfrak{h})
\ee 
where $Sw_1$ denotes the action of step $(i)$, $Sym$ the action of step $(ii)$, $Sw_2$ and $\widetilde{Sw}_2$ denotes respectively the action of step $(iii)$ and $(iv)$, $T_{U_{0}^{(1)}}$ labels instead the random tableau generated in step $(vi)$. At the end of the subroutine, a constrained random Clifford has been generated, where the randomness comes from two elements: the application of the random Clifford $U_{Sym}$ in step $(ii)(a)$ and the generation of a random Clifford $U_{0}^{(1)}$ on $n-(n_P+n_N)$ qubits.

The algorithm runs in a time $O(n^3)$. The time complexity of step $(i)$ is $O(n_P^2)$, due to the time complexity of the \ref{rou:sweeping}~\cite{berg2021SimpleMethodSampling}. Step $(ii)$ has time complexity $O((n-n_P)^3)$. Step $(iii)$ has time complexity $O((n_N)^2)$. The step $(v)$, corresponding to the generation or a random tableau, has time complexity $O((n-n_P-n_N)^2)$. While the last steps have time complexity $O(n)$. Then, one can synthesize the Clifford operator from the tableau in terms of CNOT, H, S with an overhead of $O(n^3)$ steps (see Ref.~\cite{aaronson2004ImprovedSimulationStabilizer}).
In the following, it will be discussed in more detail the action of every step on the incomplete tableau. Step $(i)$ being the iterated action of  the \ref{rou:sweeping} subroutine on the first $n_P$ pair of rows, maps $P$ to $I_{2n_P}$ (the $2n_P\times 2n_P$ identity matrix ), so step $(i)$ acts on the tableau as: 
\be
\tau_\mathfrak{h}\xrightarrow[]{Sw_1} Sw_1(\tau_\mathfrak{h})\equiv \begin{pmatrix}
I_{2n_P}& 0\\
0& N_{Sw_1}\\
0& W\\
\end{pmatrix}
\ee
where $N_{Sw_1}$ denotes the image of the matrix $N$ through the action of the sweeping $Sw_1$. The unknown part $\mathbf{0}_{2(n-n_P-n_N)}$ is untouched by any operation, being filled by zeros. 
The iterated action of the sweeping, being symplectic, preserves the commutation relations between the rows, and thus erases all the $1$ bits in the most significant qubit columns, which explains the zeros below $I_{2n_P}$. In step $(ii)$ to avoid the introduction of bias it is first applied a symplectic transformation $Sym$ on the last $2(n-n_P)$ bits of the tableau $\tau_\mathfrak{h}$. The resulting action is:
\be
Sw_1(R)\xrightarrow[]{Sym} Sym Sw_1(R)\equiv \begin{pmatrix}
I_{n_P}& 0\\
0& N_{Sym Sw_1}\\
0& \mathbf{0}_{2(n-n_P-n_N)}\\
\end{pmatrix}
\ee
where $N_{Sym Sw_1}$ denotes the action of $Sym$ on $N_{Sw_1}.$ Step $(iii)$, denoted with $Sw_2$, corresponds to the action of an iterated sweeping on $N_{Sym Sw_1}$. Looking only at the matrix $N_{Sym Sw_1}$, one obtains:
\be
N_{Sym Sw_1}\xrightarrow[]{Sw_2}N_{Sw_2 Sym  Sw_1}\equiv\begin{pmatrix}
    1&0&0&0&\ldots&0\\
    -&-&-&-&\ldots&-\\
    0&z_1&1&0&\ldots&0\\
    -&-&-&-&\ldots&-\\
\vdots&\vdots&\vdots&\vdots&\vdots&\vdots
\end{pmatrix}
\label{partsweeping}
\ee
Note that the iterated sweeping $Sw_2$, contrary to the previous cases, is unable to remove all the $1$ bits in the most significant columns. This degree of freedom is manifest in Eq.~\eqref{partsweeping} with the free $z_1$ bit: indeed, both $\bbbone_{[1]}\otimes \sigma^x$, and $\sigma^z\otimes \sigma^x$ do commute with $\sigma^x\otimes \bbbone_{[1]}$. The action of step $(iv)$, labeled with $\widetilde{Sw}_2$, is necessary to address this issue. $\widetilde{Sw}_2$ is a symplectic transformation that can be built from a Clifford circuit. The circuit, described in step $(iv)$, is made by a layer of Hadamard gates, a layer of CNOT gates, and another layer of Hadamard gates. The first layer is necessary to convert $\sigma^z$s into a $\sigma^x$s, the second layer is required to remove the redundant $\sigma^x$s, and the last one to restore the $\sigma^z$s. As a result the matrix $N_{Sw_2 Sym S_1}$ is then mapped in
\be
N_{Sw_2 Sym S_1}\xrightarrow[]{\widetilde{Sw}_2}N_{\widetilde{Sw}_2Sw_2 Sym Sw_1}\equiv\begin{pmatrix}
    1&0&0&0&\ldots&0\\
    -&-&-&-&\ldots&-\\
    0&0&1&0&\ldots&0\\
    -&-&-&-&\ldots&-\\
\vdots&\vdots&\vdots&\vdots&\vdots&\vdots
\end{pmatrix}
\ee
In step $(v)$ we complete $N_{Sw_2SymS_1}$ with $1$ bits to be the identity matrix. In step $(vi)$, the tableau is completed by the addition of a random tableau $T_{U_{0}^{(1)}}$. The resulting incomplete tableau is equal to:
\be
\left.\tau_\mathfrak{h}\right._{\widetilde{Sw}_2Sw_2 Sym Sw_1}\xrightarrow[]{\textit{Random Clifford sampling}}\begin{pmatrix}
I_{n_P}& 0&0\\
0& I_{n_N}&0\\
0&0& T_{U_{0}^{(1)}}\\
\end{pmatrix}
\ee
The final steps sample the phases of the constrained Clifford operator. At the end we obtain  
\be 
T_{U_0}=(\tau_\mathfrak{h}|\boldsymbol{\phi})=((\widetilde{Sw}_2Sw_2 Sym Sw_1)^{-1}T_{U_{0}^{(1)}}(I_{2n})|\boldsymbol{\phi})
\ee 

\subsection{Learning a Pauli string}\namedlabel{rou:leapau}{\textit{Learning a Pauli string}}
In this section, we describe the subroutine able to learn the adjoint action of a unitary operator $U$ on a given Pauli string, say $P$. We assume that $U^{\dag} P U\in\mathcal{P}$. Note that the following subroutine works also in the case $U^{\dag} P U\not\in\mathcal{P}$ and returns still a Pauli string by construction that, of course, will not correspond to the correct operator resulting from the adjoint action of $U$ on $P$. We want to stress that the strategy of the algorithm presented in Sec.~\ref{Sec:alg} is to pretend that $U^{\dag}PU\in\mathcal{P}$ for all $P\in\mathbb{P}[n-m]$ and then check the correct learning by \ref{rou:checkkill} subroutine.
\medskip
\hrule width \hsize \kern 0.5mm \hrule width \hsize
\begin{flushleft}
\begin{itemize}
    \item {\bf Input:} $n, U,P\in \mathbb{P}$
    \item {\bf Output:} \be\begin{cases}U^\dag  P U\in\mathbb{P}, \quad \text{if}\,\, U^\dag  P U\in\mathcal{P}\\
    \mathbb{P}\ni Q\neq U^\dag  P U, \quad \text{else} \end{cases}\nonumber\ee
\end{itemize}
For each $j=1,\ldots, n$, do:
\begin{enumerate}[label=(\roman*)]
\item Prepare the following states, defined on two copies of $\mathcal{H}$:
\be
\ket{\psi_{j}^{(0)}}:=\ket{\epr_j}\otimes \ket{0}^{\otimes 2} ,\quad
\ket{\psi_{j}^{(+)}}:= \ket{\epr_j}\otimes \ket{+}^{\otimes 2}
\ee
where $\ket{\epr_j}=2^{(n-1)/2}\sum_{k\neq j}\ket{kk}$ is a EPR pair but on the $j$-th qubit.
\item Evolve each branch pair with the scrambler $U$, i.e., $\ket{\psi_{j}^{(0,+)}(U)}=U\otimes U\ket{\psi_{j}^{(0,+)}}$.
\item Measure $M$ times the operator $P^{\otimes 2}$ on both $\ket{\psi_{j}^{(0,+)}(U)}$, collecting a string of $2M$ bits $\{\sigma_{1}^{(0)},\ldots, \sigma_{M}^{(0)}, \sigma_{1}^{(+)},\ldots, \sigma_{M}^{(+)}\}$. If $\sigma_{1}^{(0)}=\ldots=\sigma_{M}^{(0)}$ assign $s_{j}^{(0)}=1$ to the binary variable $s_{j}^{(0)}$, otherwise assign $s_{j}^{(0)}=0$. Analogously for $s_{j}^{(+)}$.
\begin{itemize}
    \item if $\{s_{j}^{(0)},s_{j}^{(+)}\}=\{1,1\}$, then the $j$-th component of $U^{\dag}\sigma U$ is the identity, 
    \item if $\{s_{j}^{(0)},s_{j}^{(+)}\}=\{1,0\}$ is the Pauli matrix $\sigma^z$, 
    \item if $\{s_{j}^{(0)},s_{j}^{(+)}\}=\{0,1\}$ is the Pauli matrix $\sigma^x$,
    \item if $\{s_{j}^{(0)},s_{j}^{(+)}\}=\{0,0\}$ is the Pauli matrix $\sigma^y$.
\end{itemize}
\end{enumerate}
\end{flushleft}
\hrule width \hsize \kern 0.5mm \hrule width \hsize
\medskip
The above algorithm requires $O(nM)$ queries to the unitary $U$, runs in a time $O(n^2M)$, and reveals the Pauli string image of the adjoint action of $U$ on $P$. We are interested in estimating the failure probability only for $P\in G(U_t)$. Thus, if $P\in G(U_t)$, then $U_{t}^{\dag}PU_t\in\mathbb{P}$ and we can easily estimate the probability of failure of the above algorithm. Indeed, the probability to fail the learning of $P$ is $2n\times 2^{-M}$, indeed the failure probability is given by the probability that an unbiased coin gives tail for $M$ tosses in a row, i.e., $2^{-M}$. See Fig.~\ref{fig:my_label} for a pictorial representation of the above algorithm.
\begin{figure}
    \centering
    \includegraphics[scale=.35]{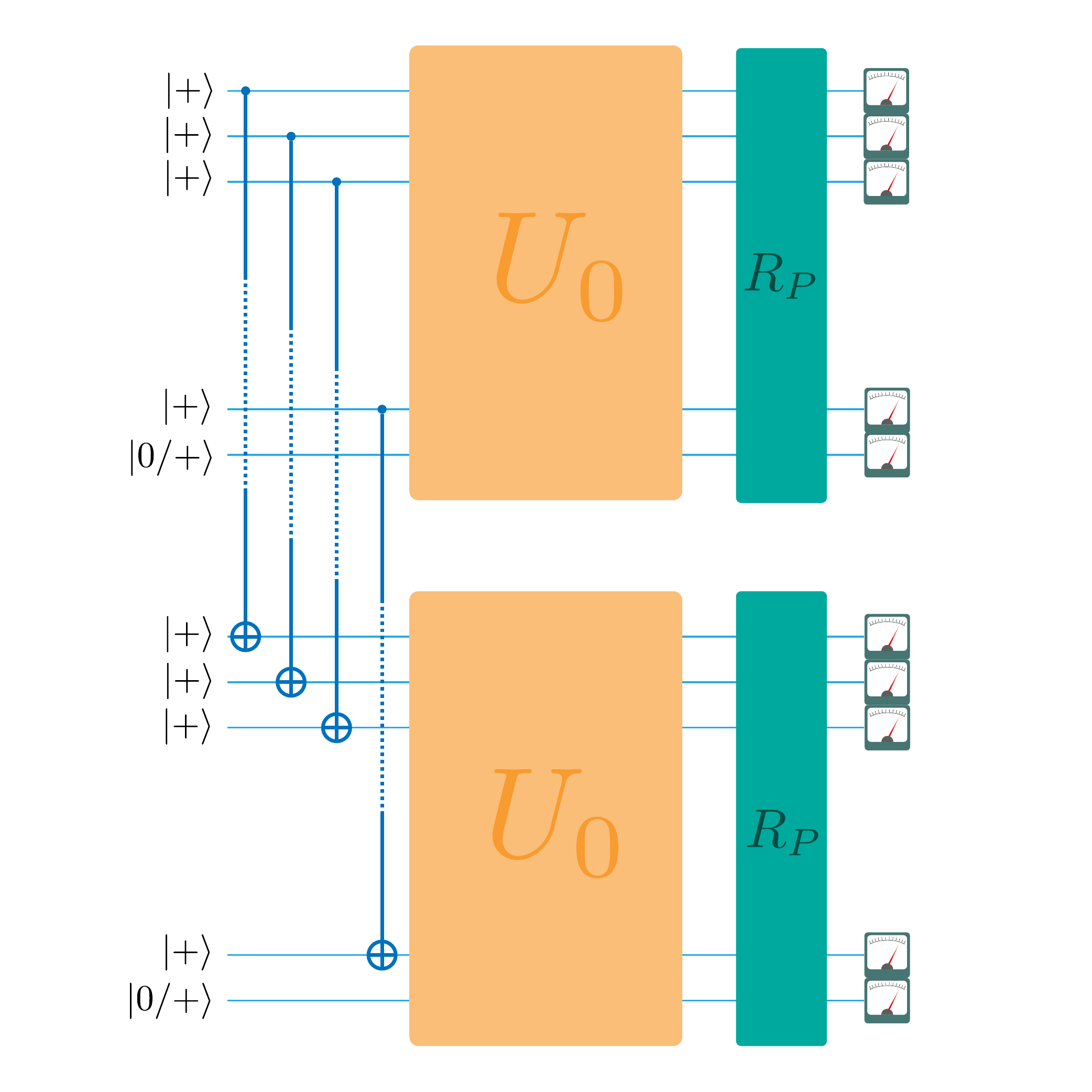}
    \caption{\raggedright Pictorial representation of \ref{rou:leapau} subroutine. The algorithm works with the quantum circuit sketched above consisting in $4$ main steps: preparation of the Bell state via $\ket{+}$ states and CNOTs; application of two copies of the scrambler $U_0$; a rotation $R_P$ conditioned on the measurement of the expectation value of a given Pauli operator $P$, e.g., $P=X\otimes Z \otimes Y \otimes \bbbone\otimes\cdots\otimes Z$ then $R_P=H\otimes\bbbone\otimes HS\otimes \bbbone\otimes\cdots\otimes \bbbone $. Finally a measurement in the computational basis. }
    \label{fig:my_label}
\end{figure}

\subsection{Verification and Removal}\namedlabel{rou:checkkill}{\textit{Verification and Removal}}
In this section, we describe the algorithm to verify whether the \ref{rou:leapau} subroutine worked correctly, or not. Denote with $P$ the input and $Q$ the output of \ref{rou:leapau} respectively, and proceed as follows.
\medskip
\hrule width \hsize \kern 0.5mm \hrule width \hsize
\begin{flushleft}
\begin{itemize}
    \item {\bf Input:} $n,P,Q\in\mathbb{P}, U_t$
    \item {\bf Output:} yes or no
\end{itemize}
\begin{enumerate}[label=(\roman*)]
    \item Construct a Bell pair $\ket{EPR}$ between two copies of $\mathcal{H}$.
    \item Evolve one branch with $U_t$ and obtain $\ket{U_t}\equiv\bbbone\otimes U_t\ket{EPR}$.
    \item Measure the expectation value $\braket{U_t|Q\otimes P|U_t}$ up to an error $\epsilon\le 2^{-2t}$;
    \item If $\braket{U_t|Q\otimes P|U_t}=\pm1$, then output yes; otherwise output no.
\end{enumerate}
\end{flushleft}
\hrule width \hsize \kern 0.5mm \hrule width \hsize
\medskip
The above discrimination works because:
\be
\braket{U_t|Q\otimes P|U_t}=\pm 1, \quad \text{iff}\,\,  U_t^\dag P U_t=\pm Q
\ee
The above algorithm requires $O(\epsilon^{-2})$ queries to $U_t$, and runs in a time $O(\epsilon^{-2})$. The key insight here is that since $U_t$ is a $t$-doped Clifford circuit, the expectation value $\braket{U_t|Q\otimes P|U_t}$ takes discrete values. In other words, there exists a minimal resolution $\delta_t$, defined as the minimum difference between two consecutive values of $\braket{U_t|Q\otimes P|U_t}$, denoted as $\delta_t\equiv\min|\braket{U_t|Q\otimes P|U_t}-\braket{U_t|Q^{\prime}\otimes P^{\prime}|U_t}|$. Thus, given $\epsilon$ such that $\epsilon\le \delta_{t}$, the learner can determine the expectation value $\braket{U_t|Q\otimes P|U_t}$ exactly. In Appendix~\ref{appD}, we present the Finite resolution Lemma~\ref{lemma:finite_resolution}, for which we bound $\delta_t\ge 2^{-bt}$, where $b\simeq 2.27$. Additionally, we expect that the average case scenario will yield even more favorable results as our Numerics in Sec.~\ref{sec:numerics} demonstrate. We therefore arrive at the conclusion that, in the worst case scenario, by selecting $b<2.5$, it is possible to determine whether $P\in G(U_t)$, meaning whether $P$ is a preserved Pauli string, with $O(2^{5t})$ queries to the $t$-doped Clifford circuit $U_t$.

\subsection{Phase check}\namedlabel{rou:phasecheck}{\textit{Phase check}}
In this section, we introduce the subroutine to learn the phase of the Pauli string, associated with the adjoint action $U^\dag PU$. Let be $\ket{epr}_{+}\equiv\frac{1}{\sqrt{2}}(\ket{00}+\ket{11})$ and $\ket{epr}_{-}\equiv\frac{1}{\sqrt{2}}(\ket{00}-\ket{11})$. The action of $\bbbone_{2}$, $\sigma^{x\otimes 2}$ or $\sigma^{z\otimes 2}$ on $\ket{epr}_{+}$ returns $\ket{epr}_{+}$; similarly the action of $\sigma^{y\otimes 2}$ on $\ket{epr}_{-}$ returns $\ket{epr}_{-}$. Let $P$ be the Pauli string of which one wants to learn the phase $s\in\{0,1\}$ of through the action of a unitary $U$. In the algorithm, it is assumed that $P(U) \equiv U^\dag PU\in\mathcal{P}$.
\medskip
\hrule width \hsize \kern 0.5mm \hrule width \hsize
\begin{flushleft}
\begin{itemize}
    \item {\bf Input:} {$n, P(U) \equiv U^\dag P U$}
    \item {\bf Output:} {$s\in\{0,1\}$}
\end{itemize}
\begin{enumerate}[label=(\roman*)]
    \item Use the \ref{rou:constrainedcliff} subroutine to build a Clifford $U_{0}$ such that $U_{0}^\dag PU_{0}=\pm P(U)$.
    \item Read each element of the Pauli string $U_{0} PU_{0}^\dag$ and build the $n$-fold state $\ket{s}$, constructed as a tensor product of $\ket{epr}_{+}$ and $\ket{epr}_{-}$ in the following way: if the $i$-th element of the Pauli string is $\bbbone_{[1]},\sigma^{x}$ or $\sigma^z$, then the $i$-th element of $\ket{s}$ is $\ket{epr}_{+}$, otherwise the $i$-th element of $\ket{s}$ is $\ket{epr}_{-}$.
    \item Evolve $\ket{s}$ with $U\otimes V$, and let $s$ be the result of a one-shot measurement of the operator $P^{\otimes 2}$. If $s=1$, then the phase is $+1$; conversely, if $s=-1$ the phase is $-1$.
\end{enumerate}
\end{flushleft}
\hrule width \hsize \kern 0.5mm \hrule width \hsize
\medskip
The algorithm needs one query to $U$, and runs in a time $O(n^3)$.

\section{Proof of Theorem~\ref{th1}}\label{App: proofs}
In order to prove Theorem~\ref{th1}, we first enunciate and prove a series of lemmas. Afterward, the proof of the theorem will  descend in a straightforward fashion. 

\begin{lemma}[Eq.~\eqref{eq:11}]\label{proofeq11}
    Let $U_t$ be  a $t$-doped Clifford circuit, the fidelity in Eq.~\eqref{fidelityunknown} can be written as:
    \be
    \mathcal{F}_{V}(U_t)=\frac{1+R}{d_{A}^{2}\Omega_{G_D}(U_t)+R^{\prime}}
    \ee
    where 
\ba
R&:=&\frac{1}{d|G_D(U_t)|}\sum_{P_D\in \overline{G_D}}\tr(P_{D}(U_t)P_{D}(V))\\
R^{\prime}&:=&\frac{1}{d|G_D(U_t)|}\sum_{P_D\in \overline{G_D},P_A}\tr(P_AP_{D}(U_t)P_AP_{D}(V))\nonumber\\
\ea
\end{lemma}

\begin{lemma}\label{lemmag}
Let $U_t$ be a $t$-doped Clifford circuit, the cardinality of $G_{[n-m]}(U_t)$, defined in Eq.~\eqref{gddefinition}, is lower bounded as $|G_{[n-m]}(U_t)|\ge 2^{2n-2m-t}$.
\end{lemma}
\begin{corollary}\label{lemmagd}
Let $U_t$ be a $t$-doped Clifford circuit, the cardinality of $G_{D}(U_t)$, defined in Eq.~\eqref{gddefinition}, is lower bounded as $|G_{D}(U_t)|\ge 2^{2|D|-t}$.
\end{corollary}
\begin{lemma}\label{lemma2}
If $\tr(P_{D}(U_t)P_{D}(V))=0$ for all $P_{D}\in \mathbb{P}(D)\setminus G_D(U_t)$, then $R=0$ and $R^{\prime}=0$. Thus
\be
\underset{V\in{\mathcal{V}}_{U_t}}{\operatorname{Pr}}[R=0,R^{\prime}=0]\ge\underset{V\in{\mathcal{V}}_{U_t}}{\operatorname{Pr}}[\tr(P_{D}(U_t)P_{D}(V))=0, \forall P_{D}\in \mathbb{P}(D)\setminus G_D(U_t)]
\ee
\end{lemma}

\begin{lemma}\label{probabilisticlemma}
For a random Clifford decoder $V\in\mathcal{V}_{U_t}^{D}$, the probability that $\tr(P_{D}(U_t)P_{D}(V))=0$ for all $P_{D}\in \mathbb{P}(D)\setminus G_D(U_t)$ is lower bounded by
\be
\underset{V\in{\mathcal{V}}_{U_t}^{D}}{\operatorname{Pr}}[\tr(P_{D}(U_t)P_{D}(V))=0, \forall P_{D}\in \mathbb{P}(D)\setminus G_D(U_t)]\ge 1-\frac{2^{t}}{d_{C}^{2}}\nonumber
\ee
\end{lemma}

\begin{lemma}\label{lemmaunitaryfidelity}
Let $\mathcal{F}_{V}(U_t)$ the fidelity defined in Eq.~\eqref{fidelityunknown}. If $V\in\mathcal{V}_{U_t}$ and $|D|=n$, $|C|=0$ then the fidelity becomes the gate fidelity
\be
\mathcal{F}_{V}(U_t)=\frac{|\tr(V^{\dag}U_t)|}{d^2}
\ee
and $\mathcal{F}_{V}(U_t)<1$ is strictly less than one if and only if $U_t$ is a non-Clifford unitary operator.
\end{lemma}

\textbf{Proof of Lemma~\ref{proofeq11}.} Rewrite the fidelity $\mathcal{F}_{V}$ in Eq. \eqref{fidelityunknown} as:
\be
\mathcal{F}_{V}(U_t)=\frac{d^{-1}\sum_{P_D}\tr(P_{D}(U_t)P_{D}(V))}{d^{-1}\sum_{P_A,P_D}\tr(P_AP_D(U_t)P_AP_D(V))}
\label{fidelityratio}
\ee
Then, let us evaluate the numerator and denominator separately, the numerator can be rewritten as:
\be
d^{-1}\sum_{P_D}\tr(P_{D}(U_t)P_{D}(V))=|G_{D}(U_t)|+d^{-1}\sum_{P_D\in \overline{G_D}(U_t)}\tr(P_{D}(U_t)P_{D}(V))
\ee
where we used the fact that $U_t^{\dag}P_DU_t=V^{\dag}P_DV$ for any $P_D\in G_{D}(U_t)$, and defined $\overline{G_D}(U_t)$ as the complement set of $G_{D}$. While for the denominator:
\ba
d^{-1}\sum_{P_A,P_D}\tr(P_AP_D(U_t)P_AP_D(V))&=&d^{-1}\!\!\!\!\!\!\!\!\sum_{P_D\in G_D,P_A}\tr(P_AP_{D}(U_t)P_AP_{D}(U_t))\nonumber\\&+&d^{-1}\!\!\!\!\!\!\!\!\sum_{P_D\in \overline{G_D},P_A}\tr(P_AP_{D}(U_t)P_AP_{D}(V))
\label{firstdenominator}
\ea
Define the following two quantities:
\ba
R&:=&\frac{1}{d|G_D(U_t)|}\sum_{P_D\in \overline{G_D}}\tr(P_{D}(U_t)P_{D}(V))\\
R^{\prime}&:=&\frac{1}{d|G_D(U_t)|}\sum_{P_D\in \overline{G_D}(U_t),P_A}\tr(P_AP_{D}(U_t)P_AP_{D}(V))\nonumber\\
\ea
Then define the truncated OTOC, similarly to Eq. \eqref{otocdef} as
\be
\otoc_{G_D}(U_t)=\frac{1}{d}\braket{\tr(P_AP_{D}(U_t)P_AP_{D}(U_t))}_{\mathbb{P}(A), G_(U_t)}
\label{truncated}
\ee
where we defined $\braket{\cdot}_{G_D}:=|G_{D}(U_t)|^{-1}\sum_{P_D\in G_D(U_t)}(\cdot)$. Note that we can write:
\be 
\mathcal{F}_{V}(U_t)=\frac{|G_D(U_t)|(1+R)}{|G_D(U_t)|(d_{A}^{2}\otoc_{G_D}(U_t)+R^{\prime})}=\frac{1+R}{d_{A}^2\otoc_{G_D}(U_t)+R^\prime}
\ee 

\textbf{Proof of Lemma~\ref{lemmag}.} First of all, let us recall the definition
\be 
G_{[n-m]}(U_t):=\{P\in \mathbb{{P}}_D\, |\,P(U_t)\equiv U_t^\dag P U_t\in \mathcal{P} \}
\ee
let us prove that it is a subgroup of the full Pauli group. It is trivial to say that $\bbbone\in G_{[n-m]}(U_t)$. Than, since $P^{-1}= P$ we have that for any $P\in G_{[n-m]}(U_t)$, then $P^{-1}\in G_{[n-m]}(U_t)$. Finally, thanks to the unitarity of $U_t$ we have that if $P,Q\in G_{[n-m]}(U_t)$ then $PQ\in G_{[n-m]}(U_t)$. Now, let us prove that $|G_{[n-m]}(U_t)|\ge 2^{2n-2m-t}$. First of all define $G(U_t)$ as:
\be 
G(U_t):=\{P\in \mathbb{{P}}\, |\,P(U_t)\in \mathcal{P} \}
\ee
It is known that~\cite{jiang2021LowerBoundTcount} $|G(U_t)|\ge 2^{2n-t}$, and it is clear that $G(U_t)\le \mathbb{P}$. Let us use the following group theory result: let $A,B\le H$ two subgroup of $H$, then:
\be
|A\cap B|\ge \frac{|A||B|}{|H|}
\ee
First, note that we can write:
\ba
G_{[n-m]}(U_t)&=&(\mathbb{P}([n-m])\cap U_t\mathbb{P}U_t^\dag)=(\mathbb{P}([n-m])\cap \mathbb{P}\cap U_t\mathbb{P}U_t^\dag)\nonumber\\
&=& (\mathbb{P}([n-m])\cap G(U_t))
\ea
We find:
\be
|G_{[n-m]}(U_t)|\ge \frac{|\mathbb{P}([n-m])||G(U_t)|}{|\mathbb{P}|}=2^{2n-2m-2t}
\ee

\textbf{Proof of Lemma~\ref{lemma2}.} Let us recall the lemma:
The proof for $R=0$ is trivial. Let us proceed to the proof for $R^{\prime}=0$. Since $V$ is a Clifford operator, then $P_{D}(V)\in \mathbb{P}$, and thus $P_{D}(V)P_{A}=\phi(P_{A},P_{D}(V)) P_{A}P_{D}(V)$, where the phase is defined as:
\be
\phi(P_{A},P_{D}(V)):=\frac{1}{d}\tr(P_{A}P_{D}(V)P_AP_{D}(V))
\ee
we can thus rewrite $R^{\prime}$ as:
\be
R^{\prime}=\sum_{P_D\in \mathbb{P}(D)\setminus G_D(U_t),P_A}\phi(P_{A},P_{D}(V)) \tr(P_{D}(U_t)P_D(V))
\ee
thus from the last equality if $\tr(P_{D}(U_t)P_{D}(V))=0$ for all $P_{D}\in\mathbb{P}(D)\setminus G_D(U_t)$, then $R^{\prime}=0$.

\textbf{Proof of Lemma~\ref{probabilisticlemma}.} Let $\mathcal{V}_{U_t}^{D}=\{V\in\mathcal{C}(n)\,|\, V^{\dag}PV=U_t^{\dag}PU_t\,, \forall P\in G_{D}(U_t)\}$ be the set of all Clifford decoders that can be found by the algorithm in Sec.~\ref{Sec:alg}. We are indeed only interested in random Clifford decoders modulo phases. 

First of all, defined $\overline{\mathfrak{g}_{D}}(U_t)$ as the set of Pauli operators such that $\braket{\mathfrak{g}_{D}\cup\overline{\mathfrak{g}_{D}}(U_t)}=\mathbb{P}(D)$. After the injection of $t$ non-Clifford gates, we have that $|\overline{\mathfrak{g}_{D}}(U_t)|\le t$. Each $P_{D}\in \mathbb{P}(D)\setminus G_D(U_t)$ can be rewritten as $P_{D}=P_{D}^{\prime} \widetilde{P}_{D}$, where $P_{D}^{\prime}\in G_D(U_t)$, while $\widetilde{P}_{D}$ belongs to the set $\overline{G}_D(U_t)$ generated by $\overline{G}_D(U_t)=\braket{\overline{\mathfrak{g}_{D}}(U_t)}\setminus\bbbone$. Since $|\mathbb{P}(D)\setminus G_D(U_t)|=d_{D}^{2}-|G_{D}(U_t)|$, we have that $|\overline{G}_D(U_t)|=d_{D}^{2}/|G_{D}(U_t)|-1$, where the $-1$ comes from the fact that $\overline{G}_D(U_t)$ does not contain the identity. For any $P_{D}\in \mathbb{P}(D)\setminus G_D(U_t)$, we thus write:
\be
\tr(P_D(U_t)P_D(V))=\pm\tr(\widetilde{P}_D(U_t)\widetilde{P}_{D}(V))
\ee
where we used the unitarity of $U_t$ and $V$ to write $P_{D}(U_t)=P_{D}^{\prime}(U_t)\widetilde{P}_D(U_t)$, and $P_{D}(V)=\pm\widetilde{P}_D(V)P_{D}^{\prime}(V)$. Thus, 
\be
\underset{V\in \mathcal{V}_{U_t}^{D}}{\operatorname{Pr}}[\tr(P_D(U_t)P_D(V))=0, \forall P_D\in \mathbb{P}(D)\setminus G_D(U_t)]=\underset{V\in \mathcal{V}_{U_t}^{D}}{\operatorname{Pr}}[\tr(\widetilde{P}_D(U_t)\widetilde{P}_D(V))=0, \forall \widetilde{P}_D\in \overline{G}_D(U_t)]
\ee
$\widetilde{P}_{D}(U_t)$ is, in general, a combination of $l$ Pauli operators, where $2\le l\le 2^{t}$, i.e., $\widetilde{P}_{D}(U_t)=\sum_{i=1}^{l}\alpha_ip_i$, where $ p_{i}\in\mathbb{P}$ and $\alpha_i=d^{-1}\tr(p_i \widetilde{P}_{D}(U_t))$. Thus, for a single $\widetilde{P}_{D}\in\overline{G}_D(U_t)$, the probability $ \operatorname{Pr}[\tr(\widetilde{P}_{D}(U_t)\widetilde{P}_{D}(V))=0]=1-\operatorname{Pr}[\tr(\widetilde{P}_{D}(U_t)\widetilde{P}_{D}(V))\neq0]$ and 
\be
\underset{V\in \mathcal{V}_{U_t}^{D}}{\operatorname{Pr}}[\tr(\widetilde{P}_{D}(U_t)\widetilde{P}_{D}(V))\neq0]=\underset{V\in \mathcal{V}_{U_t}^{D}}{\operatorname{Pr}}\left[\bigcup_{i=1}^{l}\tr(p_i\widetilde{P}_{D}(V))\neq0\right]
\ee
the above is true because if $\tr(p_{i}P_{D}(V))\neq0$ for some $i=1,\ldots,l$, then $\tr(p_{j}P_{D}(V))=0$ for any $j\neq i$. By using Fr\'echet inequality we can upper bound the above probability by:
\be
\underset{V\in \mathcal{V}_{U_t}^{D}}{\operatorname{Pr}}[\tr(\widetilde{P}_{D}(U_t)\widetilde{P}_{D}(V))\neq0]\le\sum_{i=1}^{l}\underset{V\in \mathcal{V}_{U_t}}{\operatorname{Pr}}[\tr(p_{i}\widetilde{P}_{D}(V))\neq0]
\label{b11}
\ee
    Looking at the above equation, for any given $p_i$, there are two occurring cases: either there is no $V\in\mathcal{V}_{U_t}^{D}$ such that $p_l\propto \widetilde{P}_D(V)$ for every $\widetilde{P}_D(V)$, or there exists $V\in\mathcal{V}_{U_t}^{D}$ such that $p_l\propto \widetilde{P}_D(V)$. We thus only consider the latter, being the worst case scenario for $\tr(p_{l}\widetilde{P}_{D}(V))\neq0$. Denote as $\widetilde{\mathcal{V}}_{U_t}^{D}(P):=\{V^{\dag}PV\in \mathbb{P}\,|\,V\in\mathcal{V}_{U_t}^{D}\}$ the set of images of the Pauli operator $P$ through the action of random decoders belonging to $\mathcal{V}_{U_t}^{D}$. For example, if $P\in G_{D}(U_t)$, then $\widetilde{\mathcal{V}}_{U_t}^{D}(P)=\{U_t^{\dag}PU_t\}$, i.e., $\widetilde{\mathcal{V}}_{U_t}^{D}(P)$ is just the singleton of the image of $P$ through $U_t$ by construction. Since the algorithm in Sec.~\ref{Sec:alg} is generating a decoder $V$ uniformly at random from the set $\mathcal{V}_{U_t}^{D}$, we conclude that a single $p_{i}$ with $i=1,\ldots, l$:
\be
\underset{V\in \mathcal{V}_{U_t}^{D}}{\operatorname{Pr}}[\tr(p_{i}\widetilde{P}_{D}(V))\neq0]\le\frac{1}{|\widetilde{\mathcal{V}}_{U_t}^{D}(\widetilde{P}_{D})|}
\label{eqww}
\ee
where $|\widetilde{\mathcal{V}}_{U_t}^{D}(\widetilde{P}_{D})|$ is the cardinality of the set $\widetilde{\mathcal{V}}_{U_t}^{D}(\widetilde{P}_{D})$, i.e., the number of all possible Pauli operator (not belonging to $G_D(U_t)$) resulting from the adjoint action of $V\in\mathcal{V}_{U_t}^{D}$. 
We have the following lemma
\begin{lemma}
Let $U_t$ be a $t$-doped Clifford circuit, then \be
\min_{\widetilde{P}_D\in\overline{G}_D(U_t)}|\widetilde{\mathcal{V}}_{U_t}^{D}(\widetilde{P}_D)|\ge 2^t d_{C}^2\,.
\ee
\begin{proof}
First of all note that 
\be
\min_{\widetilde{P}_D\in\overline{G}_D(U_t)}|\widetilde{\mathcal{V}}_{U_t}^{D}(\widetilde{P}_D)|= \min_{\widetilde{\sigma}\in\overline{\mathfrak{g}_{D}}(U_t)}|\widetilde{\mathcal{V}}_{U_t}^{D}(\widetilde{\sigma})|\,,
\ee
i.e., the minimum number of possible images of a generator $\widetilde{\sigma}\in\overline{\mathfrak{g}_{D}}(U_t)$ of the group $\overline{G}_D(U_t)$ is exactly the minimum number of possible images of an element of the group $\overline{G}_D(U_t)$, being $\widetilde{\sigma}\in\overline{G}_{D}(U_t)$.  Therefore, we need to compute in how many ways we can write a compatible row corresponding to the map $V\,: \widetilde{\sigma}\in \overline{\mathfrak{g}_{D}}(U_t)\mapsto \widetilde{\sigma}(V)$ in the incomplete tableau $T_{V}$. To take into account this, we need to look at the \ref{rou:constrainedcliff} subroutine (in particular Eq.~\eqref{tableau:inc}). There are two types of rows to fill: the (unpaired) ones belonging to $N$ and the ones belonging to $\mathbf{0}$. To fill a row belonging to $N$, one needs to consider the anticommutation relation with the already known row (e.g., the consecutive one). There are 
\be
2^{n_N}\times 2^{2(n-n_P-n_N)}=2^{2n-2n_P-n_N}
\ee
ways to write the resulting Pauli string. Let us explain the above counting: in the submatrix of unpaired rows $N$, there are $2n_N$ rows, and only $n_N$ of them are fixed. Thus, to fix just one of the empty rows, one has $2^{n_N}$ possibilities to write a $2n$ bit string corresponding to a valid Pauli operator. Conversely, to fill rows in the empty part of the incomplete tableau, i.e., $\mathbf{0}$, one has $2^{2n-2n_P-2n_N}$ degrees of freedom, which correspond to the number of rows contained in $\mathbf{0}$. Using the fact that $2^{2n_P+n_N}=|G_{D}(U_t)|$ by construction, one has that
\be
\min_{\widetilde{\sigma}\in\overline{\mathfrak{g}_{D}}(U_t)}|\widetilde{\mathcal{V}}_{U_t}^{D}(\widetilde{\sigma})|=\frac{d^2}{|G_{D}(U_t)|}
\ee
\end{proof}
\end{lemma}
As an immediate corollary of the above lemma, from Eq.~\eqref{eqww}, we have
\be
\forall p_i\in \{p_{1},\ldots, p_l\}\quad \underset{V\in \mathcal{V}_{U_t}}{\operatorname{Pr}}[\tr(p_{j}\widetilde{P}_{D}(V))\neq0]\le \frac{|G_{D}(U_t)|}{d^2}
\ee
For a single $\widetilde{P}_{D}\in\overline{G}_{D}(U_t)$ from Eq.~\eqref{b11} we can write
\be
\operatorname{Pr}[\tr(\widetilde{P}_{D}(U_t)\widetilde{P}_{D}(V))=0]\ge 1-\frac{l|G_{D}(U_t)|}{d^2}\,.
\ee
To have the probability for every Pauli $\widetilde{P}_{D}\in\overline{G}_{D}(U_t)$, we use the Fr\'echet bound on intersection of events:
\be
\operatorname{Pr}\left[\bigcap_{\forall \widetilde{P}_{D}\in\overline{G}_D(U_t)}(U_t)\tr(\widetilde{P}_{D}(U_t)\widetilde{P}_{D}(V))=0 \right]\ge 1-\sum_{\alpha=1}^{|\overline{G}_D(U_t)|}\frac{l_{\alpha}|G_{D}(U_t)|}{d^2}
\ee
Using the fact that $2\le l_{\alpha}\le 2^t$ for any $\alpha$ and that $|\overline{G}_D(U_t)||G_{D}(U_t)|=d_{D}^{2}-|G_{D}(U_t)|\le d_{D}^{2}$, we finally proved the statement:
\be
\operatorname{Pr}[\tr(P_{D}(U_t)P_{D}(V))=0, \forall P_{D}\in\mathbb{P}(D)\setminus G_D(U_t)]\ge 1-\frac{2^{t}d_{D}^{2}}{d^2}=1-\frac{2^{t}}{d_{C}^{2}}
\ee
\qed

\textbf{Proof of Lemma~\ref{lemmaunitaryfidelity}}
Recall Eq.~\eqref{fidelityunknown} for $\mathbb{P}(D)=\mathbb{P}$
\be
\mathcal{F}_{V}(U_t)=\frac{\braket{\tr(P(U_t)P(V))}_{P\in\mathbb{P}}}{d_{A}^{2}\braket{\tr(P_AP(U_t)P_AP(V))}_{P_A,P\in\mathbb{P}}}
\ee
Computing both average over $\mathbb{P}$, and using $\sum_{P}PAP=d\tr(A)$, we have:
\be
\mathcal{F}_{V}(U_t)=\frac{d^{-2}|\tr(U_tV^{\dag})|^2}{d_{A}^{2}d^{-2}\sum_{P_A}\tr(P_A)^2}=\frac{|\tr(U_tV^{\dag})|^2}{d^2}
\ee
Let $V\in\mathcal{V}_{U_t}$, i.e., $V^{\dag}PV=U_t^{\dag}PU_t$ for every $P\in\mathbb{P}$. Then, the unitary fidelity can be written as:
\be
\frac{|\tr(U_tV^{\dag})|^2}{d^2}=\frac{1}{d^{2}}|G(U_t)|+\frac{1}{d^{3}}\sum_{P\not\in G(U_t)}\tr(P(U_t)P(V))
\ee
now $P(U_t)$ is, at least a summation over $2$ Pauli strings and therefore $d^{-1}\tr(P(U_t)P(V))<1$. We thus obtain the following bound:
\be
\frac{|\tr(U_tV^{\dag})|^2}{d^2}<\frac{1}{d^{2}}(|G(U_t)|+|P\setminus G(U_t)|)=1
\ee

\section{Finite resolution Lemma}\label{appD}
In this section, we present the proof of the finite resolution of the expectation values of Pauli operators for the Choi state of a $t$-doped Clifford circuit $U_t$. Let us state it formally. Let us define the following set
\be
\mathcal{S}_{U_t}=\{P\in\mathbb{P}\,|\, \braket{U_t|P|U_t}\neq 0\}
\ee
the set of Pauli operators having nonzero expectation value on the Choi state $\ket{U_t}$ associated with the $t$-doped Clifford circuit $U_t$. Define
\be
\delta_{t}=\min_{\substack{P,Q\in\mathcal{S}_{U_t}\\ \braket{P}\neq \braket{P^{\prime}}}}|\braket{U_t|(P-P^{\prime})|U_t}|
\ee
then, the following lemma holds.
\begin{lemma}[Finite resolution lemma]\label{lemma:finite_resolution} Let $U_t$ be a $t$-doped Clifford circuit. Let $\ket{U_t}$ be the Choi state associated with $U_t$. Then, the following bounds hold
\be
\min_{P\in S_{U_t}}|\braket{U_t|P|U_t}|\ge  \frac{1}{3\sqrt{2}^{t-1}}\left(1-\frac{1}{\sqrt{2}}\right)^{t}\label{minbound}
\ee
while
\be
\delta_t\ge  \frac{1}{6\sqrt{2}^{t-1}}\left(1-\frac{1}{\sqrt{2}}\right)^{t}\label{proofdeltabound}
\ee
\end{lemma}

We proceed as follows: we first bound $\min_{P\in S_{U_t}}|\tr(P\st{U_t})|$ and then we bound the gap $\delta_t$, as the second will be just a trivial generalization of the first one. Before proving the statement, let us recall that the action of a $T$ gate, defined as $T=\operatorname{diag}(1,e^{-i\pi/4})$ applied on the $i$-th qubit on a Pauli operator $P$ results
\be
T_iPT_i^{\dag}=\begin{cases}
    P,\quad [P,Z]=0\\
    \frac{1}{\sqrt{2}}(P-iZ_iP),\quad \{P,Z_i\}=0
\end{cases}
\ee
Note that, to bound $\min_{P\in S_{U_t}}|\braket{U_t|P|U_t}|$, we can alternatively bound $\min_P|\tr(U_{t}PU_{t}^{\dag}\sigma)|$ for $\sigma$ being an arbitrary stabilizer state. Let us look at the action of $U_t$ on a Pauli operator. First decompose $U_{t}=\prod_{i=1}^{t}U_{1}^{(i)}$ where $U_{1}^{(i)}$ is a $(t=1)$-doped Clifford circuit. Let us set up the following notation.
\be
U_{1}^{(1)}PU_{1}^{(1)\dag}=x_{(0)}P_{(0)}+\frac{x_{(1)}}{\sqrt{2}}P_{(1)}+\frac{x_{(2)}}{\sqrt{2}}P_{(2)}\label{1111}
\ee
where $x_{(0)},x_{(1)},x_{(2)}\in \{-1,0,+1\}$. Equation~\eqref{1111} must be understood as: there is a choice of $x_{(0)},x_{(1)},x_{(2)}$ and the respective Pauli operators $P_{(0)},P_{(1)},P_{(2)}$ such that the l.h.s. is equal to the r.h.s. of Eq.~\eqref{1111}. Before generalizing to the generic $t$, it is useful to act again on $U_{1}^{(1)}PU_{1}^{(1)\dag}$ with $U_{1}^{(2)}$.
\ba
U_{1}^{(2)}U_{1}^{(1)}PU_{1}^{(1)\dag}U_{1}^{(2)\dag}&=&x_{(00)}P_{(00)}+
\frac{1}{\sqrt{2}}(x_{(10)}P_{(10)}+x_{(01)}P_{(01)}+x_{(20)}P_{(20)}+x_{(02)}P_{(02)})\nonumber\\&+&
\frac{1}{2}(x_{(11)}P_{(11)}+x_{(12)}P_{(12)}+x_{(21)}P_{(21)}+x_{(22)}P_{(22)})\label{2222}
\ea
where each variable $x_{(ij)}$ for $i=0,1,2$ can take values in $x_{(ij)}\in\{-1,0,+1\}$. As one can see, the subscript string $(ij)$ attached to each variable $x_{(ij)}$ reveals how many times a $T$-gate splits the Pauli operator $P$ in $2$ Pauli operator with the corresponding $\frac{1}{\sqrt{2}}$ factor. Again, Eq.~\eqref{2222} must be understood as: there exists a choice of the variables $x_{(ij)}$ and the respective Pauli operators $P_{(ij)}$ for which the l.h.s. and the r.h.s. of Eq.~\eqref{2222} agrees. Now, that we set up the above general and powerful notation, we can easily generalize the action to $U_t$. We have the following
\be
U_{t}PU_{t}^{\dag}=\sum_{k=0}^{t}\sum_{\pi\in S_{k}}\frac{x_{\pi(\boldsymbol{y}_k)}}{\sqrt{2}^{k}}P_{\pi(\boldsymbol{y}_k)}\label{3333}\,.
\ee
In Eq.~\eqref{3333} above, we have defined a few elements. First of all, we defined the $t$-bits string $\boldsymbol{y}_k$ with Hamming weight $k$ as
\be
\boldsymbol{y}_{k}=(\underbrace{1,1,\ldots,1}_{k},0,\ldots,0)
\ee
Next, we defined a set $S_{k}$ of operations $\pi$ that act on $\boldsymbol{y}_{k}$. $S_{k}$ is the set containing all the permutations of the $k$ $1$s in $\boldsymbol{y}_k$ into $t$ spots, combined with the operation that transforms $1\leftrightarrow2$, in accordance with Eq.~\eqref{2222}. Let us illustrate this with an example. Set $t=2$ and $k=1$, so $\boldsymbol{y}_{1}=(10)$. All the possible permutations of $(10)$, combined with the operation $1\leftrightarrow 2$, result in the strings $(10),(01),(20),(02)$. Similarly, for $t=2$ and $k=2$, $\boldsymbol{y}_2=(11)$ and the set of operations in $S_{2}$ returns $(11),(22),(21),(12)$.

It is useful to count the number of operations within $S_{k}$ for fixed $k$. The set $S_{k}$ is the combination of $\binom{t}{k}$ many ways to permute $\boldsymbol{y}_k=(1,1,\ldots, 1,0,\ldots, 0)$ times the $2^k$ different choices of either $1$ or $2$ at any site. The above simple counting thus returns:
\be
\sum_{\pi\in S_{k}}=2^{k}\binom{t}{k}
\ee
From Eq.~\eqref{3333}, we can formally compute the expectation value of $U_{t}PU_{t}^{\dag}$ with a generic stabilizer state $\sigma$ and get
\be
\tr(U_{t}PU_{t}^{\dag}\sigma)=\sum_{k=0}^{t}\sum_{\pi\in S_{k}}\frac{\tilde{x}_{\pi(\boldsymbol{y}_k)}}{\sqrt{2}^{k}}
\ee
where we defined the variables $\tilde{x}_{\pi(\boldsymbol{y}_k)}:=x_{\pi(\boldsymbol{y}_k)}\tr(P_{\pi(\boldsymbol{y}_k)}\sigma)\in\{-1,0,+1\}$ because $\tr(P_{\pi(\boldsymbol{y}_k)}\sigma)\in\{-1,0,+1\}$. Now, we set up all the necessary notation to finally prove Eq.~\eqref{minbound}. 

We are interested in computing the minimum achievable value for $\tr(U_{t}PU_{t}^{\dag}\sigma)$. Let us first multiply both sides for $\sqrt{2}^{t}$. We thus get
\ba
\sqrt{2}^{t}\tr(U_{t}PU_{t}^{\dag}\sigma)=\sum_{k=0}^{t}\sum_{\pi\in S_{k}}\sqrt{2}^{t-k}\tilde{x}_{\pi(\boldsymbol{y}_k)}=
\sum_{l=0}^{t}\sum_{\pi\in S_{t-l}}\sqrt{2}^{l}\tilde{x}_{\pi(\boldsymbol{y}_{t-l})}\nonumber
\ea
where in the second equality, we defined $l=t-k$. Let us set $t$ to be even and split odd and even terms in the sum
\ba
\sum_{l=0}^{t}\sum_{\pi\in S_{t-l}}\sqrt{2}^{l}\tilde{x}_{\pi(\boldsymbol{y}_{t-l})}=\sum_{l=0}^{t/2}2^l\sum_{\pi\in S_{t-2l}} \tilde{x}_{\pi(\boldsymbol{y}_{t-2l})}+
\sqrt{2}\sum_{l=0}^{t/2-1}2^l\sum_{\pi\in S_{t-(2l+1)}} \tilde{x}_{\pi(\boldsymbol{y}_{t-(2l+1)})}\nonumber
\ea
Define the following function of $t$
\ba
A(t)&:=&\sum_{l=0}^{t/2}2^l\sum_{\pi\in S_{t-2l}} \tilde{x}_{\pi(\boldsymbol{y}_{t-2l})}\\
B(t)&:=&\sum_{l=0}^{t/2-1}2^l\sum_{\pi\in S_{t-(2l+1)}} \tilde{x}_{\pi(\boldsymbol{y}_{t-(2l+1)})}
\ea
Note that $A(t),B(t)\in\mathbb{Z}$, i.e., they are positive and negative natural numbers, for any $t=2t^{\prime}$ for $t\in\mathbb{N}$. We can thus write 
\be
\sqrt{2}^{t}|\tr(U_{t}PU_{t}^{\dag}\sigma)|=|A(t)+\sqrt{2}B(t)|=|B(t)|\left|\sqrt{2}+\frac{A(t)}{B(t)}\right|\label{4444}
\ee
Therefore, the lower bound deals with the approximation of the algebraic number $\sqrt{2}$ by a rational number $A(t)/B(t)$. To make it explicit we can lower bound the r.h.s. of Eq.~\eqref{4444} as
\be
\sqrt{2}^{t}|\tr(U_{t}PU_{t}^{\dag}\sigma)|\ge |B(t)|\left|\sqrt{2}-\frac{|A(t)|}{|B(t)|}\right|\label{6666}
\ee
and we can invoke the Liouville Theorem (see Ref.~\cite{murty_liouville_2014}) of approximating a algebraic number $\alpha$ with two rational numbers $p,q\in\mathbb{Q}$ that reads: there exist a constant $c(\alpha)$ independent from $p,q$ such that 
\be
\left|\sqrt{2}-\frac{p}{q}\right|\ge \frac{c(\alpha)}{q^{D}}\label{5555}
\ee
where $D$ is the degree of the algebraic number $\alpha$. In the case of $\alpha=\sqrt{2}$ we have $D=2$ because $\sqrt{2}$ corresponds to the solution to the irreducible polynomial $z^2-2=0$ which has degree $2$, and $c(\sqrt{2})=\frac{1}{6}$~\cite{murty_liouville_2014}. Applying Eq.~\eqref{5555} to Eq.~\eqref{6666}, we thus get
\be
|\tr(U_{t}PU_{t}^{\dag}\sigma)|\ge \frac{1}{6|B(t)|\sqrt{2}^{t}}
\ee
We now are just left to find an upper bound to $B(t)$. We proceed with the following equality
\ba
|B(t)|&=&\left|\sum_{l=0}^{t/2-1}2^l\sum_{\pi\in S_{t-(2l+1)}} \tilde{x}_{\pi(\boldsymbol{y}_{t-(2l+1)})}\right|\le \sum_{l=0}^{t/2-1}2^l\sum_{\pi\in S_{t-(2l+1)}} |\tilde{x}_{\pi(\boldsymbol{y}_{t-(2l+1)})}|\nonumber\\&=&\sum_{l=0}^{t/2-1}2^l\sum_{\pi\in S_{t-(2l+1)}} = \sum_{l=0}^{t/2-1}2^l2^{t-(2l+1)}\binom{t}{2l+1}\nonumber\\&=&\frac{1}{\sqrt{8}}2^t\left[\left(1+\frac{1}{\sqrt{2}}\right)^t-\left(1-\frac{1}{\sqrt{2}}\right)^t\right]\le \frac{1}{\sqrt{8}}\left(1-\frac{1}{\sqrt{2}}\right)^{-t}\label{Bbound}
\ea
where in the second equality, we used the fact that $\tilde{x}_{\pi(\boldsymbol{y}_{t-(2l+1)})}\in\{-1,0,+1\}$ and in the last inequality, we used the fact that 
\be
\frac{1}{\sqrt{8}}2^t\left[\left(1+\frac{1}{\sqrt{2}}\right)^t-\left(1-\frac{1}{\sqrt{2}}\right)^t\right]\le\frac{1}{\sqrt{8}}2^t\left(1+\frac{1}{\sqrt{2}}\right)^t= \frac{1}{\sqrt{8}}\left(1-\frac{1}{\sqrt{2}}\right)^{-t}
\ee 
An analogous procedure with $t=2t^{\prime}+1$ with $t^{\prime}$ leads to the same exact bound. Therefore for any $t\in\mathbb{N}$, we find
\be
|\tr(C_{t}PC_{t}^{\dag}\sigma)|\ge \frac{\sqrt{8}}{6}\left(\frac{1}{\sqrt{2}}-\frac{1}{2}\right)^{t}
\ee
Now, let us turn to analyze the gap $\delta_t\equiv\tr[(P-P^{\prime})\psi_t]$. Using the same notation as before, we can write the adjoint action of $C_{t}$ on $P-P^{\prime}$ as follows
\be
U_{t}(P-P^{\prime})U^{\dag}_t=\sum_{k=0}^{t}\sum_{\pi\in S_{k}}\left(\frac{x_{\pi(\boldsymbol{y}_k)}}{\sqrt{2}^{k}}P_{\pi(\boldsymbol{y}_k)}+\frac{x^{\prime}_{\pi(\boldsymbol{y}_k)}}{\sqrt{2}^{k}}P^{\prime}_{\pi(\boldsymbol{y}_k)}\right)
\ee
and therefore, by repeating the same procedure, we can define 
\ba
A^{\prime}(t)&:=&\sum_{l=0}^{t/2}2^l\sum_{\pi\in S_{t-2l}} \tilde{x}_{\pi(\boldsymbol{y}_{t-2l})}+\tilde{x}^{\prime}_{\pi(\boldsymbol{y}_{t-2l})}\\
B^{\prime}(t)&:=&\sum_{l=0}^{t/2-1}2^l\sum_{\pi\in S_{t-(2l+1)}} \tilde{x}_{\pi(\boldsymbol{y}_{t-(2l+1)})}+\tilde{x}^{\prime}_{\pi(\boldsymbol{y}_{t-(2l+1)})}
\ea
and write the gap as
\be
\delta_t=\frac{1}{\sqrt{2}^t}|A^{\prime}(t)+\sqrt{2}B^{\prime}(t)|\ge \frac{|B^{\prime}(t)|}{\sqrt{2}^t}\left|\sqrt{2}-\frac{|A^{\prime}(t)|}{|B^{\prime}(t)|}\right|\ge \frac{1}{6\sqrt{2}^t|B^{\prime}(t)|}
\ee
Following the inequalities in Eq.~\eqref{Bbound}, one can find
\be
|B^{\prime}(t)|\le \frac{1}{\sqrt{2}}\left(1-\frac{1}{\sqrt{2}}\right)^{-t}
\ee
which recovers the desired result in Eq.~\eqref{proofdeltabound}
\be
\delta_t\ge \frac{\sqrt{2}}{6}\left(\frac{1}{\sqrt{2}}-\frac{1}{2}\right)^{t}
\ee
\twocolumngrid

\begin{thebibliography}{106}%
\makeatletter
\providecommand \@ifxundefined [1]{%
 \@ifx{#1\undefined}
}%
\providecommand \@ifnum [1]{%
 \ifnum #1\expandafter \@firstoftwo
 \else \expandafter \@secondoftwo
 \fi
}%
\providecommand \@ifx [1]{%
 \ifx #1\expandafter \@firstoftwo
 \else \expandafter \@secondoftwo
 \fi
}%
\providecommand \natexlab [1]{#1}%
\providecommand \enquote  [1]{``#1''}%
\providecommand \bibnamefont  [1]{#1}%
\providecommand \bibfnamefont [1]{#1}%
\providecommand \citenamefont [1]{#1}%
\providecommand \href@noop [0]{\@secondoftwo}%
\providecommand \href [0]{\begingroup \@sanitize@url \@href}%
\providecommand \@href[1]{\@@startlink{#1}\@@href}%
\providecommand \@@href[1]{\endgroup#1\@@endlink}%
\providecommand \@sanitize@url [0]{\catcode `\\12\catcode `\$12\catcode
  `\&12\catcode `\#12\catcode `\^12\catcode `\_12\catcode `\%12\relax}%
\providecommand \@@startlink[1]{}%
\providecommand \@@endlink[0]{}%
\providecommand \url  [0]{\begingroup\@sanitize@url \@url }%
\providecommand \@url [1]{\endgroup\@href {#1}{\urlprefix }}%
\providecommand \urlprefix  [0]{URL }%
\providecommand \Eprint [0]{\href }%
\providecommand \doibase [0]{http://dx.doi.org/}%
\providecommand \selectlanguage [0]{\@gobble}%
\providecommand \bibinfo  [0]{\@secondoftwo}%
\providecommand \bibfield  [0]{\@secondoftwo}%
\providecommand \translation [1]{[#1]}%
\providecommand \BibitemOpen [0]{}%
\providecommand \bibitemStop [0]{}%
\providecommand \bibitemNoStop [0]{.\EOS\space}%
\providecommand \EOS [0]{\spacefactor3000\relax}%
\providecommand \BibitemShut  [1]{\csname bibitem#1\endcsname}%
\let\auto@bib@innerbib\@empty
\bibitem [{\citenamefont {Knill}\ \emph {et~al.}(2008)\citenamefont {Knill},
  \citenamefont {Leibfried}, \citenamefont {Reichle}, \citenamefont {Britton},
  \citenamefont {Blakestad}, \citenamefont {Jost}, \citenamefont {Langer},
  \citenamefont {Ozeri}, \citenamefont {Seidelin},\ and\ \citenamefont
  {Wineland}}]{knill2008RandomizedBenchmarkingQuantum}%
  \BibitemOpen
  \bibfield  {author} {\bibinfo {author} {\bibfnamefont {E.}~\bibnamefont
  {Knill}}, \bibinfo {author} {\bibfnamefont {D.}~\bibnamefont {Leibfried}},
  \bibinfo {author} {\bibfnamefont {R.}~\bibnamefont {Reichle}}, \bibinfo
  {author} {\bibfnamefont {J.}~\bibnamefont {Britton}}, \bibinfo {author}
  {\bibfnamefont {R.~B.}\ \bibnamefont {Blakestad}}, \bibinfo {author}
  {\bibfnamefont {J.~D.}\ \bibnamefont {Jost}}, \bibinfo {author}
  {\bibfnamefont {C.}~\bibnamefont {Langer}}, \bibinfo {author} {\bibfnamefont
  {R.}~\bibnamefont {Ozeri}}, \bibinfo {author} {\bibfnamefont
  {S.}~\bibnamefont {Seidelin}}, \ and\ \bibinfo {author} {\bibfnamefont
  {D.~J.}\ \bibnamefont {Wineland}},\ }\href {\doibase
  10.1103/PhysRevA.77.012307} {\bibfield  {journal} {\bibinfo  {journal} {Phys.
  Rev. A}\ }\textbf {\bibinfo {volume} {77}},\ \bibinfo {pages} {012307}
  (\bibinfo {year} {2008})}\BibitemShut {NoStop}%
\bibitem [{\citenamefont {Wallman}\ and\ \citenamefont
  {Flammia}(2014)}]{wallman2014RandomizedBenchmarkingConfidence}%
  \BibitemOpen
  \bibfield  {author} {\bibinfo {author} {\bibfnamefont {J.~J.}\ \bibnamefont
  {Wallman}}\ and\ \bibinfo {author} {\bibfnamefont {S.~T.}\ \bibnamefont
  {Flammia}},\ }\href {\doibase 10.1088/1367-2630/16/10/103032} {\bibfield
  {journal} {\bibinfo  {journal} {New J. Phys.}\ }\textbf {\bibinfo {volume}
  {16}},\ \bibinfo {pages} {103032} (\bibinfo {year} {2014})}\BibitemShut
  {NoStop}%
\bibitem [{\citenamefont {Roth}\ \emph {et~al.}(2018)\citenamefont {Roth},
  \citenamefont {Kueng}, \citenamefont {Kimmel}, \citenamefont {Liu},
  \citenamefont {Gross}, \citenamefont {Eisert},\ and\ \citenamefont
  {Kliesch}}]{roth2018RecoveringQuantumGates}%
  \BibitemOpen
  \bibfield  {author} {\bibinfo {author} {\bibfnamefont {I.}~\bibnamefont
  {Roth}}, \bibinfo {author} {\bibfnamefont {R.}~\bibnamefont {Kueng}},
  \bibinfo {author} {\bibfnamefont {S.}~\bibnamefont {Kimmel}}, \bibinfo
  {author} {\bibfnamefont {Y.-K.}\ \bibnamefont {Liu}}, \bibinfo {author}
  {\bibfnamefont {D.}~\bibnamefont {Gross}}, \bibinfo {author} {\bibfnamefont
  {J.}~\bibnamefont {Eisert}}, \ and\ \bibinfo {author} {\bibfnamefont
  {M.}~\bibnamefont {Kliesch}},\ }\href {\doibase
  10.1103/PhysRevLett.121.170502} {\bibfield  {journal} {\bibinfo  {journal}
  {Phys. Rev. Lett,}\ }\textbf {\bibinfo {volume} {121}},\ \bibinfo {pages}
  {170502} (\bibinfo {year} {2018})}\BibitemShut {NoStop}%
\bibitem [{\citenamefont {Hosur}\ \emph {et~al.}(2016)\citenamefont {Hosur},
  \citenamefont {Qi}, \citenamefont {Roberts},\ and\ \citenamefont
  {Yoshida}}]{hosur2016ChaosQuantumChannels}%
  \BibitemOpen
  \bibfield  {author} {\bibinfo {author} {\bibfnamefont {P.}~\bibnamefont
  {Hosur}}, \bibinfo {author} {\bibfnamefont {X.-L.}\ \bibnamefont {Qi}},
  \bibinfo {author} {\bibfnamefont {D.~A.}\ \bibnamefont {Roberts}}, \ and\
  \bibinfo {author} {\bibfnamefont {B.}~\bibnamefont {Yoshida}},\ }\href
  {\doibase 10.1007/JHEP02(2016)004} {\bibfield  {journal} {\bibinfo  {journal}
  {J. High Energy Phys.}\ }\textbf {\bibinfo {volume} {2016}},\ \bibinfo
  {pages} {4} (\bibinfo {year} {2016})}\BibitemShut {NoStop}%
\bibitem [{\citenamefont {Cotler}\ \emph {et~al.}(2018)\citenamefont {Cotler},
  \citenamefont {Ding},\ and\ \citenamefont
  {Penington}}]{cotler2018OutoftimeorderOperatorsButterfly}%
  \BibitemOpen
  \bibfield  {author} {\bibinfo {author} {\bibfnamefont {J.~S.}\ \bibnamefont
  {Cotler}}, \bibinfo {author} {\bibfnamefont {D.}~\bibnamefont {Ding}}, \ and\
  \bibinfo {author} {\bibfnamefont {G.~R.}\ \bibnamefont {Penington}},\ }\href
  {\doibase 10.1016/j.aop.2018.07.020} {\bibfield  {journal} {\bibinfo
  {journal} {Ann. Phys.(NY)}\ }\textbf {\bibinfo {volume} {396}},\ \bibinfo
  {pages} {318} (\bibinfo {year} {2018})}\BibitemShut {NoStop}%
\bibitem [{\citenamefont {Xu}\ \emph {et~al.}(2020)\citenamefont {Xu},
  \citenamefont {Scaffidi},\ and\ \citenamefont
  {Cao}}]{xu2020DoesScramblingEqual}%
  \BibitemOpen
  \bibfield  {author} {\bibinfo {author} {\bibfnamefont {T.}~\bibnamefont
  {Xu}}, \bibinfo {author} {\bibfnamefont {T.}~\bibnamefont {Scaffidi}}, \ and\
  \bibinfo {author} {\bibfnamefont {X.}~\bibnamefont {Cao}},\ }\href {\doibase
  10.1103/PhysRevLett.124.140602} {\bibfield  {journal} {\bibinfo  {journal}
  {Phys. Rev. Lett,}\ }\textbf {\bibinfo {volume} {124}},\ \bibinfo {pages}
  {140602} (\bibinfo {year} {2020})}\BibitemShut {NoStop}%
\bibitem [{\citenamefont {Peruzzo}\ \emph {et~al.}(2014)\citenamefont
  {Peruzzo}, \citenamefont {McClean}, \citenamefont {Shadbolt}, \citenamefont
  {Yung}, \citenamefont {Zhou}, \citenamefont {Love}, \citenamefont
  {{Aspuru-Guzik}},\ and\ \citenamefont
  {O'brien}}]{peruzzo2014VariationalEigenvalueSolver}%
  \BibitemOpen
  \bibfield  {author} {\bibinfo {author} {\bibfnamefont {A.}~\bibnamefont
  {Peruzzo}}, \bibinfo {author} {\bibfnamefont {J.}~\bibnamefont {McClean}},
  \bibinfo {author} {\bibfnamefont {P.}~\bibnamefont {Shadbolt}}, \bibinfo
  {author} {\bibfnamefont {M.-H.}\ \bibnamefont {Yung}}, \bibinfo {author}
  {\bibfnamefont {X.-Q.}\ \bibnamefont {Zhou}}, \bibinfo {author}
  {\bibfnamefont {P.~J.}\ \bibnamefont {Love}}, \bibinfo {author}
  {\bibfnamefont {A.}~\bibnamefont {{Aspuru-Guzik}}}, \ and\ \bibinfo {author}
  {\bibfnamefont {J.~L.}\ \bibnamefont {O'brien}},\ }\href {\doibase
  10.1038/ncomms5213} {\bibfield  {journal} {\bibinfo  {journal} {Nat.
  Commun.}\ }\textbf {\bibinfo {volume} {5}},\ \bibinfo {pages} {1} (\bibinfo
  {year} {2014})}\BibitemShut {NoStop}%
\bibitem [{\citenamefont {McClean}\ \emph {et~al.}(2016)\citenamefont
  {McClean}, \citenamefont {Romero}, \citenamefont {Babbush},\ and\
  \citenamefont {{Aspuru-Guzik}}}]{mcclean2016TheoryVariationalHybrid}%
  \BibitemOpen
  \bibfield  {author} {\bibinfo {author} {\bibfnamefont {J.~R.}\ \bibnamefont
  {McClean}}, \bibinfo {author} {\bibfnamefont {J.}~\bibnamefont {Romero}},
  \bibinfo {author} {\bibfnamefont {R.}~\bibnamefont {Babbush}}, \ and\
  \bibinfo {author} {\bibfnamefont {A.}~\bibnamefont {{Aspuru-Guzik}}},\ }\href
  {\doibase 10.1088/1367-2630/18/2/023023} {\bibfield  {journal} {\bibinfo
  {journal} {New J. Phys.}\ }\textbf {\bibinfo {volume} {18}},\ \bibinfo
  {pages} {023023} (\bibinfo {year} {2016})}\BibitemShut {NoStop}%
\bibitem [{\citenamefont {Cao}\ \emph {et~al.}(2019)\citenamefont {Cao},
  \citenamefont {Romero}, \citenamefont {Olson}, \citenamefont {Degroote},
  \citenamefont {Johnson}, \citenamefont {Kieferov{\'a}}, \citenamefont
  {Kivlichan}, \citenamefont {Menke}, \citenamefont {Peropadre}, \citenamefont
  {Sawaya} \emph {et~al.}}]{cao2019QuantumChemistryAge}%
  \BibitemOpen
  \bibfield  {author} {\bibinfo {author} {\bibfnamefont {Y.}~\bibnamefont
  {Cao}}, \bibinfo {author} {\bibfnamefont {J.}~\bibnamefont {Romero}},
  \bibinfo {author} {\bibfnamefont {J.~P.}\ \bibnamefont {Olson}}, \bibinfo
  {author} {\bibfnamefont {M.}~\bibnamefont {Degroote}}, \bibinfo {author}
  {\bibfnamefont {P.~D.}\ \bibnamefont {Johnson}}, \bibinfo {author}
  {\bibfnamefont {M.}~\bibnamefont {Kieferov{\'a}}}, \bibinfo {author}
  {\bibfnamefont {I.~D.}\ \bibnamefont {Kivlichan}}, \bibinfo {author}
  {\bibfnamefont {T.}~\bibnamefont {Menke}}, \bibinfo {author} {\bibfnamefont
  {B.}~\bibnamefont {Peropadre}}, \bibinfo {author} {\bibfnamefont {N.~P.~D.}\
  \bibnamefont {Sawaya}},  \emph {et~al.},\ }\href
  {https://pubs.acs.org/doi/10.1021/acs.chemrev.8b00803} {\bibfield  {journal}
  {\bibinfo  {journal} {Chem. Rev, (Washington, DC, U.S.)}\ }\textbf {\bibinfo
  {volume} {119}},\ \bibinfo {pages} {10856} (\bibinfo {year}
  {2019})}\BibitemShut {NoStop}%
\bibitem [{\citenamefont {Huggins}\ \emph {et~al.}(2021)\citenamefont
  {Huggins}, \citenamefont {McClean}, \citenamefont {Rubin}, \citenamefont
  {Jiang}, \citenamefont {Wiebe}, \citenamefont {Whaley},\ and\ \citenamefont
  {Babbush}}]{huggins2021EfficientNoiseResilient}%
  \BibitemOpen
  \bibfield  {author} {\bibinfo {author} {\bibfnamefont {W.~J.}\ \bibnamefont
  {Huggins}}, \bibinfo {author} {\bibfnamefont {J.~R.}\ \bibnamefont
  {McClean}}, \bibinfo {author} {\bibfnamefont {N.~C.}\ \bibnamefont {Rubin}},
  \bibinfo {author} {\bibfnamefont {Z.}~\bibnamefont {Jiang}}, \bibinfo
  {author} {\bibfnamefont {N.}~\bibnamefont {Wiebe}}, \bibinfo {author}
  {\bibfnamefont {K.~B.}\ \bibnamefont {Whaley}}, \ and\ \bibinfo {author}
  {\bibfnamefont {R.}~\bibnamefont {Babbush}},\ }\href
  {https://www.nature.com/articles/s41534-020-00341-7} {\bibfield  {journal}
  {\bibinfo  {journal} {npj Quantum Information}\ }\textbf {\bibinfo {volume}
  {7}},\ \bibinfo {pages} {1} (\bibinfo {year} {2021})}\BibitemShut {NoStop}%
\bibitem [{\citenamefont {Bauer}\ \emph {et~al.}(2020)\citenamefont {Bauer},
  \citenamefont {Bravyi}, \citenamefont {Motta},\ and\ \citenamefont
  {Chan}}]{bauer2020QuantumAlgorithmsQuantum}%
  \BibitemOpen
  \bibfield  {author} {\bibinfo {author} {\bibfnamefont {B.}~\bibnamefont
  {Bauer}}, \bibinfo {author} {\bibfnamefont {S.}~\bibnamefont {Bravyi}},
  \bibinfo {author} {\bibfnamefont {M.}~\bibnamefont {Motta}}, \ and\ \bibinfo
  {author} {\bibfnamefont {G.~K.-L.}\ \bibnamefont {Chan}},\ }\href {\doibase
  10.1021/acs.chemrev.9b00829} {\bibfield  {journal} {\bibinfo  {journal}
  {Chem. Rev, (Washington, DC, U.S.)}\ }\textbf {\bibinfo {volume} {120}},\
  \bibinfo {pages} {12685} (\bibinfo {year} {2020})}\BibitemShut {NoStop}%
\bibitem [{\citenamefont {Fuchs}\ and\ \citenamefont {Van
  De~Graaf}(1999)}]{fuchs1999CryptographicDistinguishabilityMeasures}%
  \BibitemOpen
  \bibfield  {author} {\bibinfo {author} {\bibfnamefont {C.~A.}\ \bibnamefont
  {Fuchs}}\ and\ \bibinfo {author} {\bibfnamefont {J.}~\bibnamefont {Van
  De~Graaf}},\ }\href {\doibase 10.1109/18.761271} {\bibfield  {journal}
  {\bibinfo  {journal} {IEEE Trans. Inf. Theory}\ }\textbf {\bibinfo {volume}
  {45}},\ \bibinfo {pages} {1216} (\bibinfo {year} {1999})}\BibitemShut
  {NoStop}%
\bibitem [{\citenamefont {Reichardt}\ \emph {et~al.}(2013)\citenamefont
  {Reichardt}, \citenamefont {Unger},\ and\ \citenamefont
  {Vazirani}}]{reichardt2013ClassicalCommandQuantum}%
  \BibitemOpen
  \bibfield  {author} {\bibinfo {author} {\bibfnamefont {B.~W.}\ \bibnamefont
  {Reichardt}}, \bibinfo {author} {\bibfnamefont {F.}~\bibnamefont {Unger}}, \
  and\ \bibinfo {author} {\bibfnamefont {U.}~\bibnamefont {Vazirani}},\ }\href
  {\doibase 10.1038/nature12035} {\bibfield  {journal} {\bibinfo  {journal}
  {Nature}\ }\textbf {\bibinfo {volume} {496}},\ \bibinfo {pages} {456}
  (\bibinfo {year} {2013})}\BibitemShut {NoStop}%
\bibitem [{\citenamefont {Mills}\ \emph {et~al.}(2018)\citenamefont {Mills},
  \citenamefont {Pappa}, \citenamefont {Kapourniotis},\ and\ \citenamefont
  {Kashefi}}]{mills2017InformationTheoreticallySecure}%
  \BibitemOpen
  \bibfield  {author} {\bibinfo {author} {\bibfnamefont {D.}~\bibnamefont
  {Mills}}, \bibinfo {author} {\bibfnamefont {A.}~\bibnamefont {Pappa}},
  \bibinfo {author} {\bibfnamefont {T.}~\bibnamefont {Kapourniotis}}, \ and\
  \bibinfo {author} {\bibfnamefont {E.}~\bibnamefont {Kashefi}}\ }(\bibinfo
  {publisher} {Open Publishing Association},\ \bibinfo {year} {2018})\ p.\
  \bibinfo {pages} {209–221}\BibitemShut {NoStop}%
\bibitem [{\citenamefont
  {Fitzsimons}(2017)}]{fitzsimons2017PrivateQuantumComputation}%
  \BibitemOpen
  \bibfield  {author} {\bibinfo {author} {\bibfnamefont {J.~F.}\ \bibnamefont
  {Fitzsimons}},\ }\href {\doibase 10.1038/s41534-017-0025-3} {\bibfield
  {journal} {\bibinfo  {journal} {npj Quantum Information}\ }\textbf {\bibinfo
  {volume} {3}},\ \bibinfo {pages} {23} (\bibinfo {year} {2017})}\BibitemShut
  {NoStop}%
\bibitem [{\citenamefont {Fitzsimons}\ and\ \citenamefont
  {Kashefi}(2017)}]{fitzsimons2017UnconditionallyVerifiableBlind}%
  \BibitemOpen
  \bibfield  {author} {\bibinfo {author} {\bibfnamefont {J.~F.}\ \bibnamefont
  {Fitzsimons}}\ and\ \bibinfo {author} {\bibfnamefont {E.}~\bibnamefont
  {Kashefi}},\ }\href {\doibase 10.1103/PhysRevA.96.012303} {\bibfield
  {journal} {\bibinfo  {journal} {Phys. Rev. A}\ }\textbf {\bibinfo {volume}
  {96}},\ \bibinfo {pages} {012303} (\bibinfo {year} {2017})}\BibitemShut
  {NoStop}%
\bibitem [{\citenamefont {Coladangelo}\ \emph {et~al.}(2019)\citenamefont
  {Coladangelo}, \citenamefont {Grilo}, \citenamefont {Jeffery},\ and\
  \citenamefont {Vidick}}]{coladangelo2019VerifieronaLeashNewSchemes}%
  \BibitemOpen
  \bibfield  {author} {\bibinfo {author} {\bibfnamefont {A.}~\bibnamefont
  {Coladangelo}}, \bibinfo {author} {\bibfnamefont {A.~B.}\ \bibnamefont
  {Grilo}}, \bibinfo {author} {\bibfnamefont {S.}~\bibnamefont {Jeffery}}, \
  and\ \bibinfo {author} {\bibfnamefont {T.}~\bibnamefont {Vidick}},\ }in\
  \href {\doibase 10.1007/978-3-030-17659-4_9} {\emph {\bibinfo {booktitle}
  {Advances in {{Cryptology}} \textendash{} {{EUROCRYPT}} 2019}}},\ \bibinfo
  {series and number} {Lecture {{Notes}} in {{Computer Science}}},\ \bibinfo
  {editor} {edited by\ \bibinfo {editor} {\bibfnamefont {Y.}~\bibnamefont
  {Ishai}}\ and\ \bibinfo {editor} {\bibfnamefont {V.}~\bibnamefont {Rijmen}}}\
  (\bibinfo  {publisher} {{Springer International Publishing}},\ \bibinfo
  {address} {{Cham}},\ \bibinfo {year} {2019})\ pp.\ \bibinfo {pages}
  {247--277}\BibitemShut {NoStop}%
\bibitem [{\citenamefont
  {Mahadev}(2018)}]{mahadev2018ClassicalVerificationQuantuma}%
  \BibitemOpen
  \bibfield  {author} {\bibinfo {author} {\bibfnamefont {U.}~\bibnamefont
  {Mahadev}},\ }in\ \href {\doibase 10.1109/FOCS.2018.00033} {\emph {\bibinfo
  {booktitle} {2018 {{IEEE}} 59th {{Annual Symposium}} on {{Foundations}} of
  {{Computer Science}} ({{FOCS}})}}}\ (\bibinfo {year} {2018})\ pp.\ \bibinfo
  {pages} {259--267}\BibitemShut {NoStop}%
\bibitem [{\citenamefont {Gheorghiu}\ \emph {et~al.}(2019)\citenamefont
  {Gheorghiu}, \citenamefont {Kapourniotis},\ and\ \citenamefont
  {Kashefi}}]{gheorghiu2019VerificationQuantumComputation}%
  \BibitemOpen
  \bibfield  {author} {\bibinfo {author} {\bibfnamefont {A.}~\bibnamefont
  {Gheorghiu}}, \bibinfo {author} {\bibfnamefont {T.}~\bibnamefont
  {Kapourniotis}}, \ and\ \bibinfo {author} {\bibfnamefont {E.}~\bibnamefont
  {Kashefi}},\ }\href {\doibase 10.1007/s00224-018-9872-3} {\bibfield
  {journal} {\bibinfo  {journal} {Theory of Computing Systems}\ }\textbf
  {\bibinfo {volume} {63}},\ \bibinfo {pages} {715} (\bibinfo {year}
  {2019})}\BibitemShut {NoStop}%
\bibitem [{\citenamefont {Gheorghiu}\ and\ \citenamefont
  {Vidick}(2019)}]{gheorghiu2019ComputationallysecureComposableRemote}%
  \BibitemOpen
  \bibfield  {author} {\bibinfo {author} {\bibfnamefont {A.}~\bibnamefont
  {Gheorghiu}}\ and\ \bibinfo {author} {\bibfnamefont {T.}~\bibnamefont
  {Vidick}},\ }in\ \href {\doibase 10.1109/FOCS.2019.00066} {\emph {\bibinfo
  {booktitle} {2019 {{IEEE}} 60th {{Annual Symposium}} on {{Foundations}} of
  {{Computer Science}} ({{FOCS}})}}}\ (\bibinfo {year} {2019})\ pp.\ \bibinfo
  {pages} {1024--1033}\BibitemShut {NoStop}%
\bibitem [{\citenamefont {{\v S}upi{\'c}}\ and\ \citenamefont
  {Bowles}(mber)}]{supic2020SelftestingQuantumSystems}%
  \BibitemOpen
  \bibfield  {author} {\bibinfo {author} {\bibfnamefont {I.}~\bibnamefont {{\v
  S}upi{\'c}}}\ and\ \bibinfo {author} {\bibfnamefont {J.}~\bibnamefont
  {Bowles}},\ }\href {\doibase 10.22331/q-2020-09-30-337} {\bibfield  {journal}
  {\bibinfo  {journal} {Quantum}\ }\textbf {\bibinfo {volume} {4}},\ \bibinfo
  {pages} {337} (\bibinfo {year} {2020/september})}\BibitemShut {NoStop}%
\bibitem [{\citenamefont {Gisin}\ \emph {et~al.}(2002)\citenamefont {Gisin},
  \citenamefont {Ribordy}, \citenamefont {Tittel},\ and\ \citenamefont
  {Zbinden}}]{gisin2002QuantumCryptography}%
  \BibitemOpen
  \bibfield  {author} {\bibinfo {author} {\bibfnamefont {N.}~\bibnamefont
  {Gisin}}, \bibinfo {author} {\bibfnamefont {G.}~\bibnamefont {Ribordy}},
  \bibinfo {author} {\bibfnamefont {W.}~\bibnamefont {Tittel}}, \ and\ \bibinfo
  {author} {\bibfnamefont {H.}~\bibnamefont {Zbinden}},\ }\href {\doibase
  10.1103/RevModPhys.74.14} {\bibfield  {journal} {\bibinfo  {journal} {Reviews
  of modern physics}\ }\textbf {\bibinfo {volume} {74}},\ \bibinfo {pages}
  {145} (\bibinfo {year} {2002})}\BibitemShut {NoStop}%
\bibitem [{\citenamefont {Hayden}\ and\ \citenamefont
  {Preskill}(2007)}]{hayden2007BlackHolesMirrors}%
  \BibitemOpen
  \bibfield  {author} {\bibinfo {author} {\bibfnamefont {P.}~\bibnamefont
  {Hayden}}\ and\ \bibinfo {author} {\bibfnamefont {J.}~\bibnamefont
  {Preskill}},\ }\href {\doibase 10.1088/1126-6708/2007/09/120} {\bibfield
  {journal} {\bibinfo  {journal} {J. High Energy Phys.}\ }\textbf {\bibinfo
  {volume} {2007}},\ \bibinfo {pages} {120} (\bibinfo {year}
  {2007})}\BibitemShut {NoStop}%
\bibitem [{\citenamefont {Yoshida}\ and\ \citenamefont
  {Kitaev}(2017)}]{yoshida2017EfficientDecodingHaydenPreskill}%
  \BibitemOpen
  \bibfield  {author} {\bibinfo {author} {\bibfnamefont {B.}~\bibnamefont
  {Yoshida}}\ and\ \bibinfo {author} {\bibfnamefont {A.}~\bibnamefont
  {Kitaev}},\ }\href {\doibase 10.48550/arXiv.1710.03363} {\enquote {\bibinfo
  {title} {Efficient decoding for the {{Hayden-Preskill}} protocol},}\ }
  (\bibinfo {year} {2017}),\ \Eprint {http://arxiv.org/abs/1710.03363}
  {arXiv:1710.03363 [hep-th, physics:quant-ph]} \BibitemShut {NoStop}%
\bibitem [{\citenamefont {Low}(2009)}]{low2009LearningTestingAlgorithms}%
  \BibitemOpen
  \bibfield  {author} {\bibinfo {author} {\bibfnamefont {R.~A.}\ \bibnamefont
  {Low}},\ }\href {\doibase 10.1103/PhysRevA.80.052314} {\bibfield  {journal}
  {\bibinfo  {journal} {Phys. Rev. A}\ }\textbf {\bibinfo {volume} {80}},\
  \bibinfo {pages} {052314} (\bibinfo {year} {2009})}\BibitemShut {NoStop}%
\bibitem [{\citenamefont {Lai}\ and\ \citenamefont
  {Cheng}(2022)}]{lai2022LearningQuantumCircuits}%
  \BibitemOpen
  \bibfield  {author} {\bibinfo {author} {\bibfnamefont {C.-Y.}\ \bibnamefont
  {Lai}}\ and\ \bibinfo {author} {\bibfnamefont {H.-C.}\ \bibnamefont
  {Cheng}},\ }\href {\doibase 10.1109/tit.2022.3151760} {\bibfield  {journal}
  {\bibinfo  {journal} {IEEE Trans. Inf. Theory}\ }\textbf {\bibinfo {volume}
  {68}},\ \bibinfo {pages} {3951} (\bibinfo {year} {2022})}\BibitemShut
  {NoStop}%
\bibitem [{\citenamefont
  {Gottesman}(1998)}]{gottesman1998HeisenbergRepresentationQuantum}%
  \BibitemOpen
  \bibfield  {author} {\bibinfo {author} {\bibfnamefont {D.}~\bibnamefont
  {Gottesman}},\ }\href {\doibase 10.48550/arXiv.quant-ph/9807006} {\enquote
  {\bibinfo {title} {The {{Heisenberg Representation}} of {{Quantum
  Computers}}},}\ } (\bibinfo {year} {1998}),\ \Eprint
  {http://arxiv.org/abs/quant-ph/9807006} {arXiv:quant-ph/9807006} \BibitemShut
  {NoStop}%
\bibitem [{\citenamefont {Oliviero}\ \emph {et~al.}(2024)\citenamefont
  {Oliviero}, \citenamefont {Leone}, \citenamefont {Lloyd},\ and\ \citenamefont
  {Hamma}}]{oliviero2022BlackHoleComplexity}%
  \BibitemOpen
  \bibfield  {author} {\bibinfo {author} {\bibfnamefont {S.~F.~E.}\
  \bibnamefont {Oliviero}}, \bibinfo {author} {\bibfnamefont {L.}~\bibnamefont
  {Leone}}, \bibinfo {author} {\bibfnamefont {S.}~\bibnamefont {Lloyd}}, \ and\
  \bibinfo {author} {\bibfnamefont {A.}~\bibnamefont {Hamma}},\ }\href
  {\doibase 10.1103/PhysRevLett.132.080402} {\bibfield  {journal} {\bibinfo
  {journal} {Phys. Rev. Lett.}\ }\textbf {\bibinfo {volume} {132}},\ \bibinfo
  {pages} {080402} (\bibinfo {year} {2024})}\BibitemShut {NoStop}%
\bibitem [{\citenamefont {Leone}\ \emph
  {et~al.}(2021{\natexlab{a}})\citenamefont {Leone}, \citenamefont {Oliviero},
  \citenamefont {Zhou},\ and\ \citenamefont
  {Hamma}}]{leone2021QuantumChaosQuantum}%
  \BibitemOpen
  \bibfield  {author} {\bibinfo {author} {\bibfnamefont {L.}~\bibnamefont
  {Leone}}, \bibinfo {author} {\bibfnamefont {S.~F.~E.}\ \bibnamefont
  {Oliviero}}, \bibinfo {author} {\bibfnamefont {Y.}~\bibnamefont {Zhou}}, \
  and\ \bibinfo {author} {\bibfnamefont {A.}~\bibnamefont {Hamma}},\ }\href
  {\doibase 10.22331/q-2021-05-04-453} {\bibfield  {journal} {\bibinfo
  {journal} {Quantum}\ }\textbf {\bibinfo {volume} {5}},\ \bibinfo {pages}
  {453} (\bibinfo {year} {2021}{\natexlab{a}})}\BibitemShut {NoStop}%
\bibitem [{\citenamefont {Chuang}\ and\ \citenamefont
  {Nielsen}(1997)}]{chuang1997PrescriptionExperimentalDetermination}%
  \BibitemOpen
  \bibfield  {author} {\bibinfo {author} {\bibfnamefont {I.~L.}\ \bibnamefont
  {Chuang}}\ and\ \bibinfo {author} {\bibfnamefont {M.~A.}\ \bibnamefont
  {Nielsen}},\ }\href {\doibase 10.1080/09500349708231894} {\bibfield
  {journal} {\bibinfo  {journal} {Journal of Modern Optics}\ }\textbf {\bibinfo
  {volume} {44}},\ \bibinfo {pages} {2455} (\bibinfo {year}
  {1997})}\BibitemShut {NoStop}%
\bibitem [{\citenamefont {Childs}\ \emph {et~al.}(2001)\citenamefont {Childs},
  \citenamefont {Chuang},\ and\ \citenamefont
  {Leung}}]{childs2001RealizationQuantumProcess}%
  \BibitemOpen
  \bibfield  {author} {\bibinfo {author} {\bibfnamefont {A.~M.}\ \bibnamefont
  {Childs}}, \bibinfo {author} {\bibfnamefont {I.~L.}\ \bibnamefont {Chuang}},
  \ and\ \bibinfo {author} {\bibfnamefont {D.~W.}\ \bibnamefont {Leung}},\
  }\href {\doibase 10.1103/PhysRevA.64.012314} {\bibfield  {journal} {\bibinfo
  {journal} {Phys. Rev. A}\ }\textbf {\bibinfo {volume} {64}},\ \bibinfo
  {pages} {012314} (\bibinfo {year} {2001})}\BibitemShut {NoStop}%
\bibitem [{\citenamefont {Altepeter}\ \emph {et~al.}(2003)\citenamefont
  {Altepeter}, \citenamefont {Branning}, \citenamefont {Jeffrey}, \citenamefont
  {Wei}, \citenamefont {Kwiat}, \citenamefont {Thew}, \citenamefont {O'Brien},
  \citenamefont {Nielsen},\ and\ \citenamefont
  {White}}]{altepeter2003AncillaAssistedQuantumProcess}%
  \BibitemOpen
  \bibfield  {author} {\bibinfo {author} {\bibfnamefont {J.~B.}\ \bibnamefont
  {Altepeter}}, \bibinfo {author} {\bibfnamefont {D.}~\bibnamefont {Branning}},
  \bibinfo {author} {\bibfnamefont {E.}~\bibnamefont {Jeffrey}}, \bibinfo
  {author} {\bibfnamefont {T.~C.}\ \bibnamefont {Wei}}, \bibinfo {author}
  {\bibfnamefont {P.~G.}\ \bibnamefont {Kwiat}}, \bibinfo {author}
  {\bibfnamefont {R.~T.}\ \bibnamefont {Thew}}, \bibinfo {author}
  {\bibfnamefont {J.~L.}\ \bibnamefont {O'Brien}}, \bibinfo {author}
  {\bibfnamefont {M.~A.}\ \bibnamefont {Nielsen}}, \ and\ \bibinfo {author}
  {\bibfnamefont {A.~G.}\ \bibnamefont {White}},\ }\href {\doibase
  10.1103/PhysRevLett.90.193601} {\bibfield  {journal} {\bibinfo  {journal}
  {Phys. Rev. Lett,}\ }\textbf {\bibinfo {volume} {90}},\ \bibinfo {pages}
  {193601} (\bibinfo {year} {2003})}\BibitemShut {NoStop}%
\bibitem [{\citenamefont {Mohseni}\ \emph {et~al.}(2008)\citenamefont
  {Mohseni}, \citenamefont {Rezakhani},\ and\ \citenamefont
  {Lidar}}]{mohseni2008QuantumprocessTomographyResource}%
  \BibitemOpen
  \bibfield  {author} {\bibinfo {author} {\bibfnamefont {M.}~\bibnamefont
  {Mohseni}}, \bibinfo {author} {\bibfnamefont {A.~T.}\ \bibnamefont
  {Rezakhani}}, \ and\ \bibinfo {author} {\bibfnamefont {D.~A.}\ \bibnamefont
  {Lidar}},\ }\href {\doibase 10.1103/PhysRevA.77.032322} {\bibfield  {journal}
  {\bibinfo  {journal} {Phys. Rev. A}\ }\textbf {\bibinfo {volume} {77}},\
  \bibinfo {pages} {032322} (\bibinfo {year} {2008})}\BibitemShut {NoStop}%
\bibitem [{\citenamefont {Merkel}\ \emph {et~al.}(2013)\citenamefont {Merkel},
  \citenamefont {Gambetta}, \citenamefont {Smolin}, \citenamefont {Poletto},
  \citenamefont {C{\'o}rcoles}, \citenamefont {Johnson}, \citenamefont {Ryan},\
  and\ \citenamefont {Steffen}}]{merkel2013SelfconsistentQuantumProcess}%
  \BibitemOpen
  \bibfield  {author} {\bibinfo {author} {\bibfnamefont {S.~T.}\ \bibnamefont
  {Merkel}}, \bibinfo {author} {\bibfnamefont {J.~M.}\ \bibnamefont
  {Gambetta}}, \bibinfo {author} {\bibfnamefont {J.~A.}\ \bibnamefont
  {Smolin}}, \bibinfo {author} {\bibfnamefont {S.}~\bibnamefont {Poletto}},
  \bibinfo {author} {\bibfnamefont {A.~D.}\ \bibnamefont {C{\'o}rcoles}},
  \bibinfo {author} {\bibfnamefont {B.~R.}\ \bibnamefont {Johnson}}, \bibinfo
  {author} {\bibfnamefont {C.~A.}\ \bibnamefont {Ryan}}, \ and\ \bibinfo
  {author} {\bibfnamefont {M.}~\bibnamefont {Steffen}},\ }\href {\doibase
  10.1103/PhysRevA.87.062119} {\bibfield  {journal} {\bibinfo  {journal} {Phys.
  Rev. A}\ }\textbf {\bibinfo {volume} {87}},\ \bibinfo {pages} {062119}
  (\bibinfo {year} {2013})}\BibitemShut {NoStop}%
\bibitem [{\citenamefont {Roberts}\ and\ \citenamefont
  {Yoshida}(2017)}]{roberts2017ChaosComplexityDesign}%
  \BibitemOpen
  \bibfield  {author} {\bibinfo {author} {\bibfnamefont {D.~A.}\ \bibnamefont
  {Roberts}}\ and\ \bibinfo {author} {\bibfnamefont {B.}~\bibnamefont
  {Yoshida}},\ }\href {\doibase 10.1007/JHEP04(2017)121} {\bibfield  {journal}
  {\bibinfo  {journal} {J. High Energy Phys.}\ }\textbf {\bibinfo {volume}
  {2017}},\ \bibinfo {pages} {121} (\bibinfo {year} {2017})}\BibitemShut
  {NoStop}%
\bibitem [{\citenamefont {Leone}\ \emph
  {et~al.}(2021{\natexlab{b}})\citenamefont {Leone}, \citenamefont {Oliviero},\
  and\ \citenamefont {Hamma}}]{leone2021IsospectralTwirlingQuantum}%
  \BibitemOpen
  \bibfield  {author} {\bibinfo {author} {\bibfnamefont {L.}~\bibnamefont
  {Leone}}, \bibinfo {author} {\bibfnamefont {S.~F.~E.}\ \bibnamefont
  {Oliviero}}, \ and\ \bibinfo {author} {\bibfnamefont {A.}~\bibnamefont
  {Hamma}},\ }\href {\doibase 10.3390/e23081073} {\bibfield  {journal}
  {\bibinfo  {journal} {Entropy}\ }\textbf {\bibinfo {volume} {23}} (\bibinfo
  {year} {2021}{\natexlab{b}}),\ 10.3390/e23081073}\BibitemShut {NoStop}%
\bibitem [{\citenamefont {Oliviero}\ \emph
  {et~al.}(2021{\natexlab{a}})\citenamefont {Oliviero}, \citenamefont {Leone},
  \citenamefont {Caravelli},\ and\ \citenamefont
  {Hamma}}]{oliviero2021RandomMatrixTheory}%
  \BibitemOpen
  \bibfield  {author} {\bibinfo {author} {\bibfnamefont {S.~F.~E.}\
  \bibnamefont {Oliviero}}, \bibinfo {author} {\bibfnamefont {L.}~\bibnamefont
  {Leone}}, \bibinfo {author} {\bibfnamefont {F.}~\bibnamefont {Caravelli}}, \
  and\ \bibinfo {author} {\bibfnamefont {A.}~\bibnamefont {Hamma}},\ }\href
  {\doibase 10.21468/SciPostPhys.10.3.076} {\bibfield  {journal} {\bibinfo
  {journal} {SciPost Phys.}\ }\textbf {\bibinfo {volume} {10}},\ \bibinfo
  {pages} {76} (\bibinfo {year} {2021}{\natexlab{a}})}\BibitemShut {NoStop}%
\bibitem [{\citenamefont {Brown}\ and\ \citenamefont
  {Susskind}(2018)}]{brown2018SecondLawQuantum}%
  \BibitemOpen
  \bibfield  {author} {\bibinfo {author} {\bibfnamefont {A.~R.}\ \bibnamefont
  {Brown}}\ and\ \bibinfo {author} {\bibfnamefont {L.}~\bibnamefont
  {Susskind}},\ }\href {\doibase 10.1103/PhysRevD.97.086015} {\bibfield
  {journal} {\bibinfo  {journal} {Phys. Rev. D}\ }\textbf {\bibinfo {volume}
  {97}},\ \bibinfo {pages} {086015} (\bibinfo {year} {2018})}\BibitemShut
  {NoStop}%
\bibitem [{\citenamefont {Oliviero}\ \emph
  {et~al.}(2021{\natexlab{b}})\citenamefont {Oliviero}, \citenamefont {Leone},\
  and\ \citenamefont {Hamma}}]{oliviero2021TransitionsEntanglementComplexity}%
  \BibitemOpen
  \bibfield  {author} {\bibinfo {author} {\bibfnamefont {S.~F.~E.}\
  \bibnamefont {Oliviero}}, \bibinfo {author} {\bibfnamefont {L.}~\bibnamefont
  {Leone}}, \ and\ \bibinfo {author} {\bibfnamefont {A.}~\bibnamefont
  {Hamma}},\ }\href {\doibase 10.1016/j.physleta.2021.127721} {\bibfield
  {journal} {\bibinfo  {journal} {Phys. Lett. A}\ }\textbf {\bibinfo {volume}
  {418}},\ \bibinfo {pages} {127721} (\bibinfo {year}
  {2021}{\natexlab{b}})}\BibitemShut {NoStop}%
\bibitem [{\citenamefont {Yang}\ \emph {et~al.}(2017)\citenamefont {Yang},
  \citenamefont {Hamma}, \citenamefont {Giampaolo}, \citenamefont {Mucciolo},\
  and\ \citenamefont {Chamon}}]{yang2017EntanglementComplexityQuantum}%
  \BibitemOpen
  \bibfield  {author} {\bibinfo {author} {\bibfnamefont {Z.-C.}\ \bibnamefont
  {Yang}}, \bibinfo {author} {\bibfnamefont {A.}~\bibnamefont {Hamma}},
  \bibinfo {author} {\bibfnamefont {S.~M.}\ \bibnamefont {Giampaolo}}, \bibinfo
  {author} {\bibfnamefont {E.~R.}\ \bibnamefont {Mucciolo}}, \ and\ \bibinfo
  {author} {\bibfnamefont {C.}~\bibnamefont {Chamon}},\ }\href {\doibase
  10.1103/PhysRevB.96.020408} {\bibfield  {journal} {\bibinfo  {journal} {Phys.
  Rev. B}\ }\textbf {\bibinfo {volume} {96}},\ \bibinfo {pages} {020408}
  (\bibinfo {year} {2017})}\BibitemShut {NoStop}%
\bibitem [{\citenamefont {Chamon}\ \emph {et~al.}(2014)\citenamefont {Chamon},
  \citenamefont {Hamma},\ and\ \citenamefont
  {Mucciolo}}]{chamon2014EmergentIrreversibilityEntanglement}%
  \BibitemOpen
  \bibfield  {author} {\bibinfo {author} {\bibfnamefont {C.}~\bibnamefont
  {Chamon}}, \bibinfo {author} {\bibfnamefont {A.}~\bibnamefont {Hamma}}, \
  and\ \bibinfo {author} {\bibfnamefont {E.~R.}\ \bibnamefont {Mucciolo}},\
  }\href {\doibase 10.1103/PhysRevLett.112.240501} {\bibfield  {journal}
  {\bibinfo  {journal} {Phys. Rev. Lett,}\ }\textbf {\bibinfo {volume} {112}},\
  \bibinfo {pages} {240501} (\bibinfo {year} {2014})}\BibitemShut {NoStop}%
\bibitem [{\citenamefont {Leone}\ \emph
  {et~al.}(2022{\natexlab{a}})\citenamefont {Leone}, \citenamefont {Oliviero},
  \citenamefont {Piemontese}, \citenamefont {True},\ and\ \citenamefont
  {Hamma}}]{leone2022RetrievingInformationBlack}%
  \BibitemOpen
  \bibfield  {author} {\bibinfo {author} {\bibfnamefont {L.}~\bibnamefont
  {Leone}}, \bibinfo {author} {\bibfnamefont {S.~F.~E.}\ \bibnamefont
  {Oliviero}}, \bibinfo {author} {\bibfnamefont {S.}~\bibnamefont
  {Piemontese}}, \bibinfo {author} {\bibfnamefont {S.}~\bibnamefont {True}}, \
  and\ \bibinfo {author} {\bibfnamefont {A.}~\bibnamefont {Hamma}},\ }\href
  {\doibase 10.1103/PhysRevA.106.062434} {\bibfield  {journal} {\bibinfo
  {journal} {Phys. Rev. A}\ }\textbf {\bibinfo {volume} {106}},\ \bibinfo
  {pages} {062434} (\bibinfo {year} {2022}{\natexlab{a}})}\BibitemShut
  {NoStop}%
\bibitem [{\citenamefont {Holz{\"a}pfel}\ \emph {et~al.}(2015)\citenamefont
  {Holz{\"a}pfel}, \citenamefont {Baumgratz}, \citenamefont {Cramer},\ and\
  \citenamefont {Plenio}}]{holzapfel2015ScalableReconstructionUnitary}%
  \BibitemOpen
  \bibfield  {author} {\bibinfo {author} {\bibfnamefont {M.}~\bibnamefont
  {Holz{\"a}pfel}}, \bibinfo {author} {\bibfnamefont {T.}~\bibnamefont
  {Baumgratz}}, \bibinfo {author} {\bibfnamefont {M.}~\bibnamefont {Cramer}}, \
  and\ \bibinfo {author} {\bibfnamefont {M.~B.}\ \bibnamefont {Plenio}},\
  }\href {\doibase 10.1103/PhysRevA.91.042129} {\bibfield  {journal} {\bibinfo
  {journal} {Phys. Rev. A}\ }\textbf {\bibinfo {volume} {91}},\ \bibinfo
  {pages} {042129} (\bibinfo {year} {2015})}\BibitemShut {NoStop}%
\bibitem [{\citenamefont {Khatri}\ \emph {et~al.}(2019)\citenamefont {Khatri},
  \citenamefont {LaRose}, \citenamefont {Poremba}, \citenamefont {Cincio},
  \citenamefont {Sornborger},\ and\ \citenamefont
  {Coles}}]{khatri2019QuantumassistedQuantumCompiling}%
  \BibitemOpen
  \bibfield  {author} {\bibinfo {author} {\bibfnamefont {S.}~\bibnamefont
  {Khatri}}, \bibinfo {author} {\bibfnamefont {R.}~\bibnamefont {LaRose}},
  \bibinfo {author} {\bibfnamefont {A.}~\bibnamefont {Poremba}}, \bibinfo
  {author} {\bibfnamefont {L.}~\bibnamefont {Cincio}}, \bibinfo {author}
  {\bibfnamefont {A.~T.}\ \bibnamefont {Sornborger}}, \ and\ \bibinfo {author}
  {\bibfnamefont {P.~J.}\ \bibnamefont {Coles}},\ }\href {\doibase
  10.22331/q-2019-05-13-140} {\bibfield  {journal} {\bibinfo  {journal}
  {Quantum}\ }\textbf {\bibinfo {volume} {3}},\ \bibinfo {pages} {140}
  (\bibinfo {year} {2019})}\BibitemShut {NoStop}%
\bibitem [{\citenamefont {Tibbetts}\ \emph {et~al.}(2012)\citenamefont
  {Tibbetts}, \citenamefont {Brif}, \citenamefont {Grace}, \citenamefont
  {Donovan}, \citenamefont {Hocker}, \citenamefont {Ho}, \citenamefont {Wu},\
  and\ \citenamefont {Rabitz}}]{tibbetts2012ExploringTradeoffFidelity}%
  \BibitemOpen
  \bibfield  {author} {\bibinfo {author} {\bibfnamefont {K.~W.~M.}\
  \bibnamefont {Tibbetts}}, \bibinfo {author} {\bibfnamefont {C.}~\bibnamefont
  {Brif}}, \bibinfo {author} {\bibfnamefont {M.~D.}\ \bibnamefont {Grace}},
  \bibinfo {author} {\bibfnamefont {A.}~\bibnamefont {Donovan}}, \bibinfo
  {author} {\bibfnamefont {D.~L.}\ \bibnamefont {Hocker}}, \bibinfo {author}
  {\bibfnamefont {T.-S.}\ \bibnamefont {Ho}}, \bibinfo {author} {\bibfnamefont
  {R.-B.}\ \bibnamefont {Wu}}, \ and\ \bibinfo {author} {\bibfnamefont
  {H.}~\bibnamefont {Rabitz}},\ }\href {\doibase 10.1103/PhysRevA.86.062309}
  {\bibfield  {journal} {\bibinfo  {journal} {Phys. Rev. A}\ }\textbf {\bibinfo
  {volume} {86}},\ \bibinfo {pages} {062309} (\bibinfo {year}
  {2012})}\BibitemShut {NoStop}%
\bibitem [{\citenamefont {Volkoff}\ \emph {et~al.}(2021)\citenamefont
  {Volkoff}, \citenamefont {Holmes},\ and\ \citenamefont
  {Sornborger}}]{volkoff2021UniversalCompilingNo}%
  \BibitemOpen
  \bibfield  {author} {\bibinfo {author} {\bibfnamefont {T.}~\bibnamefont
  {Volkoff}}, \bibinfo {author} {\bibfnamefont {Z.}~\bibnamefont {Holmes}}, \
  and\ \bibinfo {author} {\bibfnamefont {A.}~\bibnamefont {Sornborger}},\
  }\href {\doibase 10.1103/PRXQuantum.2.040327} {\bibfield  {journal} {\bibinfo
   {journal} {PRX Quantum}\ }\textbf {\bibinfo {volume} {2}},\ \bibinfo {pages}
  {040327} (\bibinfo {year} {2021})}\BibitemShut {NoStop}%
\bibitem [{\citenamefont {Marvian}\ and\ \citenamefont
  {Lloyd}(2016)}]{marvian2016UniversalQuantumEmulator}%
  \BibitemOpen
  \bibfield  {author} {\bibinfo {author} {\bibfnamefont {I.}~\bibnamefont
  {Marvian}}\ and\ \bibinfo {author} {\bibfnamefont {S.}~\bibnamefont
  {Lloyd}},\ }\href {\doibase 10.48550/arXiv.1606.02734} {\enquote {\bibinfo
  {title} {Universal {{Quantum Emulator}}},}\ } (\bibinfo {year} {2016}),\
  \Eprint {http://arxiv.org/abs/1606.02734} {arXiv:1606.02734 [quant-ph]}
  \BibitemShut {NoStop}%
\bibitem [{\citenamefont {True}\ and\ \citenamefont
  {Hamma}(2022)}]{true2022TransitionsEntanglementComplexity}%
  \BibitemOpen
  \bibfield  {author} {\bibinfo {author} {\bibfnamefont {S.}~\bibnamefont
  {True}}\ and\ \bibinfo {author} {\bibfnamefont {A.}~\bibnamefont {Hamma}},\
  }\href {\doibase 10.22331/q-2022-09-22-818} {\bibfield  {journal} {\bibinfo
  {journal} {Quantum}\ }\textbf {\bibinfo {volume} {6}},\ \bibinfo {pages}
  {818} (\bibinfo {year} {2022})}\BibitemShut {NoStop}%
\bibitem [{\citenamefont {Piemontese}\ \emph {et~al.}(2022)\citenamefont
  {Piemontese}, \citenamefont {Roscilde},\ and\ \citenamefont
  {Hamma}}]{piemontese2022EntanglementComplexityRokhsarKivelsonsign}%
  \BibitemOpen
  \bibfield  {author} {\bibinfo {author} {\bibfnamefont {S.}~\bibnamefont
  {Piemontese}}, \bibinfo {author} {\bibfnamefont {T.}~\bibnamefont
  {Roscilde}}, \ and\ \bibinfo {author} {\bibfnamefont {A.}~\bibnamefont
  {Hamma}},\ }\href {\doibase 10.48550/arXiv.2211.01428} {\enquote {\bibinfo
  {title} {Entanglement complexity of the {{Rokhsar-Kivelson-sign}}
  wavefunctions},}\ } (\bibinfo {year} {2022}),\ \Eprint
  {http://arxiv.org/abs/2211.01428} {arXiv:2211.01428 [quant-ph]} \BibitemShut
  {NoStop}%
\bibitem [{\citenamefont
  {Kitaev}(1997)}]{kitaev1997QuantumComputationsAlgorithms}%
  \BibitemOpen
  \bibfield  {author} {\bibinfo {author} {\bibfnamefont {A.~Y.}\ \bibnamefont
  {Kitaev}},\ }\href {\doibase 10.1070/RM1997v052n06ABEH002155} {\bibfield
  {journal} {\bibinfo  {journal} {Russ. Math. Surv.}\ }\textbf {\bibinfo
  {volume} {52}},\ \bibinfo {pages} {1191} (\bibinfo {year}
  {1997})}\BibitemShut {NoStop}%
\bibitem [{\citenamefont {Nielsen}\ and\ \citenamefont
  {Chuang}(2000)}]{nielsen2000QuantumComputationQuantum}%
  \BibitemOpen
  \bibfield  {author} {\bibinfo {author} {\bibfnamefont {M.~A.}\ \bibnamefont
  {Nielsen}}\ and\ \bibinfo {author} {\bibfnamefont {I.~L.}\ \bibnamefont
  {Chuang}},\ }\href@noop {} {\emph {\bibinfo {title} {Quantum {{Computation}}
  and {{Quantum Information}}}}}\ (\bibinfo  {publisher} {{Cambridge University
  Press}},\ \bibinfo {year} {2000})\BibitemShut {NoStop}%
\bibitem [{\citenamefont {Montanaro}\ and\ \citenamefont
  {Osborne}(2010)}]{montanaro2010QuantumBooleanFunctions}%
  \BibitemOpen
  \bibfield  {author} {\bibinfo {author} {\bibfnamefont {A.}~\bibnamefont
  {Montanaro}}\ and\ \bibinfo {author} {\bibfnamefont {T.~J.}\ \bibnamefont
  {Osborne}},\ }\href {\doibase 10.4086/cjtcs.2010.001} {\bibfield  {journal}
  {\bibinfo  {journal} {Chic. J. Theor. Comput. Sci.}\ }\textbf {\bibinfo
  {volume} {16}},\ \bibinfo {pages} {1} (\bibinfo {year} {2010})}\BibitemShut
  {NoStop}%
\bibitem [{\citenamefont
  {Montanaro}(2017)}]{montanaro2017LearningStabilizerStates}%
  \BibitemOpen
  \bibfield  {author} {\bibinfo {author} {\bibfnamefont {A.}~\bibnamefont
  {Montanaro}},\ }\href {\doibase 10.48550/arXiv.1707.04012} {\enquote
  {\bibinfo {title} {Learning stabilizer states by {{Bell}} sampling},}\ }
  (\bibinfo {year} {2017}),\ \Eprint {http://arxiv.org/abs/1707.04012}
  {arXiv:1707.04012 [quant-ph]} \BibitemShut {NoStop}%
\bibitem [{\citenamefont {Levy}\ \emph {et~al.}(2024)\citenamefont {Levy},
  \citenamefont {Luo},\ and\ \citenamefont
  {Clark}}]{levy2021ClassicalShadowsQuantum}%
  \BibitemOpen
  \bibfield  {author} {\bibinfo {author} {\bibfnamefont {R.}~\bibnamefont
  {Levy}}, \bibinfo {author} {\bibfnamefont {D.}~\bibnamefont {Luo}}, \ and\
  \bibinfo {author} {\bibfnamefont {B.~K.}\ \bibnamefont {Clark}},\ }\href
  {\doibase 10.1103/PhysRevResearch.6.013029} {\bibfield  {journal} {\bibinfo
  {journal} {Phys. Rev. Res.}\ }\textbf {\bibinfo {volume} {6}},\ \bibinfo
  {pages} {013029} (\bibinfo {year} {2024})}\BibitemShut {NoStop}%
\bibitem [{\citenamefont {Holmes}\ \emph {et~al.}(2021)\citenamefont {Holmes},
  \citenamefont {Arrasmith}, \citenamefont {Yan}, \citenamefont {Coles},
  \citenamefont {Albrecht},\ and\ \citenamefont
  {Sornborger}}]{holmes2021BarrenPlateausPreclude}%
  \BibitemOpen
  \bibfield  {author} {\bibinfo {author} {\bibfnamefont {Z.}~\bibnamefont
  {Holmes}}, \bibinfo {author} {\bibfnamefont {A.}~\bibnamefont {Arrasmith}},
  \bibinfo {author} {\bibfnamefont {B.}~\bibnamefont {Yan}}, \bibinfo {author}
  {\bibfnamefont {P.~J.}\ \bibnamefont {Coles}}, \bibinfo {author}
  {\bibfnamefont {A.}~\bibnamefont {Albrecht}}, \ and\ \bibinfo {author}
  {\bibfnamefont {A.~T.}\ \bibnamefont {Sornborger}},\ }\href {\doibase
  10.1103/PhysRevLett.126.190501} {\bibfield  {journal} {\bibinfo  {journal}
  {Phys. Rev. Lett,}\ }\textbf {\bibinfo {volume} {126}},\ \bibinfo {pages}
  {190501} (\bibinfo {year} {2021})}\BibitemShut {NoStop}%
\bibitem [{\citenamefont {Kiani}\ \emph {et~al.}(2020)\citenamefont {Kiani},
  \citenamefont {Lloyd},\ and\ \citenamefont
  {Maity}}]{kiani2020LearningUnitariesGradient}%
  \BibitemOpen
  \bibfield  {author} {\bibinfo {author} {\bibfnamefont {B.~T.}\ \bibnamefont
  {Kiani}}, \bibinfo {author} {\bibfnamefont {S.}~\bibnamefont {Lloyd}}, \ and\
  \bibinfo {author} {\bibfnamefont {R.}~\bibnamefont {Maity}},\ }\href
  {https://arxiv.org/abs/2001.11897} {\enquote {\bibinfo {title} {Learning
  unitaries by gradient descent},}\ } (\bibinfo {year} {2020}),\ \Eprint
  {http://arxiv.org/abs/2001.11897} {arXiv:2001.11897} \BibitemShut {NoStop}%
\bibitem [{\citenamefont
  {Aaronson}(2007)}]{aaronson2007LearnabilityQuantumStates}%
  \BibitemOpen
  \bibfield  {author} {\bibinfo {author} {\bibfnamefont {S.}~\bibnamefont
  {Aaronson}},\ }\href {\doibase 10.1098/rspa.2007.0113} {\bibfield  {journal}
  {\bibinfo  {journal} {Proceedings of the Royal Society A: Mathematical,
  Physical and Engineering Sciences}\ }\textbf {\bibinfo {volume} {463}},\
  \bibinfo {pages} {3089} (\bibinfo {year} {2007})}\BibitemShut {NoStop}%
\bibitem [{\citenamefont
  {Aaronson}(2018)}]{aaronson2018ShadowTomographyQuantum}%
  \BibitemOpen
  \bibfield  {author} {\bibinfo {author} {\bibfnamefont {S.}~\bibnamefont
  {Aaronson}},\ }in\ \href {\doibase 10.1145/3188745.3188802} {\emph {\bibinfo
  {booktitle} {Proceedings of the 50th {{Annual ACM SIGACT Symposium}} on
  {{Theory}} of {{Computing}}}}}\ (\bibinfo  {publisher} {{Association for
  Computing Machinery}},\ \bibinfo {year} {2018})\ pp.\ \bibinfo {pages}
  {325--338--325--338}\BibitemShut {NoStop}%
\bibitem [{\citenamefont {Rocchetto}(2018)}]{rocchetto2018StabiliserStatesAre}%
  \BibitemOpen
  \bibfield  {author} {\bibinfo {author} {\bibfnamefont {A.}~\bibnamefont
  {Rocchetto}},\ }\href {\doibase 10.26421/QIC18.7-8-1} {\bibfield  {journal}
  {\bibinfo  {journal} {Quantum Information and Computation}\ }\textbf
  {\bibinfo {volume} {18}},\ \bibinfo {pages} {541} (\bibinfo {year}
  {2018})}\BibitemShut {NoStop}%
\bibitem [{\citenamefont {Rocchetto}\ \emph {et~al.}(2019)\citenamefont
  {Rocchetto}, \citenamefont {Aaronson}, \citenamefont {Severini},
  \citenamefont {Carvacho}, \citenamefont {Poderini}, \citenamefont {Agresti},
  \citenamefont {Bentivegna},\ and\ \citenamefont
  {Sciarrino}}]{rocchetto2019ExperimentalLearningQuantum}%
  \BibitemOpen
  \bibfield  {author} {\bibinfo {author} {\bibfnamefont {A.}~\bibnamefont
  {Rocchetto}}, \bibinfo {author} {\bibfnamefont {S.}~\bibnamefont {Aaronson}},
  \bibinfo {author} {\bibfnamefont {S.}~\bibnamefont {Severini}}, \bibinfo
  {author} {\bibfnamefont {G.}~\bibnamefont {Carvacho}}, \bibinfo {author}
  {\bibfnamefont {D.}~\bibnamefont {Poderini}}, \bibinfo {author}
  {\bibfnamefont {I.}~\bibnamefont {Agresti}}, \bibinfo {author} {\bibfnamefont
  {M.}~\bibnamefont {Bentivegna}}, \ and\ \bibinfo {author} {\bibfnamefont
  {F.}~\bibnamefont {Sciarrino}},\ }\href {\doibase 10.1126/sciadv.aau1946}
  {\bibfield  {journal} {\bibinfo  {journal} {Sci. Adv.}\ }\textbf {\bibinfo
  {volume} {5}},\ \bibinfo {pages} {eaau1946} (\bibinfo {year}
  {2019})}\BibitemShut {NoStop}%
\bibitem [{\citenamefont {Gollakota}\ and\ \citenamefont
  {Liang}(2022)}]{gollakota2022HardnessPAClearningStabilizer}%
  \BibitemOpen
  \bibfield  {author} {\bibinfo {author} {\bibfnamefont {A.}~\bibnamefont
  {Gollakota}}\ and\ \bibinfo {author} {\bibfnamefont {D.}~\bibnamefont
  {Liang}},\ }\href {\doibase 10.22331/q-2022-02-02-640} {\bibfield  {journal}
  {\bibinfo  {journal} {Quantum}\ }\textbf {\bibinfo {volume} {6}},\ \bibinfo
  {pages} {640} (\bibinfo {year} {2022})}\BibitemShut {NoStop}%
\bibitem [{\citenamefont {Liang}(2022)}]{liang2022CliffordCircuitsCan}%
  \BibitemOpen
  \bibfield  {author} {\bibinfo {author} {\bibfnamefont {D.}~\bibnamefont
  {Liang}},\ }\href {\doibase 10.48550/arXiv.2204.06638} {\enquote {\bibinfo
  {title} {Clifford {{Circuits}} can be {{Properly PAC Learned}} if and only if
  {$\textsf{RP}=\textsf{NP}$}},}\ } (\bibinfo {year} {2022}),\ \Eprint
  {http://arxiv.org/abs/2204.06638} {arXiv:2204.06638 [quant-ph]} \BibitemShut
  {NoStop}%
\bibitem [{\citenamefont {Zhou}\ \emph {et~al.}(2020)\citenamefont {Zhou},
  \citenamefont {Yang}, \citenamefont {Hamma},\ and\ \citenamefont
  {Chamon}}]{zhou2020SingleGateClifford}%
  \BibitemOpen
  \bibfield  {author} {\bibinfo {author} {\bibfnamefont {S.}~\bibnamefont
  {Zhou}}, \bibinfo {author} {\bibfnamefont {Z.-C.}\ \bibnamefont {Yang}},
  \bibinfo {author} {\bibfnamefont {A.}~\bibnamefont {Hamma}}, \ and\ \bibinfo
  {author} {\bibfnamefont {C.}~\bibnamefont {Chamon}},\ }\href {\doibase
  10.21468/SciPostPhys.9.6.087} {\bibfield  {journal} {\bibinfo  {journal}
  {SciPost Phys.}\ }\textbf {\bibinfo {volume} {9}},\ \bibinfo {pages} {87}
  (\bibinfo {year} {2020})}\BibitemShut {NoStop}%
\bibitem [{\citenamefont {Bravyi}\ and\ \citenamefont
  {Gosset}(2016)}]{bravyi2016ImprovedClassicalSimulation}%
  \BibitemOpen
  \bibfield  {author} {\bibinfo {author} {\bibfnamefont {S.}~\bibnamefont
  {Bravyi}}\ and\ \bibinfo {author} {\bibfnamefont {D.}~\bibnamefont
  {Gosset}},\ }\href {\doibase 10.1103/PhysRevLett.116.250501} {\bibfield
  {journal} {\bibinfo  {journal} {Phys. Rev. Lett,}\ }\textbf {\bibinfo
  {volume} {116}},\ \bibinfo {pages} {250501} (\bibinfo {year}
  {2016})}\BibitemShut {NoStop}%
\bibitem [{\citenamefont {DiVincenzo}\ \emph {et~al.}(2002)\citenamefont
  {DiVincenzo}, \citenamefont {Leung},\ and\ \citenamefont
  {Terhal}}]{divincenzo2002QuantumDataHiding}%
  \BibitemOpen
  \bibfield  {author} {\bibinfo {author} {\bibfnamefont {D.~P.}\ \bibnamefont
  {DiVincenzo}}, \bibinfo {author} {\bibfnamefont {D.~W.}\ \bibnamefont
  {Leung}}, \ and\ \bibinfo {author} {\bibfnamefont {B.~M.}\ \bibnamefont
  {Terhal}},\ }\href {\doibase 10.1109/18.985948} {\bibfield  {journal}
  {\bibinfo  {journal} {IEEE Trans. Inf. Theory}\ }\textbf {\bibinfo {volume}
  {48}},\ \bibinfo {pages} {580} (\bibinfo {year} {2002})}\BibitemShut
  {NoStop}%
\bibitem [{\citenamefont {Emerson}\ \emph {et~al.}(2003)\citenamefont
  {Emerson}, \citenamefont {Weinstein}, \citenamefont {Saraceno}, \citenamefont
  {Lloyd},\ and\ \citenamefont
  {Cory}}]{emerson2003PseudoRandomUnitaryOperators}%
  \BibitemOpen
  \bibfield  {author} {\bibinfo {author} {\bibfnamefont {J.}~\bibnamefont
  {Emerson}}, \bibinfo {author} {\bibfnamefont {Y.~S.}\ \bibnamefont
  {Weinstein}}, \bibinfo {author} {\bibfnamefont {M.}~\bibnamefont {Saraceno}},
  \bibinfo {author} {\bibfnamefont {S.}~\bibnamefont {Lloyd}}, \ and\ \bibinfo
  {author} {\bibfnamefont {D.~G.}\ \bibnamefont {Cory}},\ }\href {\doibase
  10.1126/science.1090790} {\bibfield  {journal} {\bibinfo  {journal}
  {Science}\ }\textbf {\bibinfo {volume} {302}},\ \bibinfo {pages} {2098}
  (\bibinfo {year} {2003})}\BibitemShut {NoStop}%
\bibitem [{\citenamefont {Zhu}\ \emph {et~al.}(2016)\citenamefont {Zhu},
  \citenamefont {Kueng}, \citenamefont {Grassl},\ and\ \citenamefont
  {Gross}}]{zhu2016CliffordGroupFails}%
  \BibitemOpen
  \bibfield  {author} {\bibinfo {author} {\bibfnamefont {H.}~\bibnamefont
  {Zhu}}, \bibinfo {author} {\bibfnamefont {R.}~\bibnamefont {Kueng}}, \bibinfo
  {author} {\bibfnamefont {M.}~\bibnamefont {Grassl}}, \ and\ \bibinfo {author}
  {\bibfnamefont {D.}~\bibnamefont {Gross}},\ }\href {\doibase
  10.48550/arXiv.1609.08172} {\enquote {\bibinfo {title} {The {{Clifford}}
  group fails gracefully to be a unitary 4-design},}\ } (\bibinfo {year}
  {2016}),\ \Eprint {http://arxiv.org/abs/1609.08172} {arXiv:1609.08172
  [quant-ph]} \BibitemShut {NoStop}%
\bibitem [{\citenamefont {Haferkamp}\ \emph {et~al.}(2020)\citenamefont
  {Haferkamp}, \citenamefont {{Montealegre-Mora}}, \citenamefont {Heinrich},
  \citenamefont {Eisert}, \citenamefont {Gross},\ and\ \citenamefont
  {Roth}}]{haferkamp2020QuantumHomeopathyWorks}%
  \BibitemOpen
  \bibfield  {author} {\bibinfo {author} {\bibfnamefont {J.}~\bibnamefont
  {Haferkamp}}, \bibinfo {author} {\bibfnamefont {F.}~\bibnamefont
  {{Montealegre-Mora}}}, \bibinfo {author} {\bibfnamefont {M.}~\bibnamefont
  {Heinrich}}, \bibinfo {author} {\bibfnamefont {J.}~\bibnamefont {Eisert}},
  \bibinfo {author} {\bibfnamefont {D.}~\bibnamefont {Gross}}, \ and\ \bibinfo
  {author} {\bibfnamefont {I.}~\bibnamefont {Roth}},\ }\href {\doibase
  10.48550/arXiv.2002.09524} {\enquote {\bibinfo {title} {Quantum homeopathy
  works: {{Efficient}} unitary designs with a system-size independent number of
  non-{{Clifford}} gates},}\ } (\bibinfo {year} {2020}),\ \Eprint
  {http://arxiv.org/abs/2002.09524} {arXiv:2002.09524 [math-ph,
  physics:quant-ph]} \BibitemShut {NoStop}%
\bibitem [{\citenamefont {Hinsche}\ \emph {et~al.}(2022)\citenamefont
  {Hinsche}, \citenamefont {Ioannou}, \citenamefont {Nietner}, \citenamefont
  {Haferkamp}, \citenamefont {Quek}, \citenamefont {Hangleiter}, \citenamefont
  {Seifert}, \citenamefont {Eisert},\ and\ \citenamefont
  {Sweke}}]{hinsche2022SingleGateMakes}%
  \BibitemOpen
  \bibfield  {author} {\bibinfo {author} {\bibfnamefont {M.}~\bibnamefont
  {Hinsche}}, \bibinfo {author} {\bibfnamefont {M.}~\bibnamefont {Ioannou}},
  \bibinfo {author} {\bibfnamefont {A.}~\bibnamefont {Nietner}}, \bibinfo
  {author} {\bibfnamefont {J.}~\bibnamefont {Haferkamp}}, \bibinfo {author}
  {\bibfnamefont {Y.}~\bibnamefont {Quek}}, \bibinfo {author} {\bibfnamefont
  {D.}~\bibnamefont {Hangleiter}}, \bibinfo {author} {\bibfnamefont {J.-P.}\
  \bibnamefont {Seifert}}, \bibinfo {author} {\bibfnamefont {J.}~\bibnamefont
  {Eisert}}, \ and\ \bibinfo {author} {\bibfnamefont {R.}~\bibnamefont
  {Sweke}},\ }\href {\doibase 10.48550/arXiv.2207.03140} {\enquote {\bibinfo
  {title} {A single {$T$}-gate makes distribution learning hard},}\ } (\bibinfo
  {year} {2022}),\ \Eprint {http://arxiv.org/abs/2207.03140} {arXiv:2207.03140
  [quant-ph, stat]} \BibitemShut {NoStop}%
\bibitem [{\citenamefont
  {Pietrzak}(2012)}]{pietrzak2012CryptographyLearningParity}%
  \BibitemOpen
  \bibfield  {author} {\bibinfo {author} {\bibfnamefont {K.}~\bibnamefont
  {Pietrzak}},\ }in\ \href {\doibase 10.1007/978-3-642-27660-6_9} {\emph
  {\bibinfo {booktitle} {{{SOFSEM}} 2012: {{Theory}} and {{Practice}} of
  {{Computer Science}}}}},\ \bibinfo {series and number} {Lecture {{Notes}} in
  {{Computer Science}}},\ \bibinfo {editor} {edited by\ \bibinfo {editor}
  {\bibfnamefont {M.}~\bibnamefont {Bielikov{\'a}}}, \bibinfo {editor}
  {\bibfnamefont {G.}~\bibnamefont {Friedrich}}, \bibinfo {editor}
  {\bibfnamefont {G.}~\bibnamefont {Gottlob}}, \bibinfo {editor} {\bibfnamefont
  {S.}~\bibnamefont {Katzenbeisser}}, \ and\ \bibinfo {editor} {\bibfnamefont
  {G.}~\bibnamefont {Tur{\'a}n}}}\ (\bibinfo  {publisher} {{Springer}},\
  \bibinfo {address} {{Berlin, Heidelberg}},\ \bibinfo {year} {2012})\ pp.\
  \bibinfo {pages} {99--114}\BibitemShut {NoStop}%
\bibitem [{\citenamefont {Nahum}\ \emph {et~al.}(2018)\citenamefont {Nahum},
  \citenamefont {Vijay},\ and\ \citenamefont
  {Haah}}]{nahum2018OperatorSpreadingRandom}%
  \BibitemOpen
  \bibfield  {author} {\bibinfo {author} {\bibfnamefont {A.}~\bibnamefont
  {Nahum}}, \bibinfo {author} {\bibfnamefont {S.}~\bibnamefont {Vijay}}, \ and\
  \bibinfo {author} {\bibfnamefont {J.}~\bibnamefont {Haah}},\ }\href {\doibase
  10.1103/PhysRevX.8.021014} {\bibfield  {journal} {\bibinfo  {journal} {Phys.
  Rev. X}\ }\textbf {\bibinfo {volume} {8}},\ \bibinfo {pages} {021014}
  (\bibinfo {year} {2018})}\BibitemShut {NoStop}%
\bibitem [{\citenamefont {Khemani}\ \emph {et~al.}(2018)\citenamefont
  {Khemani}, \citenamefont {Vishwanath},\ and\ \citenamefont
  {Huse}}]{khemani2018OperatorSpreadingEmergence}%
  \BibitemOpen
  \bibfield  {author} {\bibinfo {author} {\bibfnamefont {V.}~\bibnamefont
  {Khemani}}, \bibinfo {author} {\bibfnamefont {A.}~\bibnamefont {Vishwanath}},
  \ and\ \bibinfo {author} {\bibfnamefont {D.~A.}\ \bibnamefont {Huse}},\
  }\href {\doibase 10.1103/PhysRevX.8.031057} {\bibfield  {journal} {\bibinfo
  {journal} {Phys. Rev. X}\ }\textbf {\bibinfo {volume} {8}},\ \bibinfo {pages}
  {031057} (\bibinfo {year} {2018})}\BibitemShut {NoStop}%
\bibitem [{\citenamefont {Chamon}\ \emph {et~al.}(2022)\citenamefont {Chamon},
  \citenamefont {Mucciolo},\ and\ \citenamefont
  {Ruckenstein}}]{chamon2022QuantumStatisticalMechanics}%
  \BibitemOpen
  \bibfield  {author} {\bibinfo {author} {\bibfnamefont {C.}~\bibnamefont
  {Chamon}}, \bibinfo {author} {\bibfnamefont {E.~R.}\ \bibnamefont
  {Mucciolo}}, \ and\ \bibinfo {author} {\bibfnamefont {A.~E.}\ \bibnamefont
  {Ruckenstein}},\ }\href {\doibase 10.1016/j.aop.2022.169086} {\bibfield
  {journal} {\bibinfo  {journal} {Ann. Phys.(Amsterdam)}\ }\textbf {\bibinfo
  {volume} {446}},\ \bibinfo {pages} {169086} (\bibinfo {year}
  {2022})}\BibitemShut {NoStop}%
\bibitem [{\citenamefont {Page}(1993)}]{page1993AverageEntropySubsystem}%
  \BibitemOpen
  \bibfield  {author} {\bibinfo {author} {\bibfnamefont {D.~N.}\ \bibnamefont
  {Page}},\ }\href {\doibase 10.1103/PhysRevLett.71.1291} {\bibfield  {journal}
  {\bibinfo  {journal} {Phys. Rev. Lett,}\ }\textbf {\bibinfo {volume} {71}},\
  \bibinfo {pages} {1291} (\bibinfo {year} {1993})}\BibitemShut {NoStop}%
\bibitem [{\citenamefont {Yoshida}\ and\ \citenamefont
  {Yao}(2019)}]{yoshida2019DisentanglingScramblingDecoherence}%
  \BibitemOpen
  \bibfield  {author} {\bibinfo {author} {\bibfnamefont {B.}~\bibnamefont
  {Yoshida}}\ and\ \bibinfo {author} {\bibfnamefont {N.~Y.}\ \bibnamefont
  {Yao}},\ }\href {\doibase 10.1103/PhysRevX.9.011006} {\bibfield  {journal}
  {\bibinfo  {journal} {Phys. Rev. X}\ }\textbf {\bibinfo {volume} {9}},\
  \bibinfo {pages} {011006} (\bibinfo {year} {2019})}\BibitemShut {NoStop}%
\bibitem [{\citenamefont
  {Yoshida}(2021)}]{yoshida2021DecodingEntanglementStructure}%
  \BibitemOpen
  \bibfield  {author} {\bibinfo {author} {\bibfnamefont {B.}~\bibnamefont
  {Yoshida}},\ }\href {\doibase 10.48550/arXiv.2109.08691} {\enquote {\bibinfo
  {title} {Decoding the {{Entanglement Structure}} of {{Monitored Quantum
  Circuits}}},}\ } (\bibinfo {year} {2021}),\ \Eprint
  {http://arxiv.org/abs/2109.08691} {arXiv:2109.08691 [cond-mat,
  physics:hep-th, physics:quant-ph]} \BibitemShut {NoStop}%
\bibitem [{\citenamefont {Aaronson}\ and\ \citenamefont
  {Gottesman}(2004)}]{aaronson2004ImprovedSimulationStabilizer}%
  \BibitemOpen
  \bibfield  {author} {\bibinfo {author} {\bibfnamefont {S.}~\bibnamefont
  {Aaronson}}\ and\ \bibinfo {author} {\bibfnamefont {D.}~\bibnamefont
  {Gottesman}},\ }\href {\doibase 10.1103/PhysRevA.70.052328} {\bibfield
  {journal} {\bibinfo  {journal} {Phys. Rev. A}\ }\textbf {\bibinfo {volume}
  {70}},\ \bibinfo {pages} {052328} (\bibinfo {year} {2004})}\BibitemShut
  {NoStop}%
\bibitem [{\citenamefont {Leone}\ \emph
  {et~al.}(2022{\natexlab{b}})\citenamefont {Leone}, \citenamefont {Oliviero},\
  and\ \citenamefont {Hamma}}]{leone2022MagicHindersQuantum}%
  \BibitemOpen
  \bibfield  {author} {\bibinfo {author} {\bibfnamefont {L.}~\bibnamefont
  {Leone}}, \bibinfo {author} {\bibfnamefont {S.~F.~E.}\ \bibnamefont
  {Oliviero}}, \ and\ \bibinfo {author} {\bibfnamefont {A.}~\bibnamefont
  {Hamma}},\ }\href {\doibase 10.48550/arXiv.2204.02995} {\enquote {\bibinfo
  {title} {Magic hinders quantum certification},}\ } (\bibinfo {year}
  {2022}{\natexlab{b}}),\ \Eprint {http://arxiv.org/abs/2204.02995}
  {arXiv:2204.02995 [quant-ph]} \BibitemShut {NoStop}%
\bibitem [{\citenamefont {Jiang}\ and\ \citenamefont
  {Wang}(2021)}]{jiang2021LowerBoundTcount}%
  \BibitemOpen
  \bibfield  {author} {\bibinfo {author} {\bibfnamefont {J.}~\bibnamefont
  {Jiang}}\ and\ \bibinfo {author} {\bibfnamefont {X.}~\bibnamefont {Wang}},\
  }\href {\doibase 10.48550/arXiv.2103.09999} {\enquote {\bibinfo {title}
  {Lower bound the {{T-count}} via unitary stabilizer nullity},}\ } (\bibinfo
  {year} {2021}),\ \Eprint {http://arxiv.org/abs/2103.09999} {arXiv:2103.09999
  [hep-th, physics:quant-ph]} \BibitemShut {NoStop}%
\bibitem [{\citenamefont {Leone}\ \emph
  {et~al.}(2022{\natexlab{c}})\citenamefont {Leone}, \citenamefont {Oliviero},\
  and\ \citenamefont {Hamma}}]{leone2022StabilizerRenyiEntropy}%
  \BibitemOpen
  \bibfield  {author} {\bibinfo {author} {\bibfnamefont {L.}~\bibnamefont
  {Leone}}, \bibinfo {author} {\bibfnamefont {S.~F.~E.}\ \bibnamefont
  {Oliviero}}, \ and\ \bibinfo {author} {\bibfnamefont {A.}~\bibnamefont
  {Hamma}},\ }\href {\doibase 10.1103/PhysRevLett.128.050402} {\bibfield
  {journal} {\bibinfo  {journal} {Phys. Rev. Lett,}\ }\textbf {\bibinfo
  {volume} {128}},\ \bibinfo {pages} {050402} (\bibinfo {year}
  {2022}{\natexlab{c}})}\BibitemShut {NoStop}%
\bibitem [{\citenamefont {Dehaene}\ and\ \citenamefont
  {De~Moor}(2003)}]{dehaene2003CliffordGroupStabilizer}%
  \BibitemOpen
  \bibfield  {author} {\bibinfo {author} {\bibfnamefont {J.}~\bibnamefont
  {Dehaene}}\ and\ \bibinfo {author} {\bibfnamefont {B.}~\bibnamefont
  {De~Moor}},\ }\href {\doibase 10.1103/PhysRevA.68.042318} {\bibfield
  {journal} {\bibinfo  {journal} {Phys. Rev. A}\ }\textbf {\bibinfo {volume}
  {68}},\ \bibinfo {pages} {042318} (\bibinfo {year} {2003})}\BibitemShut
  {NoStop}%
\bibitem [{\citenamefont {Gosset}\ \emph {et~al.}(2021)\citenamefont {Gosset},
  \citenamefont {Grier}, \citenamefont {Kerzner},\ and\ \citenamefont
  {Schaeffer}}]{gosset2021FastSimulationPlanar}%
  \BibitemOpen
  \bibfield  {author} {\bibinfo {author} {\bibfnamefont {D.}~\bibnamefont
  {Gosset}}, \bibinfo {author} {\bibfnamefont {D.}~\bibnamefont {Grier}},
  \bibinfo {author} {\bibfnamefont {A.}~\bibnamefont {Kerzner}}, \ and\
  \bibinfo {author} {\bibfnamefont {L.}~\bibnamefont {Schaeffer}},\ }\href
  {\doibase 10.48550/arXiv.2009.03218} {\enquote {\bibinfo {title} {Fast
  simulation of planar {{Clifford}} circuits},}\ } (\bibinfo {year} {2021}),\
  \Eprint {http://arxiv.org/abs/2009.03218} {arXiv:2009.03218 [quant-ph]}
  \BibitemShut {NoStop}%
\bibitem [{\citenamefont {Bravyi}\ and\ \citenamefont
  {Maslov}(2021)}]{bravyi2021hadamard}%
  \BibitemOpen
  \bibfield  {author} {\bibinfo {author} {\bibfnamefont {S.}~\bibnamefont
  {Bravyi}}\ and\ \bibinfo {author} {\bibfnamefont {D.}~\bibnamefont
  {Maslov}},\ }\href {\doibase 10.1109/TIT.2021.3081415} {\bibfield  {journal}
  {\bibinfo  {journal} {IEEE Trans. Inf. Theory}\ }\textbf {\bibinfo {volume}
  {67}},\ \bibinfo {pages} {4546} (\bibinfo {year} {2021})}\BibitemShut
  {NoStop}%
\bibitem [{\citenamefont {Ding}\ \emph {et~al.}(2016)\citenamefont {Ding},
  \citenamefont {Hayden},\ and\ \citenamefont
  {Walter}}]{ding2016ConditionalMutualInformation}%
  \BibitemOpen
  \bibfield  {author} {\bibinfo {author} {\bibfnamefont {D.}~\bibnamefont
  {Ding}}, \bibinfo {author} {\bibfnamefont {P.}~\bibnamefont {Hayden}}, \ and\
  \bibinfo {author} {\bibfnamefont {M.}~\bibnamefont {Walter}},\ }\href
  {\doibase 10.1007/JHEP12(2016)145} {\bibfield  {journal} {\bibinfo  {journal}
  {J. High Energy Phys.}\ }\textbf {\bibinfo {volume} {2016}},\ \bibinfo
  {pages} {145} (\bibinfo {year} {2016})}\BibitemShut {NoStop}%
\bibitem [{\citenamefont {Cotler}\ \emph
  {et~al.}(2017{\natexlab{a}})\citenamefont {Cotler}, \citenamefont
  {{Gur-Ari}}, \citenamefont {Hanada}, \citenamefont {Polchinski},
  \citenamefont {Saad}, \citenamefont {Shenker}, \citenamefont {Stanford},
  \citenamefont {Streicher},\ and\ \citenamefont
  {Tezuka}}]{cotler2017BlackHolesRandom}%
  \BibitemOpen
  \bibfield  {author} {\bibinfo {author} {\bibfnamefont {J.~S.}\ \bibnamefont
  {Cotler}}, \bibinfo {author} {\bibfnamefont {G.}~\bibnamefont {{Gur-Ari}}},
  \bibinfo {author} {\bibfnamefont {M.}~\bibnamefont {Hanada}}, \bibinfo
  {author} {\bibfnamefont {J.}~\bibnamefont {Polchinski}}, \bibinfo {author}
  {\bibfnamefont {P.}~\bibnamefont {Saad}}, \bibinfo {author} {\bibfnamefont
  {S.~H.}\ \bibnamefont {Shenker}}, \bibinfo {author} {\bibfnamefont
  {D.}~\bibnamefont {Stanford}}, \bibinfo {author} {\bibfnamefont
  {A.}~\bibnamefont {Streicher}}, \ and\ \bibinfo {author} {\bibfnamefont
  {M.}~\bibnamefont {Tezuka}},\ }\href {\doibase 10.1007/JHEP05(2017)118}
  {\bibfield  {journal} {\bibinfo  {journal} {J. High Energy Phys.}\ }\textbf
  {\bibinfo {volume} {2017}},\ \bibinfo {pages} {118} (\bibinfo {year}
  {2017}{\natexlab{a}})}\BibitemShut {NoStop}%
\bibitem [{\citenamefont {Cotler}\ \emph
  {et~al.}(2017{\natexlab{b}})\citenamefont {Cotler}, \citenamefont
  {{Hunter-Jones}}, \citenamefont {Liu},\ and\ \citenamefont
  {Yoshida}}]{cotler2017ChaosComplexityRandom}%
  \BibitemOpen
  \bibfield  {author} {\bibinfo {author} {\bibfnamefont {J.}~\bibnamefont
  {Cotler}}, \bibinfo {author} {\bibfnamefont {N.}~\bibnamefont
  {{Hunter-Jones}}}, \bibinfo {author} {\bibfnamefont {J.}~\bibnamefont {Liu}},
  \ and\ \bibinfo {author} {\bibfnamefont {B.}~\bibnamefont {Yoshida}},\ }\href
  {\doibase 10.1007/JHEP11(2017)048} {\bibfield  {journal} {\bibinfo  {journal}
  {J. High Energy Phys.}\ }\textbf {\bibinfo {volume} {2017}},\ \bibinfo
  {pages} {48} (\bibinfo {year} {2017}{\natexlab{b}})}\BibitemShut {NoStop}%
\bibitem [{\citenamefont {Chen}\ and\ \citenamefont
  {Zhou}(2018)}]{chen2018OperatorScramblingQuantuma}%
  \BibitemOpen
  \bibfield  {author} {\bibinfo {author} {\bibfnamefont {X.}~\bibnamefont
  {Chen}}\ and\ \bibinfo {author} {\bibfnamefont {T.}~\bibnamefont {Zhou}},\
  }\href {\doibase 10.48550/arXiv.1804.08655} {\enquote {\bibinfo {title}
  {Operator scrambling and quantum chaos},}\ } (\bibinfo {year} {2018}),\
  \Eprint {http://arxiv.org/abs/1804.08655} {arXiv:1804.08655 [cond-mat,
  physics:hep-th, physics:quant-ph]} \BibitemShut {NoStop}%
\bibitem [{\citenamefont {Zhang}\ \emph {et~al.}(2019)\citenamefont {Zhang},
  \citenamefont {Huang}, \citenamefont {Chen} \emph
  {et~al.}}]{zhang2019InformationScramblingChaotic}%
  \BibitemOpen
  \bibfield  {author} {\bibinfo {author} {\bibfnamefont {Y.-L.}\ \bibnamefont
  {Zhang}}, \bibinfo {author} {\bibfnamefont {Y.}~\bibnamefont {Huang}},
  \bibinfo {author} {\bibfnamefont {X.}~\bibnamefont {Chen}},  \emph {et~al.},\
  }\href@noop {} {\bibfield  {journal} {\bibinfo  {journal} {Phys. Rev. B}\
  }\textbf {\bibinfo {volume} {99}},\ \bibinfo {pages} {014303} (\bibinfo
  {year} {2019})}\BibitemShut {NoStop}%
\bibitem [{\citenamefont {Yan}\ \emph {et~al.}(2020)\citenamefont {Yan},
  \citenamefont {Cincio},\ and\ \citenamefont
  {Zurek}}]{yan2020InformationScramblingLoschmidt}%
  \BibitemOpen
  \bibfield  {author} {\bibinfo {author} {\bibfnamefont {B.}~\bibnamefont
  {Yan}}, \bibinfo {author} {\bibfnamefont {L.}~\bibnamefont {Cincio}}, \ and\
  \bibinfo {author} {\bibfnamefont {W.~H.}\ \bibnamefont {Zurek}},\ }\href
  {\doibase 10.1103/PhysRevLett.124.160603} {\bibfield  {journal} {\bibinfo
  {journal} {Phys. Rev. Lett,}\ }\textbf {\bibinfo {volume} {124}},\ \bibinfo
  {pages} {160603} (\bibinfo {year} {2020})}\BibitemShut {NoStop}%
\bibitem [{\citenamefont {Mi}\ \emph {et~al.}(2021)\citenamefont {Mi},
  \citenamefont {Roushan}, \citenamefont {Quintana}, \citenamefont {Mandra},
  \citenamefont {Marshall}, \citenamefont {Neill}, \citenamefont {Arute},
  \citenamefont {Arya}, \citenamefont {Atalaya}, \citenamefont {Babbush} \emph
  {et~al.}}]{mi2021InformationScramblingComputationally}%
  \BibitemOpen
  \bibfield  {author} {\bibinfo {author} {\bibfnamefont {X.}~\bibnamefont
  {Mi}}, \bibinfo {author} {\bibfnamefont {P.}~\bibnamefont {Roushan}},
  \bibinfo {author} {\bibfnamefont {C.}~\bibnamefont {Quintana}}, \bibinfo
  {author} {\bibfnamefont {S.}~\bibnamefont {Mandra}}, \bibinfo {author}
  {\bibfnamefont {J.}~\bibnamefont {Marshall}}, \bibinfo {author}
  {\bibfnamefont {C.}~\bibnamefont {Neill}}, \bibinfo {author} {\bibfnamefont
  {F.}~\bibnamefont {Arute}}, \bibinfo {author} {\bibfnamefont
  {K.}~\bibnamefont {Arya}}, \bibinfo {author} {\bibfnamefont {J.}~\bibnamefont
  {Atalaya}}, \bibinfo {author} {\bibfnamefont {R.}~\bibnamefont {Babbush}},
  \emph {et~al.},\ }\href@noop {} {\enquote {\bibinfo {title} {Information
  scrambling in computationally complex quantum circuits},}\ } (\bibinfo {year}
  {2021}),\ \Eprint {http://arxiv.org/abs/2101.08870} {arXiv:2101.08870}
  \BibitemShut {NoStop}%
\bibitem [{\citenamefont {Zhuang}\ \emph {et~al.}(2019)\citenamefont {Zhuang},
  \citenamefont {Schuster}, \citenamefont {Yoshida},\ and\ \citenamefont
  {Yao}}]{zhuang2019ScramblingComplexityPhase}%
  \BibitemOpen
  \bibfield  {author} {\bibinfo {author} {\bibfnamefont {Q.}~\bibnamefont
  {Zhuang}}, \bibinfo {author} {\bibfnamefont {T.}~\bibnamefont {Schuster}},
  \bibinfo {author} {\bibfnamefont {B.}~\bibnamefont {Yoshida}}, \ and\
  \bibinfo {author} {\bibfnamefont {N.~Y.}\ \bibnamefont {Yao}},\ }\href
  {\doibase 10.1103/PhysRevA.99.062334} {\bibfield  {journal} {\bibinfo
  {journal} {Phys. Rev. A}\ }\textbf {\bibinfo {volume} {99}},\ \bibinfo
  {pages} {062334} (\bibinfo {year} {2019})}\BibitemShut {NoStop}%
\bibitem [{\citenamefont {Styliaris}\ \emph {et~al.}(2021)\citenamefont
  {Styliaris}, \citenamefont {Anand},\ and\ \citenamefont
  {Zanardi}}]{styliaris2021InformationScramblingBipartitions}%
  \BibitemOpen
  \bibfield  {author} {\bibinfo {author} {\bibfnamefont {G.}~\bibnamefont
  {Styliaris}}, \bibinfo {author} {\bibfnamefont {N.}~\bibnamefont {Anand}}, \
  and\ \bibinfo {author} {\bibfnamefont {P.}~\bibnamefont {Zanardi}},\ }\href
  {\doibase 10.1103/PhysRevLett.126.030601} {\bibfield  {journal} {\bibinfo
  {journal} {Phys. Rev. Lett,}\ }\textbf {\bibinfo {volume} {126}},\ \bibinfo
  {pages} {030601} (\bibinfo {year} {2021})}\BibitemShut {NoStop}%
\bibitem [{\citenamefont {Touil}\ and\ \citenamefont
  {Deffner}(2021)}]{touil2021InformationScramblingDecoherence}%
  \BibitemOpen
  \bibfield  {author} {\bibinfo {author} {\bibfnamefont {A.}~\bibnamefont
  {Touil}}\ and\ \bibinfo {author} {\bibfnamefont {S.}~\bibnamefont
  {Deffner}},\ }\href@noop {} {\bibfield  {journal} {\bibinfo  {journal} {PRX
  Quantum}\ }\textbf {\bibinfo {volume} {2}},\ \bibinfo {pages} {010306}
  (\bibinfo {year} {2021})}\BibitemShut {NoStop}%
\bibitem [{\citenamefont {Garcia}\ \emph {et~al.}(2021)\citenamefont {Garcia},
  \citenamefont {Zhou},\ and\ \citenamefont
  {Jaffe}}]{garcia2021QuantumScramblingClassical}%
  \BibitemOpen
  \bibfield  {author} {\bibinfo {author} {\bibfnamefont {R.~J.}\ \bibnamefont
  {Garcia}}, \bibinfo {author} {\bibfnamefont {Y.}~\bibnamefont {Zhou}}, \ and\
  \bibinfo {author} {\bibfnamefont {A.}~\bibnamefont {Jaffe}},\ }\href
  {\doibase 10.1103/PhysRevResearch.3.033155} {\bibfield  {journal} {\bibinfo
  {journal} {Phys. Rev. Research}\ }\textbf {\bibinfo {volume} {3}},\ \bibinfo
  {pages} {033155} (\bibinfo {year} {2021})}\BibitemShut {NoStop}%
\bibitem [{\citenamefont {Campbell}\ and\ \citenamefont
  {Browne}(2010)}]{campbell2010BoundStatesMagic}%
  \BibitemOpen
  \bibfield  {author} {\bibinfo {author} {\bibfnamefont {E.~T.}\ \bibnamefont
  {Campbell}}\ and\ \bibinfo {author} {\bibfnamefont {D.~E.}\ \bibnamefont
  {Browne}},\ }\href {\doibase 10.1103/PhysRevLett.104.030503} {\bibfield
  {journal} {\bibinfo  {journal} {Phys. Rev. Lett,}\ }\textbf {\bibinfo
  {volume} {104}},\ \bibinfo {pages} {030503} (\bibinfo {year}
  {2010})}\BibitemShut {NoStop}%
\bibitem [{\citenamefont {Campbell}\ \emph {et~al.}(2012)\citenamefont
  {Campbell}, \citenamefont {Anwar},\ and\ \citenamefont
  {Browne}}]{campbell2012MagicStateDistillationAll}%
  \BibitemOpen
  \bibfield  {author} {\bibinfo {author} {\bibfnamefont {E.~T.}\ \bibnamefont
  {Campbell}}, \bibinfo {author} {\bibfnamefont {H.}~\bibnamefont {Anwar}}, \
  and\ \bibinfo {author} {\bibfnamefont {D.~E.}\ \bibnamefont {Browne}},\
  }\href {\doibase 10.1103/PhysRevX.2.041021} {\bibfield  {journal} {\bibinfo
  {journal} {Phys. Rev. X}\ }\textbf {\bibinfo {volume} {2}},\ \bibinfo {pages}
  {041021} (\bibinfo {year} {2012})}\BibitemShut {NoStop}%
\bibitem [{\citenamefont {Veitch}\ \emph {et~al.}(2014)\citenamefont {Veitch},
  \citenamefont {Mousavian}, \citenamefont {Gottesman},\ and\ \citenamefont
  {Emerson}}]{veitch2014ResourceTheoryStabilizer}%
  \BibitemOpen
  \bibfield  {author} {\bibinfo {author} {\bibfnamefont {V.}~\bibnamefont
  {Veitch}}, \bibinfo {author} {\bibfnamefont {S.~A.~H.}\ \bibnamefont
  {Mousavian}}, \bibinfo {author} {\bibfnamefont {D.}~\bibnamefont
  {Gottesman}}, \ and\ \bibinfo {author} {\bibfnamefont {J.}~\bibnamefont
  {Emerson}},\ }\href {\doibase 10.1088/1367-2630/16/1/013009} {\bibfield
  {journal} {\bibinfo  {journal} {New J. Phys.}\ }\textbf {\bibinfo {volume}
  {16}},\ \bibinfo {pages} {013009} (\bibinfo {year} {2014})}\BibitemShut
  {NoStop}%
\bibitem [{\citenamefont {Koukoulekidis}\ and\ \citenamefont
  {Jennings}(pril)}]{koukoulekidis2022ConstraintsMagicState}%
  \BibitemOpen
  \bibfield  {author} {\bibinfo {author} {\bibfnamefont {N.}~\bibnamefont
  {Koukoulekidis}}\ and\ \bibinfo {author} {\bibfnamefont {D.}~\bibnamefont
  {Jennings}},\ }\href {\doibase 10.1038/s41534-022-00551-1} {\bibfield
  {journal} {\bibinfo  {journal} {npj Quantum Information}\ }\textbf {\bibinfo
  {volume} {8}},\ \bibinfo {pages} {1} (\bibinfo {year}
  {2022/april})}\BibitemShut {NoStop}%
\bibitem [{\citenamefont {Saxena}\ and\ \citenamefont
  {Gour}(2022)}]{saxena2022QuantifyingMultiqubitMagic}%
  \BibitemOpen
  \bibfield  {author} {\bibinfo {author} {\bibfnamefont {G.}~\bibnamefont
  {Saxena}}\ and\ \bibinfo {author} {\bibfnamefont {G.}~\bibnamefont {Gour}},\
  }\href {\doibase 10.1103/PhysRevA.106.042422} {\bibfield  {journal} {\bibinfo
   {journal} {Phys. Rev. A}\ }\textbf {\bibinfo {volume} {106}},\ \bibinfo
  {pages} {042422} (\bibinfo {year} {2022})}\BibitemShut {NoStop}%
\bibitem [{\citenamefont {Oliviero}\ \emph
  {et~al.}(2022{\natexlab{a}})\citenamefont {Oliviero}, \citenamefont {Leone},
  \citenamefont {Hamma},\ and\ \citenamefont
  {Lloyd}}]{oliviero2022MeasuringMagicQuantum}%
  \BibitemOpen
  \bibfield  {author} {\bibinfo {author} {\bibfnamefont {S.~F.~E.}\
  \bibnamefont {Oliviero}}, \bibinfo {author} {\bibfnamefont {L.}~\bibnamefont
  {Leone}}, \bibinfo {author} {\bibfnamefont {A.}~\bibnamefont {Hamma}}, \ and\
  \bibinfo {author} {\bibfnamefont {S.}~\bibnamefont {Lloyd}},\ }\href
  {\doibase 10.1038/s41534-022-00666-5} {\bibfield  {journal} {\bibinfo
  {journal} {npj Quantum Information}\ }\textbf {\bibinfo {volume} {8}},\
  \bibinfo {pages} {1} (\bibinfo {year} {2022}{\natexlab{a}})}\BibitemShut
  {NoStop}%
\bibitem [{\citenamefont {Oliviero}\ \emph
  {et~al.}(2022{\natexlab{b}})\citenamefont {Oliviero}, \citenamefont {Leone},\
  and\ \citenamefont {Hamma}}]{oliviero2022MagicstateResourceTheory}%
  \BibitemOpen
  \bibfield  {author} {\bibinfo {author} {\bibfnamefont {S.~F.~E.}\
  \bibnamefont {Oliviero}}, \bibinfo {author} {\bibfnamefont {L.}~\bibnamefont
  {Leone}}, \ and\ \bibinfo {author} {\bibfnamefont {A.}~\bibnamefont
  {Hamma}},\ }\href {\doibase 10.1103/PhysRevA.106.042426} {\bibfield
  {journal} {\bibinfo  {journal} {Phys. Rev. A}\ }\textbf {\bibinfo {volume}
  {106}},\ \bibinfo {pages} {042426} (\bibinfo {year}
  {2022}{\natexlab{b}})}\BibitemShut {NoStop}%
\bibitem [{\citenamefont {Hahn}\ \emph {et~al.}(2022)\citenamefont {Hahn},
  \citenamefont {Ferraro}, \citenamefont {Hultquist}, \citenamefont {Ferrini}
  \emph {et~al.}}]{hahn2022QuantifyingQubitMagic}%
  \BibitemOpen
  \bibfield  {author} {\bibinfo {author} {\bibfnamefont {O.}~\bibnamefont
  {Hahn}}, \bibinfo {author} {\bibfnamefont {A.}~\bibnamefont {Ferraro}},
  \bibinfo {author} {\bibfnamefont {L.}~\bibnamefont {Hultquist}}, \bibinfo
  {author} {\bibfnamefont {G.}~\bibnamefont {Ferrini}},  \emph {et~al.},\
  }\href {\doibase 10.1103/PhysRevLett.128.210502} {\bibfield  {journal}
  {\bibinfo  {journal} {Phys. Rev. Lett,}\ }\textbf {\bibinfo {volume} {128}},\
  \bibinfo {pages} {210502} (\bibinfo {year} {2022})}\BibitemShut {NoStop}%
\bibitem [{\citenamefont {Haug}\ and\ \citenamefont
  {Kim}(2023)}]{haug2023ScalableMeasuresMagic}%
  \BibitemOpen
  \bibfield  {author} {\bibinfo {author} {\bibfnamefont {T.}~\bibnamefont
  {Haug}}\ and\ \bibinfo {author} {\bibfnamefont {M.}~\bibnamefont {Kim}},\
  }\href {\doibase 10.1103/PRXQuantum.4.010301} {\bibfield  {journal} {\bibinfo
   {journal} {PRX Quantum}\ }\textbf {\bibinfo {volume} {4}},\ \bibinfo {pages}
  {010301} (\bibinfo {year} {2023})}\BibitemShut {NoStop}%
\bibitem [{\citenamefont {Goldstein}\ \emph {et~al.}(2019)\citenamefont
  {Goldstein}, \citenamefont {Lebowitz}, \citenamefont {Tumulka},\ and\
  \citenamefont {Zangh{\`i}}}]{goldstein2019GibbsBoltzmannEntropy}%
  \BibitemOpen
  \bibfield  {author} {\bibinfo {author} {\bibfnamefont {S.}~\bibnamefont
  {Goldstein}}, \bibinfo {author} {\bibfnamefont {J.~L.}\ \bibnamefont
  {Lebowitz}}, \bibinfo {author} {\bibfnamefont {R.}~\bibnamefont {Tumulka}}, \
  and\ \bibinfo {author} {\bibfnamefont {N.}~\bibnamefont {Zangh{\`i}}},\ }in\
  \href {\doibase 10.1142/9789811211720_0014} {\emph {\bibinfo {booktitle}
  {Statistical {{Mechanics}} and {{Scientific Explanation}}}}}\ (\bibinfo
  {publisher} {{WORLD SCIENTIFIC}},\ \bibinfo {year} {2019})\ pp.\ \bibinfo
  {pages} {519--581}\BibitemShut {NoStop}%
\bibitem [{\citenamefont {Van Den~Berg}(2021)}]{berg2021SimpleMethodSampling}%
  \BibitemOpen
  \bibfield  {author} {\bibinfo {author} {\bibfnamefont {E.}~\bibnamefont {Van
  Den~Berg}},\ }in\ \href {\doibase 10.1109/QCE52317.2021.00021} {\emph
  {\bibinfo {booktitle} {2021 IEEE International Conference on Quantum
  Computing and Engineering (QCE)}}}\ (\bibinfo {year} {Broomfield, CO, USA,
  2021})\ pp.\ \bibinfo {pages} {54--59}\BibitemShut {NoStop}%
\bibitem [{\citenamefont {Murty}\ and\ \citenamefont
  {Rath}(2014)}]{murty_liouville_2014}%
  \BibitemOpen
  \bibfield  {author} {\bibinfo {author} {\bibfnamefont {M.~R.}\ \bibnamefont
  {Murty}}\ and\ \bibinfo {author} {\bibfnamefont {P.}~\bibnamefont {Rath}},\
  }in\ \href {\doibase 10.1007/978-1-4939-0832-5_1} {\emph {\bibinfo
  {booktitle} {Transcendental {{Numbers}}}}},\ \bibinfo {editor} {edited by\
  \bibinfo {editor} {\bibfnamefont {M.~R.}\ \bibnamefont {Murty}}\ and\
  \bibinfo {editor} {\bibfnamefont {P.}~\bibnamefont {Rath}}}\ (\bibinfo
  {publisher} {{Springer}},\ \bibinfo {address} {{New York, NY}},\ \bibinfo
  {year} {2014})\ pp.\ \bibinfo {pages} {1--6}\BibitemShut {NoStop}%
\end{thebibliography}
%

\end{document}